\documentclass[a4paper,11pt]{article}

\usepackage{jheppub}
\makeatletter
\let\@fpheader      % To remove the JHEP header
\makeatother

\usepackage[utf8]{inputenc}
\usepackage[american]{babel}
\usepackage{xcolor}
\usepackage{subcaption}
\usepackage{multicol}
\usepackage{booktabs}
\usepackage{bigdelim}

\usepackage{hyperref}
\hypersetup{%
    colorlinks,%
    breaklinks,%
    citecolor=blue,urlcolor=blue,linkcolor=blue,%
    hypertexnames=false,%
    pdftitle={The isospin-3 three-particle K-matrix at NLO in ChPT},%
    pdfauthor={J. Baeza-Ballesteros, J. Bijnens, T. Husek, F. Romero-Lopez, S. R. Sharpe, M. Sjo}%
}
\makeatletter
\renewcommand\HyPsd@CatcodeWarning[1]{}     % To remove some warnings
\makeatother

\usepackage{cleveref}
\crefname{equation}{eq.}{eqs.}
\crefname{section}{sec.}{secs.}
\Crefname{equation}{Eq.}{Eqs.}
\Crefname{section}{Sec.}{Secs.}
\newcommand{\rcite}[1]{ref.~\cite{#1}}
\newcommand{\rrcite}[1]{refs.~\cite{#1}}

\usepackage{glossaries-extra}
\setabbreviationstyle[acronym]{long-short}
\glssetcategoryattribute{acronym}{nohyperfirst}{true}

\let\Re\relax\let\Im\relax
\DeclareMathOperator{\Re}{Re}
\DeclareMathOperator{\Im}{Im}

\usepackage{pgf,tikz}
\usetikzlibrary{%
    calc,
    decorations.pathreplacing,
    decorations.markings,
    decorations.pathmorphing,
    shapes.geometric,
    external%
}
\usepackage{pgfplots}
\pgfplotsset{compat=1.17}   % To remove some warnings
\newcommand{\arrowscale}{1.2}
\newcommand{\diagramxscale}{2}
\newcommand{\diagramyscale}{1.4}
\usepgfplotslibrary{fillbetween}
\usetikzlibrary{decorations.text}
\tikzset{
    on each segment/.style={
        decorate,
        decoration={
              show path construction,
              moveto code={},
              lineto code={
                \path [#1]
                (\tikzinputsegmentfirst) -- (\tikzinputsegmentlast);
              },
              curveto code={
                    \path [#1] (\tikzinputsegmentfirst)
                    .. controls
                    (\tikzinputsegmentsupporta) and (\tikzinputsegmentsupportb)
                    ..
                    (\tikzinputsegmentlast);
              },
              closepath code={
                    \path [#1]
                    (\tikzinputsegmentfirst) -- (\tikzinputsegmentlast);
              },
        },
    },
    mid arrow/.style={postaction={decorate,decoration={
        markings,
        mark=at position #1 with {\arrow[scale=\arrowscale]{stealth}}
      }}},
    mid arrow/.default=0.55,
    mid barrow/.style={postaction={decorate,decoration={
        markings,
        mark=at position #1 with {\arrowreversed[scale=\arrowscale]{stealth}}
      }}},
    mid barrow/.default=0.45,
    prop/.style={thick,join=round},
    dprop/.style={prop,postaction={on each segment={mid arrow=#1}}},
    bdprop/.style={prop,postaction={on each segment={mid barrow=#1}}},
    sketch onshell prop/.style={very thick},
    sketch offshell prop/.style={sketch onshell prop, densely dotted},
    sketch blob/.style={draw=black,thick, fill=#1, anchor=center},
    sketch onshell blob/.style={sketch blob=#1,
                                shape=regular polygon, regular polygon sides=4},
    sketch offshell blob/.style={sketch blob=#1,
                                shape=ellipse}
}
\newcommand{\makeexternallegcoordinates}{%
    \coordinate (k1) at (+1,+1);%
    \coordinate (k2) at (+1, 0);%
    \coordinate (k3) at (+1,-1);%
    \coordinate (p1) at (-1,+1);%
    \coordinate (p2) at (-1, 0);%
    \coordinate (p3) at (-1,-1);%
    }
\newcommand{\makeexternallegs}{%
    \makeexternallegcoordinates
    \draw (k1) node [right] {$k_1$};%
    \draw (k2) node [right] {$k_2$};%
    \draw (k3) node [right] {$k_3$};%
    \draw (p1) node [left] {$p_1$};%
    \draw (p2) node [left] {$p_2$};%
    \draw (p3) node [left] {$p_3$};%
    }
\newcommand{\makeexternallegsshifted}{%
    \makeexternallegcoordinates
    \draw (k1) node [right] {$k_3$};%
    \draw (k2) node [right] {$k_1$};%
    \draw (k3) node [right] {$k_2$};%
    \draw (p1) node [left] {$p_1$};%
    \draw (p2) node [left] {$p_2$};%
    \draw (p3) node [left] {$p_3$};%
    }

\definecolor{plotI}{HTML}{0077BB}
\definecolor{plotII}{HTML}{33BBEE}
\definecolor{plotIII}{HTML}{009988}
\definecolor{plotIV}{HTML}{EE7733}
\definecolor{plotV}{HTML}{CC3311}

\newcommand{\sketchoperatorscale}{1.6}
\definecolor{LOcolor_}{HTML}{6699CC}
\definecolor{NLOcolor_}{HTML}{EE99AA}
\colorlet{LOcolor}{LOcolor_!70}
\colorlet{NLOcolor}{NLOcolor_!70}

\definecolor{fitorange}{HTML}{FFA500}
\definecolor{fitblue}{HTML}{0000FF}
\definecolor{fitgray}{HTML}{808080}

\pgfmathsetlengthmacro\MajorTickLength{
      \pgfkeysvalueof{/pgfplots/major tick length} * 0.4
    }
\pgfmathsetlengthmacro\MinorTickLength{
      \pgfkeysvalueof{/pgfplots/minor tick length} * 0.3
    }

\tikzset{
    LOdash/.style={dash pattern=on 3pt off 2.5pt},
    fitdash/.style={dash pattern=on 1pt off 1.3pt},
    zeroline/.style={mark=none, black, ultra thin},   
    verticalline/.style={thin, black, dash pattern=on 3pt off 3.5pt},
    KXline/.style={thick},
    DXline/.style={ultra thick},
    LOline/.style={KXline, black,   LOdash},
    NLOline/.style={KXline, fitgray},
    NLOupperline/.style={line width = 0pt, fitgray},
    NLOlowerline/.style={line width = 0pt, fitgray},
    Fitline/.style={KXline, fitorange, fitdash},
    Fitupperline/.style={line width = 0pt, fitorange},
    Fitlowerline/.style={line width = 0pt, fitorange},
    NLO2line/.style={NLOline},
    NLO2upperline/.style={NLOupperline},
    NLO2lowerline/.style={NLOlowerline},
    NLOAline/.style={NLOline, LOdash},
    NLOAupperline/.style={NLOupperline},
    NLOAlowerline/.style={NLOlowerline},
    numericline/.style={KXline, red},
    threshline/.style={KXline, blue, dash pattern=on 3pt off 2.5pt},
    D0line/.style={DXline, plotI,   solid},
    D1line/.style={DXline, plotII,  dash pattern=on 8pt off 2pt on 2pt off 2pt},
    D2line/.style={DXline, plotIII, dash pattern=on 6pt off 2pt on 2pt off 2pt on 2pt off 2pt},
    DAline/.style={DXline, plotIV,  dash pattern=on 4.4pt off 4.4pt},
    DBline/.style={DXline, plotV,   dash pattern=on 2pt off 2pt},
    allwaveline/.style={DXline, black},
    swaveline/.style={D1line},
    dwaveline/.style={DAline},
    gwaveline/.style={DBline},
}
\pgfkeys{/pgf/number format/1000 sep={}}
\pgfplotsset{
    general plot/.style={
        set layers=axis on top,
        width=1.0\textwidth,
        height=0.75\textwidth,
        minor tick num=4,
        every tick/.style={
                semithick,
            },
        major tick length=\MajorTickLength,
        minor tick length=\MinorTickLength,
        y tick label style={
                /pgf/number format/.cd,
                fixed,
             fixed zerofill,
                precision=1,
                /tikz/.cd
                },
            every tick label/.append style={font=\footnotesize},
            axis line style={black, semithick},
        xticklabel style={inner sep=1.5pt},
        yticklabel style={inner sep=1.5pt},
            },
    fit plot/.style={
        general plot,
        ylabel style = {rotate=-90}, ylabel shift=-1.1ex,
        legend style={font=\scriptsize}
    },
    legend image code/.code={
        \draw [#1] (0pt,0pt) -- (15pt,0pt);
    },
    filled legend/.style={legend image code/.code={
        \draw [#1] (0pt,0pt) -- (15pt,0pt);
        \path [draw=none, fill=#1, opacity=0.3] (0pt,-4pt) rectangle (15pt,4pt);
    }},
    double filled legend/.style={legend image code/.code={
        \draw [#1] (0pt,5pt) -- (15pt,5pt);
        \path [draw=none, fill=#1, opacity=0.3] (0pt,1pt) rectangle (15pt,9pt);
        \draw [#1, solid] (0pt,-5pt) -- (15pt,-5pt);
        \path [draw=none, fill=#1, opacity=0.3] (0pt,-1pt) rectangle (15pt,-9pt);
    }},
    legend with mark/.style={legend image code/.code={
        \draw [#1, thick, solid] (7.5pt,4pt) -- (7.5pt,-4pt);
        \draw [#1, thick, solid, fill=white] (7.5pt,0pt) circle[radius=1.5pt];
    }},
    filled legend with mark/.style={legend image code/.code={
        \draw [#1] (0pt,0pt) -- (15pt,0pt);
        \path [draw=none, fill=#1, opacity=0.3] (0pt,-4pt) rectangle (15pt,4pt);
        \draw [#1, thick, solid] (7.5pt,4pt) -- (7.5pt,-4pt);
        \draw [#1, thick, solid, fill=white] (7.5pt,0pt) circle[radius=1.5pt];
    }},
    wide legend/.style={legend image code/.code={
        \draw [#1] (0pt,0pt) -- (22pt,0pt);
    }},
    lattice data/.style={
        thick,
        scatter, scatter/use mapped color={draw=#1,fill=white},
        only marks, mark=*,
        mark options={scale=0.7},
        error bars/.cd, y dir = both, x dir = both, y explicit, y explicit, error bar style={color=#1, thick, mark size = 0pt},
        /pgfplots/.cd
    },
    text along plot/.style 2 args={
        mark=none, draw=none, 
        decoration={text along path, text align=center, raise=1ex, 
            text={#1}, 
            text align={left indent=#2} },
        postaction={decorate}}
}

\newcommand{\tikzineq}[2][]{%
	\tikz[%
		anchor=base,%
		baseline={([yshift=-1.0ex]current bounding box.center)},%
		#1]{#2}%
	}

\newcommand{\breit}{\diamond} % frame where integrals are nice
\newcommand{\breitint}{\int^{(\breit)}}

\newcommand{\looptools}{{\em LoopTools}}
\newcommand{\cpp}{C\texttt{++}}

\newcommand{\cD}{\mathcal{D}}

\newcommand{\cK}{\mathcal{K}}
\newcommand{\cL}{\mathcal{L}}
\newcommand{\cM}{\mathcal{M}}
\newcommand{\cO}{\mathcal{O}}
\newcommand{\cQ}{\mathcal{Q}}
\newcommand{\cR}{\mathcal{R}}
\newcommand{\cS}{\mathcal{S}}
\newcommand{\cT}{\mathcal{T}}

\renewcommand{\d}{\mathrm{d}}
\newcommand{\bm}{\boldsymbol}
\newcommand{\bmh}[1]{\hat{\bm #1}}

\newcommand{\rr}{\mathrm{r}}
\newcommand{\OO}{\mathrm{O}}

\newcommand{\uu}{{(u,u)}}
\newcommand{\df}{\mathrm{df}}
\newcommand{\Mdf}{\cM_{\df,3}}
\newcommand{\Kdf}{\cK_{\df,3}}
\newcommand{\Kdfuu}{\Kdf^\uu}

\newcommand{\iso}{{\mathrm{iso}}}
\newcommand{\on}{{\mathrm{on}}}
\newcommand{\off}{{\mathrm{off}}}

\newcommand{\m}{M_\pi}
\newcommand{\Mpi}{\m}
\newcommand{\Mphys}{M_{\pi,\text{phys}}}
\newcommand{\F}{F_\pi}
\newcommand{\Fpi}{\F}
\newcommand{\Fphys}{F_{\pi,\text{phys}}}

\newcommand{\MF}[1]{\bigg(\frac{\Mpi}{\Fpi}\bigg)^{\!#1}}

\newcommand{\Kiso}{{\cK_0}}
\newcommand{\Kisoone}{{\cK_1}}
\newcommand{\Kisotwo}{{\cK_2}}
\newcommand{\KA}{{\cK_\mathrm{A}}}  
\newcommand{\KB}{{\cK_\mathrm{B}}}

\newcommand{\Diso}{{\cD_0}}
\newcommand{\Disoone}{{\cD_1}}
\newcommand{\Disotwo}{{\cD_2}}
\newcommand{\DA}{{\cD_\mathrm{A}}}
\newcommand{\DB}{{\cD_\mathrm{B}}}

\newcommand{\DeltaA}{{\Delta_\mathrm{A}}}
\newcommand{\DeltaB}{{\Delta_\mathrm{B}}}

\newcommand{\lrI}{\ell_1^\rr}
\newcommand{\lrII}{\ell_2^\rr}
\newcommand{\lrIII}{\ell_3^\rr}
\newcommand{\lrIV}{\ell_4^\rr}

\newcommand{\elliso}{{\ell_{(0)}^\rr}}
\newcommand{\ellisoone}{{\ell_{(1)}^\rr}}
\newcommand{\ellisotwo}{{\ell_{(2)}^\rr}}
\newcommand{\ellA}{{\ell_{(\mathrm{A})}^\rr}}
\newcommand{\ellB}{{\ell_{(\mathrm{B})}^\rr}}

\newcommand{\qk}{q_{2,k}^*}
\newcommand{\Ek}{E_{2,k}^*}
\newcommand{\qksq}{q_{2,k}^{*2}}
\newcommand{\Eksq}{E_{2,k}^{*2}}
\newcommand{\qp}{q_{2,p}^*}
\newcommand{\Ep}{E_{2,p}^*}
\newcommand{\qpsq}{q_{2,p}^{*2}}

\newcommand{\tij}[1]{\tilde t_{#1}}

\newacronym{CMF}{CMF}{center-of-momentum frame}
\newcommand{\LO}{\mathrm{LO}}
\newcommand{\NLO}{\mathrm{NLO}}
\newcommand{\OPE}{\mathrm{OPE}}
\newcommand{\nonOPE}{{\mathrm{non}\text{-}\OPE}} % Puttng most of it in mathrm rather than text ensures is stays non-bold in headers
\newcommand{\BH}{\mathrm{BH}}

\newcommand{\DuuBH}{\cD^{\uu\mathrm{BH}}}
\newcommand{\MuuBH}{\cM_{3}^{\uu\mathrm{BH}}}
\newcommand{\pp}{p_+}
\newcommand{\kk}{k_+}
\newcommand{\bpp}{\bm p_+}
\newcommand{\bkk}{\bm k_+}
\newcommand{\ppkk}{\Omega}

\newcommand{\principal}{\mathcal P}
\newcommand{\family}{\mathcal{F}}

\newcommand{\fakeblanton}{\cite{Blanton:2021llb}}

\preprint{{\small LU TP 23-03, MIT-CTP/5541}}

\title{The isospin-3 three-particle $K$-matrix at NLO in ChPT}

\author[a]{Jorge Baeza-Ballesteros,}
\author[b]{Johan Bijnens,}
\author[b,c]{Tom\'a\v s Husek,}
\author[d]{Fernando Romero-L\'opez,}
\author[e]{Stephen R.\ Sharpe,}
\author[b]{and Mattias Sj\"o}

\affiliation[a]{IFIC, CSIC-Universitat de València, 46980 Paterna, Spain}
\affiliation[b]{Department of Physics, Lund University, Box 118, SE 22100 Lund, Sweden}
\affiliation[c]{Institute of Particle and Nuclear Physics, Charles University,\\
	V Hole\v sovi\v ck\'ach 2, 180 00 Prague, Czech Republic}
\affiliation[d]{CTP, Massachusetts Institute of Technology, Cambridge, MA 02139, USA}
\affiliation[e]{Physics Department, University of Washington, Seattle, WA 98195-1560, USA}

\emailAdd{jorge.baeza@ific.uv.es}
\emailAdd{johan.bijnens@hep.lu.se}
\emailAdd{husek@ipnp.mff.cuni.cz}
\emailAdd{fernando@mit.edu}
\emailAdd{srsharpe@uw.edu}
\emailAdd{mattias.sjo@hep.lu.se}

\abstract{
The three-particle $K$-matrix, $\mathcal{K}_{\mathrm{df},3}$, is a scheme-dependent quantity that parametrizes short-range three-particle interactions in the relativistic-field-theory three-particle finite-volume formalism. In this work, we compute its value for systems of three pions at maximal isospin through next-to-leading order (NLO) in Chiral Perturbation Theory (ChPT). We compare the values to existing lattice QCD results and find that the agreement between lattice QCD data and ChPT in the first two coefficients of the threshold expansion of $\mathcal{K}_{\mathrm{df},3}$ is significantly improved with respect to leading order once NLO effects are incorporated.
}


\begin{document}

\maketitle
\flushbottom

\section{Introduction}

Lattice QCD provides a systematically improvable approach to calculate strongly-in\-ter\-act\-ing processes, including several that are inaccessible in experiment.
One example is the scattering of three particles, e.g., \mbox{$3 \pi^+\to 3\pi^+$}.
Experimentally, such processes can be very difficult due to the challenge of creating and scattering three beams and because all hadrons (except the nucleons) are short-lived.
By contrast, lattice QCD provides a method for obtaining multihadron scattering amplitudes and allows one to treat the strong force in isolation, thus rendering the lightest mesons (and some other hadrons) stable due to the absence of weak and electromagnetic interactions.

The extraction of scattering amplitudes from lattice QCD is a very active topic of research (see \rrcite{Briceno:2017max,Hansen:2019nir,Rusetsky:2019gyk,Horz:2022glt,Mai:2021lwb,Mai:2022eur,Romero-Lopez:2021zdo,Romero-Lopez:2022usb} for recent reviews), and, in particular, three-particle processes have recently received a lot of attention.
The formalism to extract these amplitudes from the three-particle finite-volume spectrum computed on the lattice has been developed over the last decade~\cite{Beane:2007qr,Detmold:2008gh,Briceno:2012rv,Polejaeva:2012ut,Hansen:2014eka,Hansen:2015zga,Briceno:2017tce,Konig:2017krd,Hammer:2017uqm,Hammer:2017kms,Mai:2017bge,Briceno:2018mlh,Briceno:2018aml,Blanton:2019igq,Pang:2019dfe,Jackura:2019bmu,Briceno:2019muc,Romero-Lopez:2019qrt,Hansen:2020zhy,Blanton:2020gha,Blanton:2020jnm,Pang:2020pkl,Romero-Lopez:2020rdq,Blanton:2020gmf,Muller:2020vtt,Blanton:2021mih,Muller:2021uur,Blanton:2021eyf,Jackura:2022gib,Muller:2022oyw,Draper:2023xvu}, using three main approaches, and has been applied to results from lattice simulations for a number of scattering amplitudes~\cite{Beane:2007es,Detmold:2011kw,Mai:2018djl,Horz:2019rrn,Blanton:2019vdk,Mai:2019fba,Culver:2019vvu,Fischer:2020jzp,Hansen:2020otl,NPLQCD:2020ozd,Alexandru:2020xqf,Brett:2021wyd,Blanton:2021llb,Mai:2021nul}.
So far, the system that has been most extensively explored is that of three pions at maximal isospin, i.e., \mbox{${3\pi^+ \to 3 \pi^+}$} scattering.
Several of these works (\rrcite{Blanton:2019vdk, Fischer:2020jzp, Hansen:2020otl, Blanton:2021llb}) use the relativistic-field-theory (RFT) three-particle finite-volume formalism, which parametrizes short-range three-body interactions via an intermediate cutoff-dependent quantity, the three-particle $K$-matrix, $\Kdf$.
As explained in \rcite{Hansen:2015zga}, $\Kdf$ is related to the physical three-particle amplitude, $\cM_3$, via integral equations.

An alternative approach to QCD is the use of effective field theories, with Chiral Perturbation Theory (ChPT)~\cite{Weinberg:1978kz,Gasser:1983yg} being the paradigm for meson dynamics at low energies.
Besides its many phenomenological applications, the synergy between ChPT and lattice QCD is indeed frequently exploited in the literature.
ChPT expressions allow one to address the quark-mass dependence, discretization effects, and finite-volume effects for certain quantities.
To highlight one example, the $\pi\pi$ scattering lengths can be very well constrained by combining lattice QCD results at heavier-than-physical pion masses with ChPT extrapolations to the real-world value of the mass; see the dedicated chapter in the FLAG report~\cite{FLAG:2021npn} and \rrcite{ETM:2015bzg,Mai:2019pqr,Fu:2017apw}.
Given the recent progress in three-particle scattering amplitudes \cite{Bijnens:2021hpq,Bijnens:2022zsq}, one hopes that a similar path can be followed for three-pion quantities.
So far, however, comparisons between ChPT and lattice QCD results for three-pion systems have only been qualitative.

The leading-order (LO) ChPT prediction for $\Kdf$ for the $3\pi^+$ system was determined in \rcite{Blanton:2019vdk}.
When compared to lattice QCD results from \rrcite{Blanton:2019vdk,Fischer:2020jzp,Blanton:2021llb}, however, a significant disagreement was observed.
This finding was surprising, given how well the two-particle counterpart, the maximal-isospin $\pi\pi$ $K$-matrix, is described by LO ChPT.
It is thus important to understand the cause of this discrepancy.
One source could be systematic errors in the lattice QCD calculation, since extracting $\Kdf$ from the finite-volume spectrum is numerically challenging as the shifts in the finite-volume energy levels from their free values are primarily determined by two-particle interactions.
Another source could be the importance of higher-order ChPT corrections.
In this work, we address the latter possibility by determining the next-to-leading-order (NLO) predictions for $\Kdf$.

The first step in this direction was carried out in \rrcite{Bijnens:2021hpq,Bijnens:2022zsq}, where the three-meson scattering amplitude, $\cM_3$, was determined to NLO for a number of mesonic effective theories, including the one relevant for pions, i.e., ChPT with two quark flavors.
It is, however, not obvious how to connect $\cM_3$ to $\Kdf$, since their relation, based on integral equations, needs to be inverted.
The aim of this work is to combine the results of \rcite{Bijnens:2021hpq} with the RFT approach to provide the NLO ChPT prediction for $\Kdf$.
We focus here on the case of three pions at maximal isospin, where most of the lattice QCD data is available.

The derivation of the RFT formalism in \rcite{Hansen:2014eka} leads to $\Kdf$ having the key properties of being real, smooth, and invariant under the same symmetries as $\cM_3$ (i.e.\ Lorentz, parity, and time-reversal symmetries).
In particular, all the branch cuts present in $\cM_3$ due to unitarity (two- and three-particle cuts), as well as the single-particle pole due to one-particle exchange (OPE), are absent in
$\Kdf$ by construction.%
\footnote{
    The absence of the OPE pole is the reason why this is denoted a df $=$ ``divergence-free'' quantity.}
An important aspect of an NLO ChPT calculation is that it can provide a check of these properties (and the RFT derivation) much more extensively than the LO result.

Since $\Kdf$ is smooth, it can be expanded about threshold, constrained only by the above-mentioned symmetries.
Such a ``threshold expansion'' is the three-particle analog of the effective-range expansion for the two-particle phase shift (or $K$-matrix).
It has been worked out for the $3\pi^+$ scattering amplitude up to quadratic order as an expansion in relativistic invariants.
As described in \rcite{Blanton:2019igq}, at this order there are only five unknown constants%
\footnote{
    In \rcite{Blanton:2019igq}, the constants are called $\cK_{\df,3}^{\iso}$, $\cK_{\df,3}^{\iso,1}$, $\cK_{\df,3}^{\iso,2}$, $\cK_{\df,3}^{(2,A)}$ and $\cK_{\df,3}^{(2,B)}$, respectively, where ``iso'' marks the coefficients of the isotropic terms.
    We have chosen to use an abbreviated notation here.}
in $\Kdf$, i.e., $\Kiso$, $\Kisoone$, $\Kisotwo$, $\KA$, and $\KB$; see \cref{eq:Kdfthrexp}.
The leading two orders, $\Kiso$ and $\Kisoone$, give rise only to isotropic terms, i.e., those that are independent of the angles between particles.
 Angular dependence enters through the $\KA$ and $\KB$ terms.
 At LO in ChPT, \rcite{Blanton:2019vdk} finds that all terms but $\Kiso$ and $\Kisoone$ vanish.
 At NLO, all five terms are expected to be nonzero, and our aim here is to determine their values.

Several technical complications need to be addressed to obtain the NLO result for $\Kdf$.
First, a relation between $\Kdf$ and $\cM_3$ that is valid at NLO in ChPT has to be established.
As we will show, both at LO and NLO, the relation between $\Kdf$ and $\cM_3$ is algebraic and linear, which simplifies the subsequent calculation.
Second, $\Kdf$ depends, in general, on a cutoff function, and is thus unphysical.
This cutoff function appears in the subtraction to cancel the aforementioned divergences in $\cM_3$, and evaluating its contribution requires a tailored numerical approach.
Third, a strategy to isolate the different terms in the threshold expansion of $\Kdf$ is needed, both in the contributions that are analytical as well as in those that can only be evaluated numerically.
Here, we succeeded in having an essentially fully analytical result checked by numerical calculations.

Once the NLO results for the different terms in the threshold expansion of $\Kdf$ have been worked out, we can compare to lattice QCD data.
In particular, we use the results of \rcite{Blanton:2021llb}, which provides values for different coefficients of the threshold expansion at three values of the pion mass.
As we will see, the agreement in $\cK_0$ and $\cK_1$ between lattice QCD results and ChPT is significantly improved once NLO effects are incorporated.
It is, however, interesting to note that NLO effects seem to be rather large, in particular in $\Kisoone$.

The remainder of this paper is structured as follows.
In \cref{sec:kdfchpt}, we provide the necessary background to compute $\Kdf$ in ChPT at NLO.
\Cref{sec:summary} then presents the central results of this paper, while we leave the technical part of the calculation to \cref{sec:calculation}.
Finally, some conclusions are presented in \cref{sec:conclusions}.
This paper contains 5 appendices, detailing the cutoff dependence of $\Kdf$ (\cref{app:Hdependence}) and the loop integrals in $\cM_3$ (\cref{app:loops}), verifying the cancellation of imaginary parts in $\Kdf$ (\cref{app:noimag}), and supplementing \cref{sec:threxpand-M3} (\cref{app:families}) and \cref{sec:bullhead} (\cref{app:MminusD}).

\section{The three-particle $K$-matrix from ChPT}
\label{sec:kdfchpt}

In this section, we provide the necessary background to compute $\Kdf$ at NLO in ChPT.
\Cref{sec:rolekdf} introduces $\Kdf$ and describes its role in the three-particle formalism, and  \cref{sec:M3Kdf} establishes its connection to $\cM_3$ at NLO in ChPT.
Then, an explicit calculation of $\Kdf$ at LO is provided in \cref{sec:explicitLO}, and the strategy to follow at NLO is outlined in \cref{sec:outlineNLO}.
We defer all technical details of the computation to \cref{sec:calculation}.

\subsection{The role of $\Kdf$ in the three-particle formalism}
\label{sec:rolekdf}

The three-particle finite-volume formalism for identical scalar particles without two-to-three transitions was derived in \rcite{Hansen:2014eka}.
In (isospin-symmetric) QCD, it applies for three pions or three kaons at maximal isospin, $I=3$.
We will focus here on the former case.
The central equation of the formalism is the quantization condition, whose solutions correspond to the energy levels $E_n$ of a three-pion system with total three-momentum $\bm P$ in a box of side $L$,
\begin{equation}
    \det\Bigl[ F_3^{-1}(E, \bm P, L) + \Kdf(E^*) \Bigr] = 0 \qquad\text{at}\qquad E = E_n\,.\label{eq:qc3}
\end{equation}
This is valid in the energy range  where only three-pion intermediate states can go on shell, i.e., \mbox{$\m< E^* < 5\m $},%
\footnote{
    In the following, $\m$ will denote the renomalized mass of the pions, i.e., $p_i^2=\m^2$ for on-shell pions of momenta $p_i$.
    In amplitude calculations, this is called the ``physical'' mass to distinguish it from the non-renormalized mass appearing in the Lagrangian.
    In the lattice community, the ``physical'' mass typically refers to the real-world value $\Mphys\approx139.570$\;MeV.
    In this work we will use the latter convention, and so $\m$ will in general be different from $\Mphys$.}
with $E^*$ the total energy in the center-of-mass frame (CMF).
$F_3$ is a quantity that depends on the volume, kinematic functions and two-particle interactions, and $\Kdf$ is a real, Lorentz-invariant and smooth function of $E^*$ that parametrizes short-range three-particle interactions.
Both quantities are matrices in a space that describes the kinematics of three on-shell particles, and the determinant is taken over those indices.
In particular, the choice in the RFT formalism is to describe the three-particle system as composed by a pair of particles with angular momentum indices $\ell$ and $m$, usually called the \textit{interacting pair} or \textit{dimer}, plus a third particle with three-momentum $\bm k$, called the {\textit{spectator}}.
Note that $\Kdf$ is a scheme-dependent quantity, with the scheme being determined by the choice of a cutoff function applied to the momentum of the spectator particle.
This function ensures that the matrices appearing in the quantization conditions have finite size.
For more details on the implementation of the formalism, we refer the reader to \rrcite{Briceno:2018mlh,Blanton:2019igq,Romero-Lopez:2019qrt,Blanton:2021eyf}.

 The key feature of \cref{eq:qc3} needed for this work is that, given a set of 3$\pi^+$ finite-volume energy levels, $\Kdf$ can be extracted by fitting the predicted spectrum to the measured one.
 For this, one needs a parametrization of $\Kdf$ in terms of few independent quantities.
 A systematic approach is to expand $\Kdf$ in terms of relativistic invariants organized by the distance to the three-particle threshold.
 To reduce the number of independent parameters, one can use the fact that $\Kdf$ has the same symmetries as the scattering amplitude, i.e., parity, time-reversal and particle-exchange symmetries.
 This leads to the threshold expansion worked out in \rcite{Blanton:2019igq}.
 As explained in that work, only five independent terms contribute to the expansion up to quadratic order:
\begin{equation}
    \m^2\Kdf = \Kiso + \Kisoone \Delta + \Kisotwo \Delta^2 + \KA \DeltaA + \KB \DeltaB + \cO(\Delta^3)\,,
    \label{eq:Kdfthrexp}
\end{equation}
where the following kinematic quantities have been defined:
\begin{equation}
    \begin{gathered}
        \Delta \equiv -\frac{1}{2}\sum_{i,j} \tij{ij}= \frac{P^2 - 9\m^2}{9\m^2}\,,
        \\ 
        \DeltaA \equiv \sum_i\left(\Delta_i^2+\Delta_i^{\prime 2}\right)-\Delta^2\,, 
        \qquad
        \DeltaB \equiv \sum_{i,j} \tij{ij}^2 -\Delta^2\,.
    \end{gathered}
    \label{eq:Deltas}
\end{equation}
Here, $P=(E,\bm P)$ is the total four-momentum of the system, and we define
\begin{equation}
    \begin{gathered}
        \tij{ij} \equiv \frac{\left(p_i-k_j\right)^2}{9\m^2}\,, 
        \\
        \Delta_j \equiv \sum_i \tij{ij} +\Delta = \frac{(P-k_j)^2 - 4\m^2}{9\m^2}\,,
        \\
        \Delta'_i \equiv \sum_j \tij{ij} +\Delta = \frac{(P-p_i)^2 - 4\m^2}{9\m^2}\,.
    \end{gathered}
    \label{eq:Deltas_}
\end{equation}
We choose $k_1,k_2,k_3$ to be the incoming momenta, and $p_1,p_2,p_3$ the outgoing ones, so that 
$P=k_1+k_2+k_3=p_1+p_2+p_3$ .
We reiterate that $\cK_X$ with $X = 0,1,2,\mathrm{A},\mathrm{B}$ are unknown, dimensionless constants to be determined.
As noted in the introduction, the only terms that lead to nontrivial dependence on the relative angles in the initial or final three-particle state are $\DeltaA$ and $\DeltaB$.
Moreover, only $\DeltaB$ leads to contributions with overall angular momentum differing from zero.

Once the coefficients of \cref{eq:Kdfthrexp} are determined from lattice QCD simulation, one has to connect the scheme-dependent $\Kdf$ to the physical scattering amplitude, $\cM_3$.
The relation between both quantities was derived in \rcite{Hansen:2015zga} and involves integral equations.
In this paper, we use that relation extensively, so as to obtain $\Kdf$ at a consistent order in ChPT provided the corresponding prediction for $\cM_3$.
Therefore, we reproduce here the key results from \rcite{Hansen:2015zga}.
For numerical solutions of the integral equations, see \rrcite{Jackura:2020bsk,Dawid:2023jrj}.

We begin by recalling the kinematic variables used in the RFT approach.
As already anticipated, the configuration of three on-shell particles is described by singling out one as a spectator with three-momentum $\bm k$, and boosting the interacting pair to their CMF, in which the two particles of the pair have three-momentum $\bm a_k^*$ and $-\bm a_k^*$, respectively.%
\footnote{
    Here $*$ indicates that quantities are expressed in the CMF of the interacting pair, and the subscript is used to emphasize that the quantity is expressed in the CMF of the pair for which the spectator has that particular momentum.}
The magnitude of these momenta is denoted $\qk \equiv |\bm a_k^*|$ and is given by
\begin{equation}
    \qksq = \frac14\big({\Eksq - 4 \m^2}\big)\,, \qquad\text{with} \quad 
    \Eksq = (P-k)^2\,,
    \label{eq:pair-magn}
\end{equation}
where $\Ek$ is the energy of the pair in their rest frame.
Expressing the initial-state kinematics this way, and expressing the final state analogously as a spectator and an interacting pair with three-momenta $\bm p$ and $\pm\bm a'^*_p$, respectively, the three-particle amplitude can be written as a function of the spectator momenta and the \emph{directions} of the pair momenta,
\begin{equation}
    \cM_3(\bm p,\bmh a'^*_p; \bm k, \bmh a^*_k)\,,
\end{equation}
since magnitudes are fixed by \cref{eq:pair-magn}; the hats denote unit vectors.
Here and in what follows, the dependence on $P$ is left implicit.
This description is somehow redundant, as it involves 10 (11 if we include $P^2$) variables, while there are only 8 independent kinematic quantities describing a general three-to-three process.
In particular, the rotational invariance of the amplitude in, say, the overall CMF is not taken into account.%
\footnote{
    A convenient set of 8 parameters, from which the 11-parameter set $\{P^2, \bm p,\bmh a'^*_p, \bm k, \bmh a^*_k\}$ is easy to obtain, is $\{E^*,|\bm p|,|\bm k|, \psi, \bmh a'^*_p, \bmh a^*_k\}$, where $E$ is the total CMF energy and $\psi$ is the angle between $\bm p$ and $\bm k$ in the overall CMF.
    The remaining three degrees of freedom correspond to rotations of the full system.}
Nevertheless, these variables are the natural choice in the RFT approach.

A further step is to decompose the angular dependence in the pair CMFs into spherical harmonics:%
\footnote{
    Later in this paper, we discuss the real and imaginary parts of quantities like $\cM_3$.
    When doing so, the fact that, in the standard basis, the spherical harmonics are complex should be ignored.
    One can show that the spherical harmonics arise as overall factors in the unitarity-like relations, and it is the imaginary part of the remainder of the expression that matters.
    To avoid this issue in numerical calculations, one can use the real spherical harmonics, as is done in present implementations of the three-particle quantization condition (see, e.g., \rcite{Blanton:2019igq}).}
\begin{equation}
    \cM_3(\bm p,\bmh a'^*_p; \bm k, \bmh a^*_k)
        = \sum_{\ell' m' \ell m} 4\pi Y^*_{\ell' m'}(\bmh a'^*_p) \cM_3(\bm p, \bm k)_{\ell' m'; \ell m} Y_{\ell m}(\bmh a^*_k)\,.
    \label{eq:sphharm}
\end{equation}
This is needed because, as will be seen shortly, some of the subtractions that appear in the definition of the divergence-free version of $\cM_3$ depend on the pair angular momenta.
Subsequent relations will be written for $\cM_3(\bm p, \bm k)_{\ell' m';\ell m}$, 
with the angular momentum indices treated as matrix indices and often left implicit.
Other three-particle quantities entering the following equations, such as $\Kdf$ itself, are also written in this hybrid notation.
For two-particle quantities, in which the spectator is unchanged, we follow \rcite{Hansen:2015zga} and label them with a single spectator momentum,%
\footnote{
    An alternative notation involving two variables and an adjustment of factors of the energy, $2\omega$, has been used in some subsequent works, e.g., \rcite{Blanton:2020gmf}.}
e.g.,
\begin{equation}
    \cM_2(\bm p)_{\ell' m';\ell m} = \delta_{\ell' \ell} \delta_{m' m} \cM_{2,\ell}(\qp)\,,
    \label{eq:M2def}
\end{equation}
and similarly for the two-particle phase-space factor,
\begin{equation}
    \rho(\bm p)_{\ell' m';\ell m} = \delta_{\ell' \ell} \delta_{m' m}\, \bar \rho(\qp)\,,
    \qquad
    \bar \rho(q_{2,p}^*) \equiv -i\,\frac{\qp}{16 \pi \Ep}\,,
    \label{eq:rhodef}
\end{equation}
which we only define above threshold, i.e., for $\Ep\geq2\m$, as this is all we need in this work.
A full definition of $\rho(\bm p)$ is given in \rcite{Hansen:2015zga}, including its subthreshold behavior, where it is real and cutoff-dependent.

We are now ready to define the divergence-free three-particle amplitude, $\Mdf$:
\begin{equation}
    \Mdf(\bm p,\bm k) = \cM_3(\bm p,\bm k) - \cS\Big\{ \cD^\uu(\bm p, \bm k)\Big\}\,.
    \label{eq:Mdfdef}
\end{equation}
Here, $\cS$ indicates symmetrization over choices of initial and final spectators (9 terms in total), while $\cD^\uu$ is the unsymmetrized  subtraction term, with $\uu$ indicating unsymmetrized for both final and initial states.
This symmetrization procedure is explained in detail in \rcite{Hansen:2015zga}, and we give only specific examples below.
It is important to keep in mind that $\cM_3$ and $\Kdf$ are, by definition, fully symmetrized quantities.

The subtraction term solves an integral equation that can be expanded in powers of $\cM_2$ to the order we need as
\begin{equation}
    \cD^\uu(\bm p, \bm k) = 
        - \cM_2(\bm p) G^\infty(\bm p, \bm k) \cM_2(\bm k)
        + \int_r \cM_2(\bm p) G^\infty(\bm p, \bm r) \cM_2(\bm r) G^\infty(\bm r, \bm k) \cM_2(\bm k)
        + \ldots\,,
    \label{eq:Duu-expand}
\end{equation}
where $\int_r \equiv \int d^3 r/[2\omega_r (2\pi)^3]$, with $\omega_r =\sqrt{\bm r^2+\m^2}$, and%
\footnote{
    Comparing to eq.~(81) of \rcite{Hansen:2015zga}, we note that here we use the relativistic form of the denominator, which is needed to obtain a relativistically invariant $\Kdf$.}
\begin{equation}
    G^\infty(\bm p, \bm k)_{\ell' m';\ell m} =
        \biggl(\frac{ k_p^*}{\qp}\biggr)^{\!\ell'}
        \frac{4\pi Y_{\ell' m'}(\bmh k_p^*) H(x_p) H(x_k) Y^*_{\ell m}(\bmh p_k^*)}{b_{pk}^2-\m^2 + i\epsilon}
        \biggl(\frac{ p_k^*}{\qk}\biggr)^{\!\ell}
        \,.
    \label{eq:Ginfdef}
\end{equation}
Here, $b_{pk} \equiv P - p - k$, and $H(x)$ is a smooth cutoff function that is 0 when $x\leq 0$ and $1$ when $x\geq 1$, with $x_k \equiv (P-k)^2/(4\m^2)$
and similarly for $x_p$.%
\footnote{
    In the range $0<x<1$, we let
    \begin{equation*}
        H(x) = \exp\Bigl[-\tfrac1x\, \exp\bigl(-\tfrac{1}{1-x}\bigr)\Bigr]
    \end{equation*}
    in accordance with, for example, eqs.~(28) and (29) of \rcite{Hansen:2014eka}, but any smooth function that interpolates between the constant values may be used.
    Some other choices of $H(x)$ are studied in \cref{app:Hdependence}.}
Note that $k_p^*$ refers to the magnitude of $\bm k$ taken in the CMF of the pair associated with $p$, and analogously for $p_k^*$.
We stress that the cutoff functions do not violate Lorentz symmetry, because $x_k$ and $x_p$ are both Lorentz invariant.

The factors of $k_p^*/\qp$ and $p_k^*/\qk$ in $G^\infty$ result from the analysis of the power-law volume-dependent contributions to the \emph{finite-volume} correlator.
In particular, it is important in that analysis that the dependence on $\bm k_p^*$ and $\bm p_k^*$ is smooth near threshold, and this requires the presence of the harmonic polynomials, e.g., $(k_p^*)^\ell Y_{\ell m}(\bmh k_p^*)$, rather than the spherical harmonics alone.
We stress that the factors of $k_p^*/\qp$ and $p_k^*/\qk$ imply that the sum over angular momentum indices cannot be performed analytically.

Some features of $\cD^\uu$ will be important below and we comment on them here.
First, it depends only on the \emph{on-shell} two-particle amplitudes.
Second, while in general one needs to keep the entire infinite series in $\cD^\uu$ to determine $\Mdf$, when working at some fixed order in a perturbative expansion such as ChPT, only a finite subset of the terms in \cref{eq:Duu-expand} appear.
Finally, we stress that $\cD^\uu$ depends on the choice of cutoff function, and is thus not a physical quantity.

With $\Mdf$ in hand, the full relation to $\Kdf$ is given by%
\footnote{
    Here and below, we have rearranged some factors of $2 \omega_r$ relative to \rcite{Hansen:2015zga} in order to simplify the notation.}
\begin{equation}
    \Mdf(\bm p, \bm  k) 
        = \cS\bigg\{\int_s \int_r 
            \cL^\uu(\bm p, \bm s) \cT(\bm s, \bm r) 
            \cR^\uu(\bm r, \bm k) \bigg\}\,,
    \label{eq:MdftoKdf}
\end{equation}
where
\begin{equation}
    \cL^\uu(\bm p, \bm k) 
        \equiv  \biggl[ \frac13 - \cM_2(\bm p) \rho(\bm p) \biggr] \bar\delta(\bm p - \bm k) 
        - \cD^\uu(\bm p, \bm k) \rho(\bm k)\,,
    \label{eq:Ldef}
\end{equation}
\begin{equation}
    \cR^\uu(\bm p, \bm k)  
        \equiv \bar\delta(\bm p - \bm k) \biggl[ \frac13 - \rho(\bm p) \cM_2(\bm p)  \biggr] 
        - \rho(\bm p) \cD^\uu(\bm p, \bm k)\,,
    \label{eq:Rdef}
\end{equation}
with $\bar\delta(\bm p - \bm k) \equiv 2\omega_k (2\pi)^3 \delta^{(3)}(\bm p - \bm k)$, and, finally,
\begin{equation}
    \cT(\bm p, \bm k) 
        \equiv \Kdf(\bm p, \bm k)
        - \int_s \int_r \Kdf(\bm p, \bm s) \rho(\bm s) \cL^\uu(\bm s, \bm r) \Kdf(\bm r, \bm k)
        + \dots
    \label{eq:Tdef}
\end{equation}
The last equation shows the first two terms in the expansion of the integral equation for $\cT$ in powers of $\Kdf$.

\subsection{Relation between $\cM_3$ and $\Kdf$ in ChPT}
\label{sec:M3Kdf}

ChPT describes the low-energy regime of QCD in terms of mesonic degrees of freedom.
It allows a perturbative determination  of mesonic observables  in terms of momenta and masses and as a function of some a priori unknown parameters---the so-called low-energy constants (LECs).
These can be determined from experiment or first principles (the latter usually via matching to lattice QCD).
We refer the unfamiliar reader to \rrcite{Scherer:2012xha,Pich:2018ltt} for an introduction to ChPT and to \rrcite{Bijnens:2021hpq,Bijnens:2022zsq} for brief summaries of what is needed for NLO three-meson scattering.
In this work, we restrict ourselves to the case of two-flavor ChPT.

Regarding the integral equations that are part of the RFT formalism, we can implement the usual ChPT power counting by expanding in powers of $1/\F^2$, where $\F$ is the pion decay constant.%
\footnote{
    Regarding what is considered ``physical'', $\F$ is treated the same way as $\m$.
    Its real-world value is $\Fphys\approx92.2$\;MeV.}
From here on, we will focus on the $3\pi^+$ system, and so $\cM_2$ and $\cM_3$ will refer to the two-particle $I=2$ and three-particle $I=3$ scattering amplitudes, respectively.
We note that $\cM_2=\cO(1/\F^2)$ and $\cM_3 = \cO(1/\F^4)$.
Since $\cL$ and $\cR$ begin at $\cO(1)$,
\begin{equation} \label{eq:LRLO}
    \cL^{\uu\LO}(\bm p, \bm k) 
        = \frac13 \bar\delta(\bm p - \bm k)
        = \cR^{\uu\LO}(\bm p, \bm k)\,,
\end{equation}
we have that $\Kdf = \cO(1/\F^4)$, and thus that $\cT^\LO = \Kdf^\LO$.
Putting this together and noting that the symmetrization of $\Kdf/9$ simply yields $\Kdf$, the LO version of \cref{eq:MdftoKdf} reduces to
\begin{align}
    \Kdf^\LO(\bm p, \bm k) &= \Mdf^\LO(\bm p, \bm k)
        = \cM_3^\LO(\bm p, \bm k) 
        - \cS\Big\{ \cD^{\uu\LO}(\bm p, \bm k) \Big\}\,,
    \label{eq:KdfLO}\\
    \cD^{\uu\LO}(\bm p, \bm k) &= - \cM_2^\LO(\bm p) G^\infty(\bm p, \bm k) \cM_2^\LO(\bm k)\,.
    \label{eq:DuuLO}
\end{align}
The second equation can be further simplified by noting that $\cM_2^\LO$ is purely $s$-wave, so we can make the replacement 
\begin{equation}
    G^\infty(\bm p, \bm k) \longrightarrow G_{ss}^\infty(\bm p, \bm k)\,, 
    \qquad
    G_{ss}^\infty(\bm p, \bm k)_{\ell' m'; \ell m} 
        \equiv \delta_{\ell' 0} \delta_{m' 0} \delta_{\ell 0} \delta_{m 0}\,
        \frac{H(x_p)H(x_k)}{b_{pk}^2 - \m^2 + i\epsilon}\,.
    \label{eq:Gss}
\end{equation}
We can also rewrite the expression for $\Kdf$ as 
\begin{equation}
    \Kdf^\LO(\bm p, \bm k) = \cS\Big\{ \Kdf^{\uu\LO}(\bm p, \bm k) \Big\}
        = \cS\Big\{ \cM_3^{\uu\LO}(\bm p, \bm k) - \cD^{\uu\LO}(\bm p, \bm k) \Big\}\,,
    \label{eq:KdfLOn}
\end{equation}
where $\cM_3^\uu$ is the unsymmetrized amplitude.
This form was used to perform the LO calculation of $\Kdf$ in \rcite{Blanton:2019vdk}, which we reproduce below in \cref{sec:explicitLO}.
We note that the subtraction produces a divergence-free quantity that is automatically real.

Moving to NLO, we need to keep terms up to $\cO(1/\F^6)$.
We note that the second term in the right-hand side of \cref{eq:Tdef} is $\cO(1/\F^8)$, so we still have $\cT = \Kdf$ at NLO.
Since $\cD^\uu = \cO(1/\F^4)$, the NLO expressions for $\cL^\uu$ and $\cR^\uu$ are equal and given by
\begin{equation}
    \cL^{\uu\NLO}(\bm p, \bm k) = 
        - \cM_2^\LO(\bm p) \rho(\bm p) \bar\delta(\bm p - \bm k)
        = \cR^{\uu\NLO}(\bm p, \bm k)\,.
\end{equation}
Here, we are adopting the notation, also used below, that NLO indicates the next-to-leading-order contribution alone rather than the sum of LO and NLO contributions.
Applying these results to \cref{eq:MdftoKdf}, we find
\begin{equation}
    \Mdf^\NLO(\bm p, \bm k) 
        = \Kdf^\NLO(\bm p, \bm k)
        - \frac{1}{3} \cS\Big\{
            \Kdf^\LO(\bm p, \bm k) \rho(\bm k) \cM_2^\LO(\bm k) 
            +  \cM_2^\LO(\bm p) \rho(\bm p) \Kdf^\LO(\bm p, \bm k)
            \Big\}\,.
\end{equation}
Using the equality of $\Kdf$ and $\Mdf$ at LO, this can be reorganized into
\begin{equation}
    \Kdf^\NLO(\bm p, \bm k) 
        = \Mdf^\NLO(\bm p, \bm k)  
        + \frac{1}{3} \cS\Big\{
            \Mdf^\LO(\bm p, \bm k) \rho(\bm k) \cM_2^\LO(\bm k) 
            +  \cM_2^\LO(\bm p) \rho(\bm p) \Mdf^\LO(\bm p, \bm k) 
            \Big\}\,.
    \label{eq:inteqNLO}
\end{equation}

The final quantity that we need is the NLO part of the subtraction term $\cD^\uu$, which is given by
\begin{multline}
    \cD^{\uu\NLO}(\bm p, \bm k) 
        = - \cM_2^\LO(\bm p) G^\infty(\bm p, \bm k) \cM_2^\NLO(\bm k)
        - \cM_2^\NLO(\bm p) G^\infty(\bm p, \bm k) \cM_2^\LO(\bm k) 
        \\
        + \int_r \cM_2^\LO(\bm p) G_{ss}^\infty(\bm p, \bm r) \cM_2^\LO(\bm r) G_{ss}^\infty(\bm r, \bm k) \cM_2^\LO(\bm k)\,.
    \label{eq:DuuNLO}
\end{multline}
Note that we have made the replacement $G^\infty \to G^\infty_{ss}$ in the final term since there all two-particle interactions are LO and thus purely $s$-wave.
Accordingly, this replacement cannot be made in the other two terms, since $\cM_2^\NLO$ contains all (even) partial waves.

The final equation we need to completely specify the NLO contribution to $\Kdf$ is
\begin{equation}
    \Mdf^\NLO(\bm p,\bm k) 
        = \cM_3^\NLO(\bm p,\bm k) 
        - \cS\Big\{ \cD^{\uu\NLO}(\bm p, \bm k) \Big\} \,.
    \label{eq:MdfdefNLO}
\end{equation}
The procedure is thus, given $\cM_3^\NLO$, to subtract $\cD^{\uu\NLO}$, \cref{eq:DuuNLO}, after symmetrization, and then add in the ``$\rho$ terms'' on the right-hand side of \cref{eq:inteqNLO}.
In fact, the latter are purely imaginary since both LO amplitudes are real, and thus, given the fact that $\Kdf$ is real, we obtain a simplified result
\begin{equation}
    \Kdf^\NLO(\bm p, \bm k) = \Re\Mdf^\NLO(\bm p, \bm k) \,.
    \label{eq:master}
\end{equation}
In principle, we do not need to calculate the $\rho$ terms.
However, an important cross-check on the formalism and the calculations can be obtained by showing explicitly that the imaginary part of $\Kdf$ vanishes based on the unitarity of off-shell amplitudes.
This is presented in \cref{app:noimag}.

\subsection{A note on off-shell conventions}
\label{sec:off-shell}
The subtraction term $\cD$ naturally separates into a part that cancels the OPE poles, namely the symmetrization of the first line of \cref{eq:DuuNLO}, and a remainder.
Below, we will find it useful to similarly separate $\cM_3$ into an OPE and a non-OPE part.
The same separation of $\Mdf$ and $\Kdf$ follows from this.
However, this separation (unlike that of $\cD$) is not unique.

Feynman diagrams can be categorized as either OPE (e.g., \cref{fig:FeynmanLO:OPE,fig:OPENLO}) or non-OPE (e.g., \cref{fig:FeynmanLO:contact,fig:nosubNLO,fig:BHNLO}), but the contribution of each diagram in ChPT depends on the parametrization of the Nambu--Goldstone manifold.%
\footnote{
    See appendix B of \rcite{Bijnens:2022zsq} for an in-depth discussion on parametrizations.}
This dependence must cancel when all contributions are summed into a physical amplitude, but separating based on diagrams does introduce parametrization dependence into the OPE and non-OPE parts.

Alternatively, one may view the OPE part as two four-point amplitudes with one leg off-shell, with the OPE propagator joining the off-shell legs.
The off-shell four-point amplitude can be modified by the addition of a smooth function of the kinematic variables that vanishes on shell, and the remainder of the complete amplitude is deferred to the non-OPE part.
Thus, we say that the separation is determined by an \emph{off-shell convention}.
Note that each parametrization will naturally give rise to a specific off-shell convention, as can be seen in \cref{app:MminusD}.

Here, we follow the off-shell convention of \rcite{Bijnens:2021hpq}, where the off-shell amplitude is defined by directly replacing the on-shell Mandelstam variables by their off-shell counterparts in a particular form of the on-shell four-pion amplitude.%
\footnote{
    This form is presented in eqs.~(18) and~(23) of \rcite{Bijnens:2021hpq} and repeated (for maximum isospin) in \cref{eq:A4piplus,eq:A4NLO} here. 
    Ultimately, it comes from the form used in \rcite{Bijnens:2011fm}, which is not entirely arbitrary but rather based on crossing symmetry considerations.}
In this approach, the off-shell amplitude is unique up to the freedom to rewrite the on-shell amplitude using $s+t+u=4\Mpi^2$, which does not hold off-shell.
This prescription leads to the same OPE and non-OPE parts independently of the underlying initial parametrization, and both parts are separately scale-independent.
Nevertheless, we stress that the separation into parts depends on a choice, and that only their sum is physical; also, contributions from individual diagrams remain parametrization dependent.
This is in contrast to the subtraction terms, which are unique up to the choice of cutoff function, since they are built from on-shell quantities.

\subsection{Explicit calculation of $\Kdf^\LO$ for $3\pi^+$ scattering}
\label{sec:explicitLO}

To illustrate the subtractions needed to obtain a divergence-free quantity, we work through the calculation of $\Mdf^\LO=\Kdf^\LO$ for the $3\pi^+$ system, the results of which were first presented in \rcite{Blanton:2019vdk};
see also the calculation of the corresponding quantities for $\pi^+\pi^+K^+$ and $\pi^+ K^+ K^+$ systems in \rcite{Blanton:2021eyf}.
This calculation also illustrates how intermediate results may depend on the convention used for off-shell amplitudes, as described in the previous section, although the final result should be (and is) independent of such conventions.

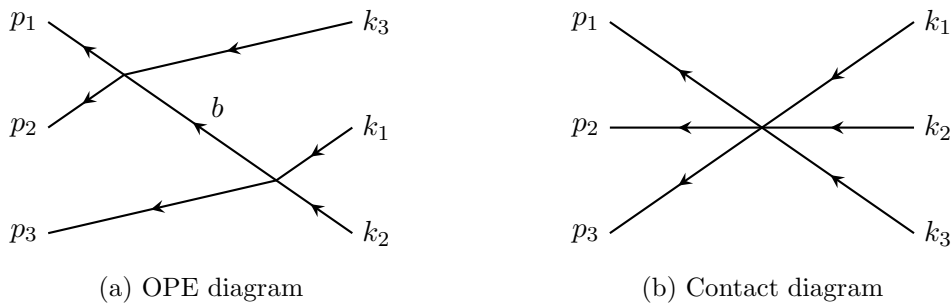
\begin{figure}[htp]
    \centering
    \begin{subfigure}{0.48\textwidth} 
        \centering
        \begin{tikzpicture}[xscale=\diagramxscale,yscale=\diagramyscale]
            \makeexternallegsshifted
            \coordinate (v1) at (+.5,-.5);
            \coordinate (v2) at (-.5,+.5);
            \draw[dprop] (k1) -- (v2) -- (p1);
            \draw[dprop] (k2) -- (v1) -- (v2) node [midway, above right] {$b$} -- (p2);
            \draw[dprop] (k3) -- (v1) -- (p3);
        \end{tikzpicture}
        \caption{OPE diagram}
        \label{fig:FeynmanLO:OPE}
    \end{subfigure}
    \begin{subfigure}{0.48\textwidth}
        \centering
        \begin{tikzpicture}[xscale=\diagramxscale,yscale=\diagramyscale]
            \makeexternallegs
            \coordinate (v) at (0,0);
            \draw[dprop] (k1) -- (v) -- (p1);
            \draw[dprop] (k2) -- (v) -- (p2);
            \draw[dprop] (k3) -- (v) -- (p3);
        \end{tikzpicture}
        \caption{Contact diagram}
        \label{fig:FeynmanLO:contact}
    \end{subfigure}
    \caption{
        Feynman diagrams contributing to $\cM_3$ at LO for maximal isospin.
        For diagram (a), there are an additional eight diagrams corresponding to the symmetrization of initial and final momenta.}
    \label{fig:FeynmanLO}
\end{figure}

To calculate $\Mdf^\LO$, we use \cref{eq:KdfLOn}.
Only two diagrams contribute to $\cM_3^\LO$, which are shown in \cref{fig:FeynmanLO}, and only the first of them, the OPE diagram, requires a subtraction, that is given by \cref{eq:DuuLO}.
We consider the OPE contributions first.

The results are simplified by the fact that $\cM_2^\LO$ is purely $s$-wave, so we need only keep the $\ell'=\ell=0$ contribution to $\Kdf$.
We obtain (with the $s$-wave limitation indicated by the subscripts $s$ and using the momentum labeling of \cref{fig:FeynmanLO})
\begin{multline}
    \cK_{\df,3, s}^{\uu\LO,\OPE}(\bm p_3, \bm k_1) \\
        = -\cM_{2,\off}^\LO(\bm p_3) \frac1{b^2-\m^2+i\epsilon} \cM_{2,\off}^\LO(\bm k_3)
        + \cM_{2s}^\LO(\bm p_3) G_{ss}^\infty(\bm p_3, \bm k_3) \cM_{2s}^\LO(\bm k_3)\,.
\end{multline}
Here, $\cM_{2,\off}^\LO$ is the two-particle amplitude with a single leg off shell.
Since both $p_3$ and $k_3$ are on shell, the $H$ functions in $G_{ss}^\infty$ both equal unity, and we can combine the terms to find the following result:
\begin{multline}
    \cK_{\df,3, s}^{\uu\LO,\OPE}(\bm p_3, \bm k_3) 
        =
        -\delta\cM_{2,\off}^\LO(\bm p_3)\frac1{b^2-\m^2+i\epsilon} \delta \cM_{2,\off}^\LO(\bm k_3)\\
        -\delta\cM_{2,\off}^\LO(\bm p_3) \frac1{b^2-\m^2+i\epsilon} \cM_{2s}^\LO(\bm k_3)
        -  \cM_{2s}^\LO(\bm p_3) \frac1{b^2-\m^2+i\epsilon} \delta\cM_{2,\off}^\LO(\bm k_3)\,,
    \label{eq:KdfLOOPE}
\end{multline}
where we define the difference between the off- and on-shell amplitudes,
\begin{equation}
    \delta\cM_{2,\off}^\LO(\bm p) =
    \cM_{2,\off}^\LO(\bm p)  - \cM_{2s}^\LO(\bm p) \,.
\end{equation}
As we will see explicitly below, this quantity is proportional to $b^2-\m^2$ and thus cancels the poles appearing in \cref{eq:KdfLOOPE}.

Using the results and notation of \rcite{Bijnens:2021hpq}, the $I=2$ $\pi\pi$ off-shell amplitude is
\begin{align}
    \cM_{2,\off}^\LO(\bm p_3) 
        &= A^{(2)}(t_2,u_2,s_2) + A^{(2)}(u_2,s_2,t_2)
    \label{eq:A4piplus}
        \notag\\
        &= \frac1{\F^2} (t_2 + u_2 - 2 \m^2)
        \\
        &= \frac1{\F^2} \bigl[-s_2 + 2 \m^2 + (b^2-\m^2)\bigr] = \frac1{\F^2} \bigl[-2 p_1\cdot p_2 + (b^2-\m^2)\bigr]\,,\notag
\end{align}
where we use the subscript $2$ on the Mandelstam variables to indicate that these are two-particle quantities, while $(b^2-\Mpi^2)$ is the off-shellness of one of the legs.
For example, using the labeling of momenta given in \cref{fig:FeynmanLO:OPE} and focusing on the left vertex, we have $s_2= (p_1+p_2)^2$, $t_2 = (k_3-p_1)^2$, and $u_2= (k_3-p_2)^2$, with $s_2+t_2+u_2 = 3 \m^2+b^2$.
The on-shell amplitude $\cM_{2s}^\LO$ is then obtained by setting $b^2=\m^2$.
Given these results, \cref{eq:KdfLOOPE} yields
\begin{equation}
    \F^4 \cK_{\df,3, s}^{\uu {\LO,\OPE}}(\bm p_3, \bm k_3) 
        =  2 p_1\cdot p_2 + 2 k_1\cdot k_2 - (b^2-\m^2) \,.
\end{equation}
To symmetrize, we use the following relations:
\begin{equation}
    \cS\{ 2 p_1\cdot p_2 \} = \cS\{2 k_1\cdot k_2 \} = 3P^2-9\m^2\,,
    \qquad
    \cS\{b^2-\m^2\} = 9 \m^2 -P^2\,,
\end{equation}
and we arrive at the final OPE contribution to $\Mdf^\LO$,
\begin{align}
    \F^4 \Mdf^{\LO,\OPE} &=  6(P^2 - 3\m^2)  - (9\m^2 - P^2)  = 7P^2 - 27 \m^2\,.
\end{align}

The contribution from the contact term (six-point vertex) of \cref{fig:FeynmanLO:contact} can be read off from eqs.~(30), (33), and~(34) of \rcite{Bijnens:2021hpq} which, however, describes the interaction of six pion fields $\pi^{f_1}(p_1)\pi^{f_2}(p_2)\pi^{f_3}(p_3)\pi^{f_4}(p_4)\pi^{f_5}(p_5)\pi^{f_6}(p_6)$, with flavor indices $f_i=1,2,3$, in the ``all-incoming'' convention.
In this language, our amplitude corresponds to the interaction of $\pi^+(k_1)\pi^+(k_2)\pi^+(k_3)\pi^-(-p_1)\pi^-(-p_2)\pi^-(-p_3)$.
The connection between our conventions and those of \rcite{Bijnens:2021hpq} is given by using $\pi^\pm=(\pi^1\pm i\pi^2)/\sqrt2$ and replacing $\{p_1,\ldots,p_6\}$ by $\{k_1,k_2,k_3,-p_1,-p_2,-p_3\}$.
Thus,
\begin{align}
    \Mdf^{\LO,6\mathrm{pt}} 
        &= A^{(2)}(k_1, -p_1, k_2, -p_2, k_3, -p_3)
        + A^{(2)}(k_1, -p_2, k_2, -p_1, k_3, -p_3)
        \notag\\
        &+ A^{(2)}(k_1, -p_1, k_2, -p_3, k_3, -p_2)
        + A^{(2)}(k_1, -p_2, k_2, -p_3, k_3, -p_1)
        \notag\\
        &+ A^{(2)}(k_1, -p_3, k_2, -p_1, k_3, -p_2)
        + A^{(2)}(k_1, -p_3, k_2, -p_2, k_3, -p_1)\,,
    \label{eq:6pt}
\end{align}
with
\begin{equation}
    \F^4 A^{(2)}(k_1, -p_1, k_2, -p_2, k_3, -p_3) 
        = -2 k_1\cdot p_1 -2 k_2 \cdot p_2 - 2 k_3 \cdot p_3 + 3 \m^2\,.
\end{equation}
Here, no subtraction or symmetrization is needed.
Summing all six terms, we find
\begin{align}
    \F^4 \Mdf^{\LO,6\mathrm{pt}} &= -4 P^2 +18 \Mpi^2 \,.
\end{align}

Combining the OPE and contact terms, we finally obtain
\begin{equation}
    \F^4 \Mdf^\LO = 3P^2 -9 \m^2 =  \m^2 ( 18 + 27 \Delta)\,,
    \label{eq:result-LO}
\end{equation}
which agrees with the result of \rcite{Blanton:2019vdk}.
However, we note that the values for the separate OPE and contact contributions are not the same as in that reference, due to our different off-shell conventions.

\subsection{Procedure to calculate $\Kdf$ at NLO}

\label{sec:outlineNLO}

\begin{figure}[bhtp]
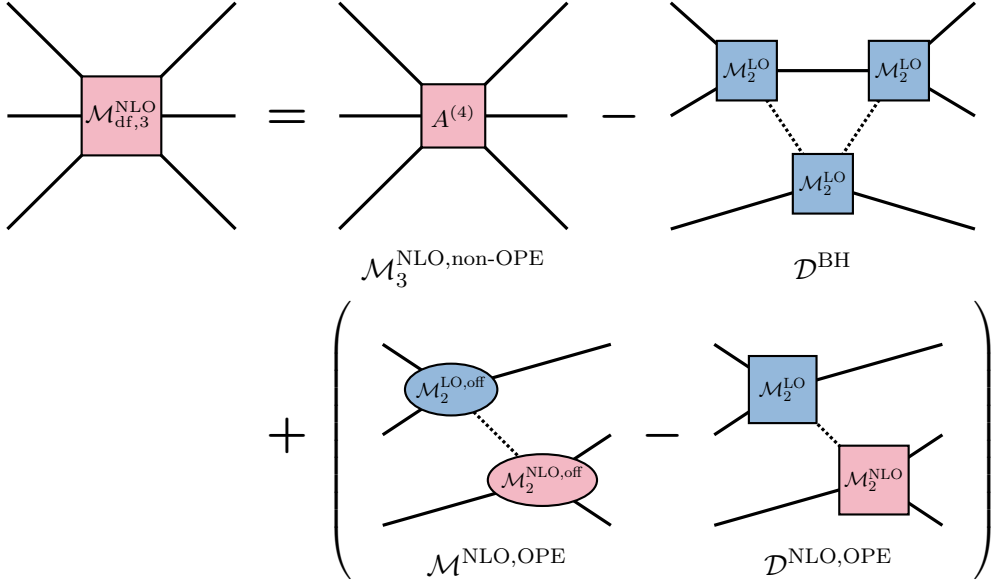

    % Please ignore various errors in the editor here -
    % my somewhat arcane nesting of math and not-math confuses Overleaf,
    % but the LaTeX engine has no issue with it.
    \begin{align*}
            \tikzineq[xscale=1.5,yscale=1.5]{
                \makeexternallegcoordinates
                \coordinate (v) at (0,0);
                \draw[sketch onshell prop] (k1) -- (v);
                \draw[sketch onshell prop] (k2) -- (v);
                \draw[sketch onshell prop] (k3) -- (v);
                \draw[sketch onshell prop] (v) -- (p1);
                \draw[sketch onshell prop] (v) -- (p2);
                \draw[sketch onshell prop] (v) -- (p3);
                \node[sketch onshell blob=NLOcolor, inner sep=-3pt]
                    (M) at (0,0) {\footnotesize $\Mdf^\NLO$};
            }
        &\quad\scalebox{\sketchoperatorscale}{$\bm=$}\quad
        \underset{\rule{0pt}{1.5em}\displaystyle\cM_3^{\NLO,\nonOPE}}{%
            \tikzineq[xscale=1.5,yscale=1.5]{
                \makeexternallegcoordinates
                \coordinate (v) at (0,0);
                \draw[sketch onshell prop] (k1) -- (v);
                \draw[sketch onshell prop] (k2) -- (v);
                \draw[sketch onshell prop] (k3) -- (v);
                \draw[sketch onshell prop] (v) -- (p1);
                \draw[sketch onshell prop] (v) -- (p2);
                \draw[sketch onshell prop] (v) -- (p3);
                \node[sketch onshell blob=NLOcolor, inner sep=0pt]
                    (M) at (0,0) {\footnotesize $A^{(4)}$};
            }
        }
        \quad\scalebox{\sketchoperatorscale}{$\bm-$}\quad
        \underset{\rule{0pt}{1.5em}\displaystyle\cD^\BH}{%
            \tikzineq[xscale=2,yscale=1.5]{
                \makeexternallegcoordinates
                \coordinate (v1) at (+.5,+.4);
                \coordinate (v2) at (-.5,+.4);
                \coordinate (v3) at (  0,-.6);
                \draw[sketch onshell prop] (k1) -- (v1);
                \draw[sketch onshell prop] (k2) -- (v1);
                \draw[sketch onshell prop] (k3) -- (v3);
                \draw[sketch onshell prop] (v2) -- (p1);
                \draw[sketch onshell prop] (v2) -- (p2);
                \draw[sketch onshell prop] (v3) -- (p3);
                \draw[sketch onshell prop] (v1) -- (v2);
                \draw[sketch offshell prop] (v1) -- (v3);
                \draw[sketch offshell prop] (v2) -- (v3);
                \foreach \i in {1,2,3}
                    \node[sketch onshell blob=LOcolor, inner sep=-1pt]
                        (M\i) at (v\i) {\scalebox{.7}{$\cM_2^\LO$}};
            }
        }
        \\
        &\quad\scalebox{\sketchoperatorscale}{$\bm+$}\;\,
        \left(\quad
        \underset{\rule{0pt}{1.5em}\displaystyle\cM^{\NLO,\OPE}}{%
            \tikzineq[xscale=1.5,yscale=1.2]{
                \makeexternallegcoordinates
                \coordinate (v1) at (+.4,-.5);
                \coordinate (v2) at (-.4,+.5);
                \draw[sketch onshell prop] (k1) -- (v2);
                \draw[sketch onshell prop] (k2) -- (v1) -- (k3);
                \draw[sketch onshell prop] (p3) -- (v1);
                \draw[sketch onshell prop] (p1) -- (v2) -- (p2);
                \draw[sketch offshell prop] (v1) -- (v2);
                \node[sketch offshell blob=NLOcolor, inner ysep=2pt, inner xsep=-1pt]
                    (NLO) at (v1) {\scalebox{.7}{$\cM_2^{\NLO,\off}$}};
                \node[sketch offshell blob=LOcolor, inner ysep=2pt, inner xsep=-1pt]
                    (LO) at (v2) {\scalebox{.7}{$\cM_2^{\LO,\off}$}};
            }
        }
        \quad\scalebox{\sketchoperatorscale}{$\bm-$}\quad
        \underset{\rule{0pt}{1.5em}\displaystyle\cD^{\NLO,\OPE}}{%
            \tikzineq[xscale=1.5,yscale=1.2]{
                \makeexternallegcoordinates
                \coordinate (v1) at (+.4,-.5);
                \coordinate (v2) at (-.4,+.5);
                \draw[sketch onshell prop] (k1) -- (v2);
                \draw[sketch onshell prop] (k2) -- (v1) -- (k3);
                \draw[sketch onshell prop] (p3) -- (v1);
                \draw[sketch onshell prop] (p1) -- (v2) -- (p2);
                \draw[sketch offshell prop] (v1) -- (v2);
                \node[sketch onshell blob=LOcolor, inner sep=0pt]
                    (LO) at (v2) {\scalebox{.7}{$\cM_2^{\LO}$}};
                \node[sketch onshell blob=NLOcolor, inner sep=-2pt]
                    (NLO) at (v1) {\scalebox{.7}{$\cM_2^{\NLO}$}};
            }
        }
        \quad\right)
    \end{align*}
    \caption{
        Sketch of \cref{eq:sketch}.
        Solid lines represent on-shell pions, while dotted lines are off-shell propagators.
        Square boxes indicate fully on-shell amplitudes, while oval boxes have one leg off shell (factors of $G^\infty$ ensure only on-shell amplitudes are needed  in $\cD$).
        Finally, blue and pink filling indicate, respectively, LO and NLO quantities.
        We leave implicit that we take only the real parts of all quantities.}
    \label{fig:sketch}
\end{figure}

Thanks to \cref{eq:master}, the determination of our main result, $\Kdf^\NLO$, is equivalent to calculating $\Re\Mdf^\NLO$.
We subdivide the calculation  into multiple pieces according to the following schematic equation, which is also represented graphically in \cref{fig:sketch}:%
\footnote{
    Colorblind- and monochrome-safe colors, courtesy of P.~Tol (\url{https://personal.sron.nl/~pault/}), are used in \cref{fig:sketch,fig:K_NLO_OPE_l,fig:alpha}.}
\begin{equation}
    \Re\Mdf^\NLO = \Re\cM_3^{\NLO,\nonOPE} - \Re\cD^{\BH} + \Re\Bigl\{\cM_3^{\NLO,\OPE}-\cD^{\NLO,\OPE}\Bigr\}\,.
    \label{eq:sketch}
\end{equation}
Employing this separation, the one-particle-reducible diagrams contribute to the term $\cM_3^{\NLO,\OPE}$, while the remainder of the full $I=3$ amplitude is denoted $\cM_3^{\NLO,\nonOPE}$.
Recall from \cref{sec:off-shell} that $\cM_3^{\NLO,\nonOPE}$ and $\cM_3^{\NLO,\OPE}$, while parametrization- and scale-independent, depend on our choice of off-shell convention.
    
We find that the real part of $\cM_3^{\NLO,\nonOPE}$ is smooth at threshold, so it can be expanded directly without first subtracting any divergences, unlike for the imaginary part.
Since the non-OPE contribution is smooth, the corresponding part of the subtraction must also be smooth, and can be calculated individually.
This contribution, given by the third term in \cref{eq:DuuNLO}, is the subtraction corresponding to the ``bull's head'' (BH) triangle diagram shown below in \cref{fig:BHNLO:regular}.
In \cref{eq:sketch}, $\cD^{\BH}$ is the symmetrization of this term.
It is the only piece that depends on the cutoff function $H(x)$.

Calculating each of these pieces leads to distinctive difficulties and methods of solution.
Since the full calculation is rather long and technical, we first present the main results in \cref{sec:summary}, and then describe the full calculation in \cref{sec:calculation}:
In \cref{sec:genthreshKdf}, we provide a general form of the threshold expansion of $\Kdf$;
\cref{sec:threxpand-M3} then covers the full threshold expansion of $\Re\cM_3^{\NLO,\nonOPE}$, \cref{sec:bullhead} deals with $\Re\cD^{\BH}$, and \cref{sec:OPE} covers the $\Re\bigl\{\cM_3^{\NLO,\OPE}-\cD^{\NLO,\OPE}\bigr\}$ term.
All pieces can be computed analytically except $\Re\cD^{\BH}$.
Since the real part is applied term-by-term in \cref{eq:sketch}, each of \cref{sec:threxpand-M3,sec:bullhead,sec:OPE} gives a real, finite result.
We explicitly verify the cancellation of imaginary parts in \cref{app:noimag}.

\section{Summary of results}
\label{sec:summary}

We now present, in \cref{sec:expressions}, the threshold expansion of $\cK_\text{df,3}$ at NLO in ChPT for the 3$\pi$ system at maximal isospin, which constitutes the main result of this work.
Some of the coefficients are then compared against lattice results from \rcite{Blanton:2021llb} in \cref{sec:comparisonlattice}.
Finally, we compare the threshold expansion to the full NLO result for a particular kinematic configuration in \cref{sec:validity}.

\subsection{Complete results}
\label{sec:expressions}

Including both LO and NLO contributions from ChPT, i.e., combining \cref{eq:result-LO,eq:fullnonOPE,eq:fullBH,eq:fullOPE}, the results are
\begin{subequations}
    \begin{alignat}{2}
        \Kiso 
            &= \MF4 18
            &\;+\;& \MF6 \biggl[- 3\kappa(35 + 12\log3) - \Diso + 111L + \elliso\biggr]\,, 
            \label{eq:resultsK0}\\
        \Kisoone 
            &= \MF4 27   
            &\;+\;& \MF6 \biggl[ - \frac{\kappa}{20}(1999 + 1920\log3) - \Disoone + 384L + \ellisoone \biggr]\,,
            \label{eq:resultsK1}\\
        \Kisotwo 
            &= 
            &&  \MF6 \biggl[\frac{207\kappa}{1400}(2923 - 420\log3) -\Disotwo + 360L +  \ellisotwo\biggr]\,,
            \label{eq:resultsK2}\\
        \KA 
            &=
            && \MF6 \biggl[\frac{9\kappa}{560}(21809 - 1050\log3) - \DA - 9L + \ellA\biggr]\,,
            \label{eq:resultsKA}\\
        \KB 
            &= 
            && \MF6 \biggl[\frac{27\kappa}{1400}(6698 - 245\log3) - \DB + 54L + \ellB\biggr]\,.
            \label{eq:resultsKB}
    \end{alignat}
    \label{eq:results}%
\end{subequations}
Here, $\kappa \equiv 1/(16\pi^2)$, $L\equiv\kappa\log(\Mpi^2/\mu^2)$, $\cD_X$ are cutoff-dependent numerical constants related to the bull's head subtraction (see \cref{sec:bullhead}),%
\footnote{
    All digits shown are exact, and we have computed the values numerically to much higher precision, with the first twenty digits being verified by at least two independent methods.
    Higher-precision values and codes for calculating them are available upon request.}
\begin{equation}
    \begin{gathered}
        \Diso       \approx -0.0563476589\,,\qquad
        \Disoone    \approx 0.129589681\,,\qquad
        \Disotwo    \approx 0.432202370\,,\\
        \DA         \approx 9.07273890\cdot10^{-4}\,,\qquad
        \DB         \approx 1.62394747\cdot10^{-4}\,,
    \end{gathered}
    \label{eq:Dthrexp}
\end{equation}
and we have defined the following linear combinations of LECs:
\begin{equation}
    \begin{gathered}
        \elliso     = -288\lrI-432\lrII-36\lrIII+72\lrIV\,, \qquad
        \ellisoone  = -612\lrI-1170\lrII+108\lrIV\,, \\
        \ellisotwo  = -432\lrI-864\lrII\,, \qquad
        \ellA       = 27\lrI+\frac{27}{2}\lrII\,, \qquad
        \ellB       = -162\lrI-81\lrII\,.
    \end{gathered}
    \label{eq:LECthrexp}
\end{equation}
A numerical comparison of the different contributions to each coefficient at the physical point is given in \cref{tab:breakdown}.
\begin{table}[tb]
\setlength{\tabcolsep}{7.5pt}
    \centering
    \begin{tabular}{l*{5}{r@{.}l}}
        \toprule
            &   \multicolumn{2}{c}{$\big(\frac{\Fpi}{\Mpi}\big)^{\!6}\Kiso$}
            &   \multicolumn{2}{c}{$\big(\frac{\Fpi}{\Mpi}\big)^{\!6}\Kisoone$}
            &   \multicolumn{2}{c}{$\big(\frac{\Fpi}{\Mpi}\big)^{\!6}\Kisotwo$}
            &   \multicolumn{2}{c}{$\big(\frac{\Fpi}{\Mpi}\big)^{\!6}\KA$}
            &   \multicolumn{2}{c}{$\big(\frac{\Fpi}{\Mpi}\big)^{\!6}\KB$}
        \\
        \midrule
        non-OPE
            &   $-$2&04(28) &   $-$3&75(61) &   1&43(37)    &   3&00(14)    &   0&25(28)    \\
        OPE 
            &   0&50(53)    &   $-$1&8(1.0)&   $-$5&11(58) &   $-$2&76(15) &   $-$0&22(37) \\
        BH,  excl.\ $\cD_X$
            &   $-$1&16234   &   $-$3&35289   &   $-$1&67334   &   1&97425      &   0&08225     \\
        BH, only $\cD_X$
            &   0&05635      &   $-$0&12959   &   $-$0&43220   &   $-$0&00091   &   $-$0&00016  \\
        \cmidrule{1-11}
        Total NLO
            &   $-$2&65(26) &   $-$9&04(46) &   $-$5&79(24) &   2&212(16)   &   0&118(93)   \\
        \bottomrule
        \toprule
            &   \multicolumn{2}{c}{$\Kiso$}
            &   \multicolumn{2}{c}{$\Kisoone$}
            &   \multicolumn{2}{c}{$\Kisotwo$}
            &   \multicolumn{2}{c}{$\KA$}
            &   \multicolumn{2}{c}{$\KB$}
            \\
        \midrule
        LO  
            &   94&5186     &   141&778     &   \multicolumn{1}{r@{\phantom{.}}}{0}&
                                            &   \multicolumn{1}{r@{\phantom{.}}}{0}&
                                            &   \multicolumn{1}{r@{\phantom{.}}}{0}&   
                                           \\
        NLO 
            &   $-$31&9(3.1)   &  $-$108&8(5.5)   &   $-$69&6(2.9)   &   26&62(19)   &   1&4(1.1) \\
        \midrule
        Total   
            &   62&6(3.1)  &   34&0(5.5)      &   $-$69&6(2.9)   &   26&62(19)   &   1&4(1.1) \\
        \bottomrule
    \end{tabular}
    \caption[]{
        Numerical comparison of the different contributions to $\Kdf$ presented in \cref{eq:results}, evaluated at the physical point, $\Mphys=139.57$\;MeV, $\Fphys=92.2$\;MeV.
        Errors inherited from the LECs, as listed in \cref{eq:LECref}, are given where applicable; the BH (bull's head subtraction) numbers are exact up to rounding.
        The top part of the table covers the NLO contributions, with factors of $\Mpi/\Fpi$ removed.
        In the bottom part, these factors are included so that LO and NLO can be compared; the uncertainty of $\Mpi/\Fpi$ is not taken into account.
        The LO piece comes from \cref{eq:result-LO}.
        Of the NLO pieces, non-OPE comes from \cref{eq:fullnonOPE}, OPE from \cref{eq:fullOPE}, and BH from \cref{eq:fullBH}, with the numerical residues $\cD_X$, also given in \cref{eq:Dthrexp}, separated out.}
    \label{tab:breakdown}
\end{table}
Above, $\ell_i^\rr\equiv \ell_i^\rr(\mu)$ are scale-dependent LECs, with $\mu$ being the renormalization scale.
Different scales are related via
\begin{equation}
    \ell_i^\rr(\mu_2)=\ell_i^\rr(\mu_1)+\frac{\gamma_i\kappa}{2}\log\frac{\mu_1^2}{\mu_2^2}\,,
\end{equation}
with
\begin{equation}
    \gamma_1=1/3\,,\qquad\gamma_2=2/3\,,\qquad\gamma_3=-1/2\,,\qquad\gamma_4=2\,.
\end{equation}
In combination with the $L$ terms, this ensures the scale independence of the results in \cref{eq:results} and thus of $\Kdf$.
Often, scale-independent variants of the LECs, $\bar \ell_i$, are used.
They are related to $\ell_i^\rr$ via
\begin{equation}\label{eq:lbar}
    \ell^\rr_i(\mu)
    =\kappa\,\frac{\gamma_i}{2}\Bigg(\bar{\ell}_i+\log\frac{\Mphys^2}{\mu^2}\Bigg)\,,
\end{equation} 
where $\Mphys\approx139.57$\;MeV is the real-world pion mass.
The $\bar \ell_i$ are fairly well known, either from phenomenology or from lattice QCD.

\subsection{Comparison to lattice results}
\label{sec:comparisonlattice}

We are now in position to compare our results to lattice determinations of $\Kdf$.
Several works~\cite{Blanton:2019vdk, Fischer:2020jzp, Hansen:2020otl, Blanton:2021llb} have applied the RFT formalism to the study of three pions at maximal isospin, and in all cases similar qualitative disagreement with LO ChPT predictions was found.
We will see that this can be explained by NLO ChPT contributions.
In particular, we compare to \rcite{Blanton:2021llb}, which studied the scattering process for pion masses of 200, 280 and 340\;MeV.
Note this is the only work in which $\Kdf$ has been (partially) determined to quadratic order.
The disagreement with LO ChPT in $\Kiso$ and $\Kisoone$ is seen also in \rcite{Fischer:2020jzp} and is resolved similarly at NLO, but we do not include this in the plots due to its large uncertainties.

We take the following reference values for the scale-independent LECs,
\begin{equation}
    \bar\ell_1 = -0.4(6)\,,\qquad \bar\ell_2 = 4.3(1)\,,\qquad \bar\ell_3 = 3.07(64)\,,\qquad \bar\ell_4 = 4.02(45)\,,
    \label{eq:LECref}
\end{equation}
where $\bar\ell_1$ and $\bar\ell_2$ are determined by combining experiment, ChPT and dispersion relations \cite{Colangelo:2001df}, while $\bar\ell_3$ and $\bar\ell_4$ come from the averaged $N_\mathrm{f}=2+1$ lattice QCD results~\cite{FLAG:2021npn}, based on \rrcite{MILC:2010hzw, Borsanyi:2012zv, Budapest-Marseille-Wuppertal:2013vij, Boyle:2015exm, Beane:2011zm}.
We also take into account correlations between $\bar \ell_1$ and $\bar \ell_2$ using the covariance matrix from \rcite{Colangelo:2001df}:
\begin{equation}
    \mathrm{Cov}(\bar\ell_1,\bar\ell_2)=
    \begin{pmatrix}
    0.35 & -0.033\\
    -0.033 & 0.012
    \end{pmatrix}.
\end{equation}

Although the results in \cref{eq:results} are independent of the choice of $\mu$, it is customary in lattice computations to choose $\mu$ such that the results depend only on $\m/\F$.
In our case this can be achieved by taking $\mu=4\pi\F$ in the $L$ terms of \cref{eq:results}, and approximating it as $\mu\approx 4\pi\Fphys$ in \cref{eq:lbar}.
This approximation only impacts the results for $\Kdf$ at NNLO since $\F$ is independent of $\m$ at LO.
With this choice, for example, \cref{eq:resultsK0} can be rewritten as
\begin{equation}
    \Kiso 
        = \MF4 18
        \;+\; \MF6 \biggl[ - 3\kappa(35 + 12\log3) - \Diso + 111\kappa\log\frac{\xi}{\xi_\text{phys}} + \kappa\bar{\ell}_{(0)}\biggr]\,, 
\end{equation}
with $\bar{\ell}_{(0)}=-48\bar\ell_1-144\bar\ell_2+9\bar\ell_3+72\bar\ell_4$, $\xi\equiv\m^2/(4\pi\F)^2$ and $\xi_\text{phys}\equiv\Mphys^2/(4\pi\Fphys)^2$.

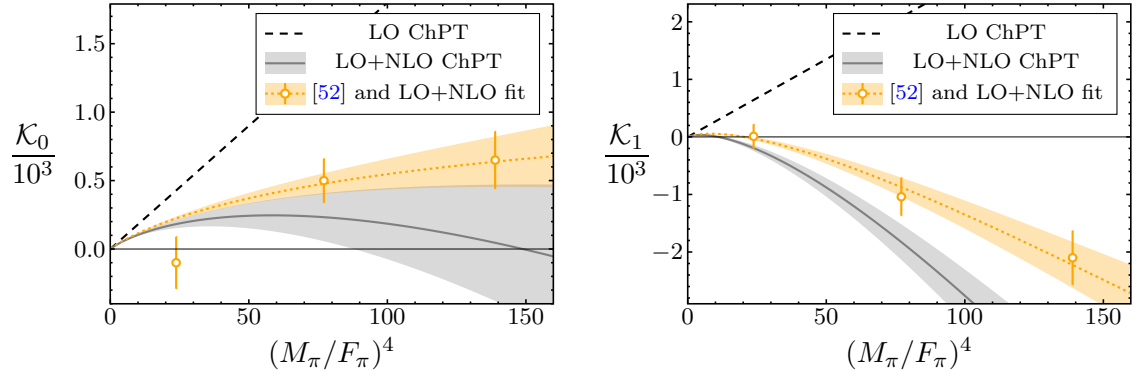
\begin{figure}[t!]
    \centering
    \begin{subfigure}{0.49\textwidth}
    \centering
        \begin{tikzpicture}
            \begin{axis}[
         		fit plot,
                xlabel={$(\m/\F)^4$}, 
                ylabel={$\displaystyle\frac{\Kiso}{10^3}$}, ylabel shift=0.2ex,
                ymin=-.4, ymax=1.79, ytick distance=.5, restrict y to domain=-1:2,
                xmin=-0.01, xmax=159.9, xtick distance=50, restrict x to domain=-1:180,
                legend pos = north east
                ]
    
                \addplot[LOline, mark=none] 
                    table[x index=0, y index=1, col sep=tab] 
                    {FitData/K0data.txt};

                \addplot[NLOline, mark=none, filled legend] 
                    table[x index=0, y index=2, col sep=tab] 
                    {FitData/K0data.txt};
                \addplot[NLOupperline, mark=none, name path global=NLOU,draw=none] 
                    table[x index=0, y index=3, col sep=tab] 
                    {FitData/K0data.txt};
                \addplot[NLOlowerline, mark=none, name path global=NLOL,draw=none] 
                    table[x index=0, y index=4, col sep=tab] 
                    {FitData/K0data.txt};
                    
                \addplot[Fitline, mark=none, filled legend with mark] 
                    table[x index=0, y index=5, col sep=tab] 
                    {FitData/K0data.txt};
                \addplot[Fitupperline, mark=none, name path global=FitU,draw=none] 
                    table[x index=0, y index=6, col sep=tab] 
                    {FitData/K0data.txt};
                \addplot[Fitlowerline, mark=none, name path global=FitL,draw=none] 
                    table[x index=0, y index=7, col sep=tab] 
                    {FitData/K0data.txt};
                    
                \addplot [fitorange, opacity=0.3] fill between[of=FitU and FitL];
                \addplot [fitgray, opacity=0.3] fill between[of=NLOU and NLOL];
                
    			\addplot [lattice data=fitorange]
               	table [x index = 0, y index = 1, x error index = 2, y error index = 3]
               	{FitData/K0points.txt};        
               	
               	\addplot[zeroline] coordinates {(-1,0) (170,0)};
                        
                \legend{LO ChPT,LO+NLO ChPT,,,{\fakeblanton\ and LO+NLO fit}}
    		
    		\end{axis}
        \end{tikzpicture}
    \end{subfigure}%
    \hspace{1ex}
    \begin{subfigure}{0.49\textwidth}
    \centering
        \begin{tikzpicture}
            \begin{axis}[
         		fit plot,
                xmin=-0.01, xmax=159.9, xtick distance=50, restrict x to domain=-1:180,     
                ymin=-2.9, ymax=2.31, ytick distance=1, restrict y to domain=-4:4,
                xlabel={$(\m/\F)^4$},       
                ylabel={$\displaystyle\frac{\Kisoone}{10^3}$},
                legend pos = north east,
                y tick label style={
                /pgf/number format/.cd,
                fixed,
                precision=0,
                /tikz/.cd
                },
                ]
    
                \addplot[LOline, mark=none] 
                    table[x index=0, y index=1, col sep=tab] 
                    {FitData/K1data.txt};
                    
                \addplot[NLOline, mark=none, filled legend] 
                    table[x index=0, y index=2, col sep=tab] 
                    {FitData/K1data.txt};
                \addplot[NLOupperline, mark=none, name path global=NLOU,draw=none] 
                    table[x index=0, y index=3, col sep=tab] 
                    {FitData/K1data.txt};
                \addplot[NLOlowerline, mark=none, name path global=NLOL,draw=none] 
                    table[x index=0, y index=4, col sep=tab] 
                    {FitData/K1data.txt};
                    
                \addplot[Fitline, mark=none, filled legend with mark] 
                    table[x index=0, y index=5, col sep=tab] 
                    {FitData/K1data.txt};
                \addplot[Fitupperline, mark=none, name path global=FitU,draw=none] 
                    table[x index=0, y index=6, col sep=tab] 
                    {FitData/K1data.txt};
                \addplot[Fitlowerline, mark=none, name path global=FitL,draw=none] 
                    table[x index=0, y index=7, col sep=tab] 
                    {FitData/K1data.txt};
                    
                \addplot [fitorange, opacity=0.3] fill between[of=FitU and FitL];
                \addplot [fitgray, opacity=0.3] fill between[of=NLOU and NLOL];
                
                \addplot [scatter, only marks, mark=*, scatter/use mapped color={draw=fitorange,fill=orange}, mark options={scale=0.7}]
               	table [x index = 0, y index = 1]
               	{FitData/K1points.txt};
    
    			\addplot [lattice data=fitorange]
               	table [x index = 0, y index = 1, x error index = 2, y error index = 3]
               	{FitData/K1points.txt};  
               	
               	\addplot[zeroline] coordinates {(-1,0) (170,0)};             
    
                \legend{LO ChPT,LO+NLO ChPT,,,{\fakeblanton\ and LO+NLO fit}}
            \end{axis}
           
        \end{tikzpicture}
    \end{subfigure}

    \caption[]{
        LO (dashed black line) and NLO (grey line and band) ChPT predictions for $\Kiso$ (left) and $\Kisoone$ (right) as functions of $(\m/\Fpi)^4$, using LECs from \rrcite{Colangelo:2001df, FLAG:2021npn} [see \cref{eq:LECref}].
        These are compared to lattice results from \rcite{Blanton:2021llb} (orange points); for reference, the physical point is at $(\Mphys/\Fphys)^4\approx5.25$.
        We also present the best fit to the lattice data (dotted orange line and orange band).}
\label{fig:fitK0K1}
\end{figure}

In \cref{fig:fitK0K1}, we compare the ChPT predictions for $\Kiso$ and $\Kisoone$ including NLO contributions to lattice results.
We also show the LO predictions for comparison.
We observe how the agreement with the lattice results is vastly improved by the inclusion of NLO terms.
For $\Kiso$ the addition of the NLO term leads to smaller values that are closer to the lattice results, while for $\Kisoone$ the correction produces a change of sign (for all except very small pion masses) that brings the sign and rough magnitude into agreement with that of the lattice results.
A conservative interpretation of these results could be that, since the NLO corrections are so large, particularly for $\Kisoone$, the convergence of the chiral expansion is poor in the regime where lattice results have been obtained and the ChPT results cannot be trusted.
A more optimistic interpretation is that the NLO results are, for some reason, larger than the LO contributions, but are nevertheless reliable.
This could be because new classes of diagrams, such as the bull's-head diagram, appear at NLO.
In either view, however, the discrepancy between lattice results and LO ChPT is resolved.

Due to the similarity between the ChPT prediction and the lattice results, as an exercise we perform a fit to the lattice data for $\Kiso$ and $\Kisoone$ to estimate how well the values of the LECs in \cref{eq:LECthrexp} can be constrained.
We obtain the following results:
\begin{equation}
    \begin{alignedat}{2}
        \elliso     &= 1.55(11)\,,    & \qquad \chi^2/\text{dof} &= 2.93/2\,,\\
        \ellisoone  &= 4.09(25)\,,    & \qquad \chi^2/\text{dof} &= 0.36/2\,,\\
    \end{alignedat}
\end{equation}
which are not that dissimilar to those computed using \cref{eq:LECref},
\begin{equation}
    \begin{aligned}
        \elliso     &= 1.19(25)\,, \qquad
        \ellisoone   &= 2.71(46)\,.
    \end{aligned}
\end{equation}
These fits are also shown in \cref{fig:fitK0K1}.

\begin{figure}[t!]
    \centering
    \begin{subfigure}[t]{0.47\textwidth}
    \centering
        \begin{tikzpicture}
            \begin{axis}[
         		fit plot,
                xmin=-0., xmax=1900., xtick distance=500, restrict x to domain=-1:2800,         
                ymin=-5.2, ymax=5.2, ytick distance=2.5, restrict y to domain=-5.3:5.3,
                xlabel={$(\m/\F)^6$},    
                ylabel={$\displaystyle\frac{\mathcal{K}_X}{10^3}$},
                legend pos = south west
                ]
                
                \addplot[NLOAline, mark=none, double filled legend] 
                    table[x index=0, y index=1, col sep=tab] 
                    {FitData/K2Adata.txt};
                \addplot[NLOAupperline, mark=none, name path global=NLOAU,draw=none] 
                    table[x index=0, y index=2, col sep=tab] 
                    {FitData/K2Adata.txt};
                \addplot[NLOAlowerline, mark=none, name path global=NLOAL,draw=none] 
                    table[x index=0, y index=3, col sep=tab] 
                    {FitData/K2Adata.txt};
                    
                \addplot[NLO2line, mark=none] 
                    table[x index=0, y index=4, col sep=tab,] 
                    {FitData/K2Adata.txt};
                \addplot[NLO2upperline, mark=none, name path global=NLO2U,draw=none] 
                    table[x index=0, y index=5, col sep=tab] 
                    {FitData/K2Adata.txt};
                \addplot[NLO2lowerline, mark=none, name path global=NLO2L,draw=none] 
                    table[x index=0, y index=6, col sep=tab] 
                    {FitData/K2Adata.txt};
    
                \addplot [fitgray, opacity=0.3] fill between[of=NLOAU and NLOAL];
                \addplot [fitgray, opacity=0.3] fill between[of=NLOAL and NLOAU];
                \addplot [fitgray, opacity=0.3] fill between[of=NLO2U and NLO2L];

    			\addplot[text along plot={$\KA$}{0cm}] 
                    table[x index=0, y index=1, col sep=tab] 
                    {FitData/K2Adata.txt};
                    
                \addplot[text along plot={$\Kisotwo$}{0cm}] 
                    table[x index=0, y index=4, col sep=tab] 
                    {FitData/K2Adata.txt};
                     
               	\addplot[zeroline] coordinates {(0,0) (2000,0)};      
    
                \legend{NLO ChPT}
    		\end{axis}
        \end{tikzpicture}
    \end{subfigure}
    \hspace{2ex}
    \begin{subfigure}[t]{0.47\textwidth}
    \centering
        \begin{tikzpicture}
            \begin{axis}[
                fit plot,
                xmin=0, xmax=1900, xtick distance=500, restrict x to domain=-1:2600, 
                ymin=-3.5, ymax=2, ytick distance=1, restrict y to domain=-4:4,
                xlabel={$(\m/\F)^6$},            
                ylabel={$\displaystyle\frac{\KB}{10^3}$}, ylabel shift=0.ex,
                legend pos = south west,
                y tick label style={
                /pgf/number format/.cd,
                fixed,
                precision=0,
                /tikz/.cd
                },
                x tick label style={
                /pgf/number format/.cd,
                fixed,
                precision=0,
                /tikz/.cd
                },
                ]
    
                \addplot[NLOline, mark=none, filled legend] 
                    table[x index=0, y index=1, col sep=tab] 
                    {FitData/KBdata.txt};
                \addplot[NLOupperline, mark=none, name path global=NLOU,draw=none] 
                    table[x index=0, y index=2, col sep=tab] 
                    {FitData/KBdata.txt};
                \addplot[NLOlowerline, mark=none, name path global=NLOL,draw=none] 
                    table[x index=0, y index=3, col sep=tab] 
                    {FitData/KBdata.txt};
               
                \addplot [fitgray, opacity=0.3] fill between[of=NLOU and NLOL];
                    
    			\addplot [lattice data=fitblue, legend with mark=fitblue]
               	table [x index = 0, y index = 1, x error index = 2, y error index = 3]
               	{FitData/KBpoints.txt};  
               	
               	\addplot[zeroline] coordinates {(0,0) (2100,0)};           
    
                \legend{NLO ChPT,,,,{\fakeblanton}}
            \end{axis}
        \end{tikzpicture}
    \end{subfigure}
    
    \caption[]{
        NLO ChPT predictions for $\Kisotwo$ (left panel, solid line), $\KA$ (left panel, dashed line) and $\KB$ (right panel, solid line) as a function of $(\m/\Fpi)^6$, using LECs from \rrcite{Colangelo:2001df, FLAG:2021npn} [see \cref{eq:LECref}].
        In the case of $\KB$, we compare to lattice results from \rcite{Blanton:2021llb} (blue points); for reference, the physical point is at $(\Mphys/\Fphys)^6\approx12.0$.
        Note that all three coefficients vanish at LO in ChPT.}
    \label{fig:fitKB}
\end{figure}
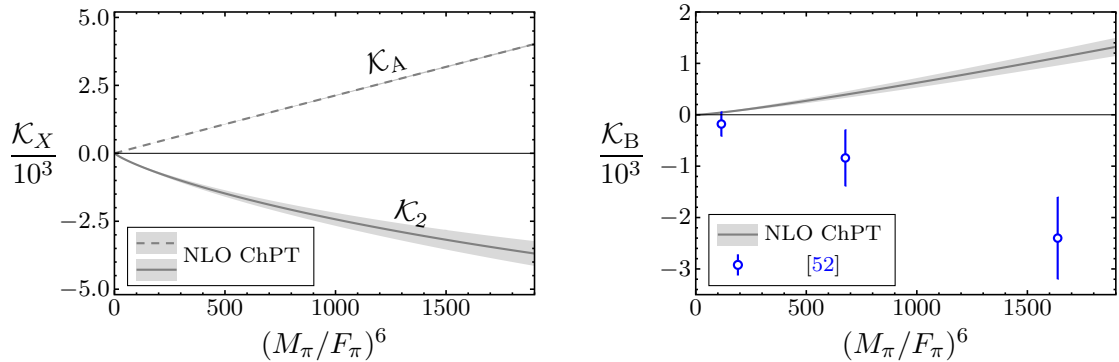

We show the predictions for $\Kisotwo$, $\KA$ and $\KB$ in \cref{fig:fitKB}.
Here there are only NLO contributions, since these quantities vanish at LO in ChPT.
In the case of $\KB$, we also compare the expectations to results from \rcite{Blanton:2021llb}.
This time, however, we observe a much larger discrepancy, with the ChPT prediction taking the opposite sign to the lattice results, although the magnitude is roughly correct.
This discrepancy is superficially similar to that between $\Kisoone$ and its leading nonzero prediction in ChPT, so it is possible that it is resolved by NNLO terms.

\subsection{Range of validity of the threshold expansion}
\label{sec:validity}

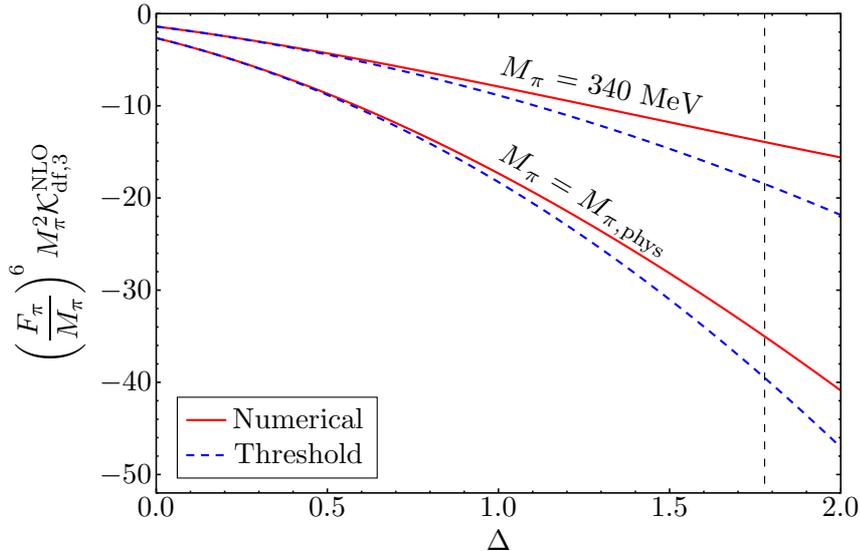
\begin{figure}[htb!]
    \newcommand{\subfigurehspace}{.7\columnwidth}
    \centering
    \begin{minipage}{\subfigurehspace}
    \centering\hspace{5ex}
        \begin{tikzpicture}[trim axis left]
            \begin{axis}[
         		general plot,
                xmin=-0., xmax=2., xtick distance=0.5, restrict x to domain=-1:3,
                ymin=-52, ymax=0, ytick distance=10, restrict y to domain=-60:19,
                xlabel={$\Delta$},          
                ylabel={$\left(\displaystyle\frac{\Fpi}{\Mpi}\right)^{\!6}\m^2\Kdf^\text{NLO}$},
                legend pos = south west,
                every tick label/.append style={font=\normalsize},
                legend style={font=\normalsize},
                            y tick label style={
                /pgf/number format/.cd,
                fixed,
             fixed zerofill,
                precision=0,
                /tikz/.cd
                },
            x tick label style={
                /pgf/number format/.cd,
                fixed,
             fixed zerofill,
                precision=1,
                /tikz/.cd
                },
                ]
    
                \addplot[numericline, mark=none] 
                    table[x index=0, y index=1, col sep=tab] 
                    {FitData/data340.txt};
                \addplot[threshline, mark=none] 
                    table[x index=0, y index=2, col sep=tab] 
                    {FitData/data340.txt};
                \addplot[numericline, mark=none] 
                    table[x index=0, y index=1, col sep=tab] 
                    {FitData/dataPP.txt};
                \addplot[threshline, mark=none] 
                    table[x index=0, y index=2, col sep=tab] 
                    {FitData/dataPP.txt};

    			\addplot[text along plot={$M{_\pi}$ = 340 MeV}{1.7cm}] 
                    table[x index=0, y index=1, col sep=tab] 
                    {FitData/data340rough.txt};
                    
                \addplot[text along plot={$M{_\pi}$ = $M{_\pi}{_,}{_\text{phys}}$}{0.7cm}] 
                    table[x index=0, y index=1, col sep=tab] 
                    {FitData/dataPPrough.txt};
                     
               	\addplot[verticalline] coordinates {(1.77778,19) (1.77778,-55)};  
               	
               	\legend{Numerical, Threshold}   
    
    		\end{axis}
        \end{tikzpicture}
    \end{minipage}
    \caption[]{
        Comparison between numerical results and the threshold expansion  for $\Kdf$, evaluated for momentum family 1 (see \cref{tab:families}).
        The comparison is presented for $\Mpi=\Mphys$  and $\Mpi=340$\;MeV, the latter corresponding to the heaviest pion mass used in \rcite{Blanton:2021llb}.
        The dashed vertical line indicates the inelastic threshold, which occurs at $E^*=5 \m$.}
    \label{fig:K_NLO}
\end{figure}

\begin{figure}[t!]
    \newcommand{\subfigurehspace}{.47\columnwidth}
    \centering
    \begin{subfigure}[t]{\subfigurehspace}
    \centering
    \begin{tikzpicture}
        \begin{axis}[
     		general plot,
            xmin=-0., xmax=2., xtick distance=0.5, restrict x to domain=-1:3,   
            ymin=-16, ymax=0, ytick distance=5, restrict y to domain=-20:19,
            xlabel={$\Delta$},         
            ylabel={$\left(\frac{\Fpi}{\Mpi}\right)^{\!6}\m^2\Kdf^\text{NLO,non-OPE}$}, ylabel shift=-0.8ex,
            legend pos = south west,
                        y tick label style={
                /pgf/number format/.cd,
                fixed,
             fixed zerofill,
                precision=0,
                /tikz/.cd
                },
            x tick label style={
                /pgf/number format/.cd,
                fixed,
             fixed zerofill,
                precision=1,
                /tikz/.cd
                },
            ]

            \addplot[numericline, mark=none] 
                table[x index=0, y index=1, col sep=tab] 
                {FitData/dataC340.txt};
            \addplot[threshline, mark=none] 
                table[x index=0, y index=2, col sep=tab,] 
                {FitData/dataC340.txt};

           	\addplot[verticalline] coordinates {(1.77778,19) (1.77778,-20)};
           	
           	\legend{Numerical, Threshold}   

		\end{axis}
    \end{tikzpicture}
    \end{subfigure}%
\hfill
    \begin{subfigure}[t]{\subfigurehspace}
    \centering
    \begin{tikzpicture}
        \begin{axis}[
     		general plot,
            xmin=-0., xmax=2., xtick distance=0.5, restrict x to domain=-1:3,
            ymin=1, ymax=20, ytick distance=5, restrict y to domain=-20:25,
            xlabel={$\Delta$},             
            ylabel={$\left(\frac{\Fpi}{\Mpi}\right)^{\!6}\m^2\mathcal{D}^\text{BH}$}, ylabel shift=-1.3ex,
            legend pos = north west,
            y tick label style={
                /pgf/number format/.cd,
                fixed,
             fixed zerofill,
                precision=0,
                /tikz/.cd
                },
            x tick label style={
                /pgf/number format/.cd,
                fixed,
             fixed zerofill,
                precision=1,
                /tikz/.cd
                },
            ]
      
            \addplot[numericline, mark=none] 
                table[x index=0, y index=1, col sep=tab] 
                {FitData/dataBH340.txt};
            \addplot[threshline, mark=none] 
                table[x index=0, y index=2, col sep=tab,] 
                {FitData/dataBH340.txt};

           	\addplot[verticalline] coordinates {(1.77778,25) (1.77778,-20)};
           	
           	\legend{Numerical, Threshold}   

		\end{axis}
    \end{tikzpicture}
    \end{subfigure}%
\par\medskip\medskip
    \begin{subfigure}[t]{\subfigurehspace}
    \centering
    \begin{tikzpicture}
        \begin{axis}[
     		general plot,
            xmin=-0., xmax=2., xtick distance=0.5, restrict x to domain=-1:3,
            ymin=1, ymax=19, ytick distance=5, restrict y to domain=-20:25,
            xlabel={$\Delta$}, 
            ylabel={$\left(\frac{\Fpi}{\Mpi}\right)^{\!6}\m^2\Kdf^\text{NLO,OPE}$}, ylabel shift=-0.8ex,
            legend pos = north west,
                        y tick label style={
                /pgf/number format/.cd,
                fixed,
             fixed zerofill,
                precision=0,
                /tikz/.cd
                },
            x tick label style={
                /pgf/number format/.cd,
                fixed,
             fixed zerofill,
                precision=1,
                /tikz/.cd
                },
            ]

            \addplot[numericline, mark=none] 
                table[x index=0, y index=1, col sep=tab] 
                {FitData/dataOPE340.txt};
            \addplot[threshline, mark=none] 
                table[x index=0, y index=2, col sep=tab,] 
                {FitData/dataOPE340.txt};

           	\addplot[verticalline] coordinates {(1.77778,25) (1.77778,-20)};    
           	
           	\legend{Numerical, Threshold}   

		\end{axis}
    \end{tikzpicture}
    \end{subfigure}
    \caption[]{
        Comparison between numerical results and the threshold expansion  for the different terms in the right-hand side of \cref{eq:sketch}, evaluated for momentum family 1 (see \cref{tab:families}).
        The panels correspond to the terms in the right-hand side of \cref{eq:sketch}: the non-OPE (top left), the BH subtraction (top right) and the OPE (bottom) contributions to $\Kdf$.
        The black vertical line indicates the inelastic threshold, which occurs at $E^*=5 \m$.
        The comparison is made for $\Mpi=340$\;MeV, corresponding to the heaviest pion mass used in \rcite{Blanton:2021llb}.}
    \label{fig:K_NLOparts}
\end{figure}
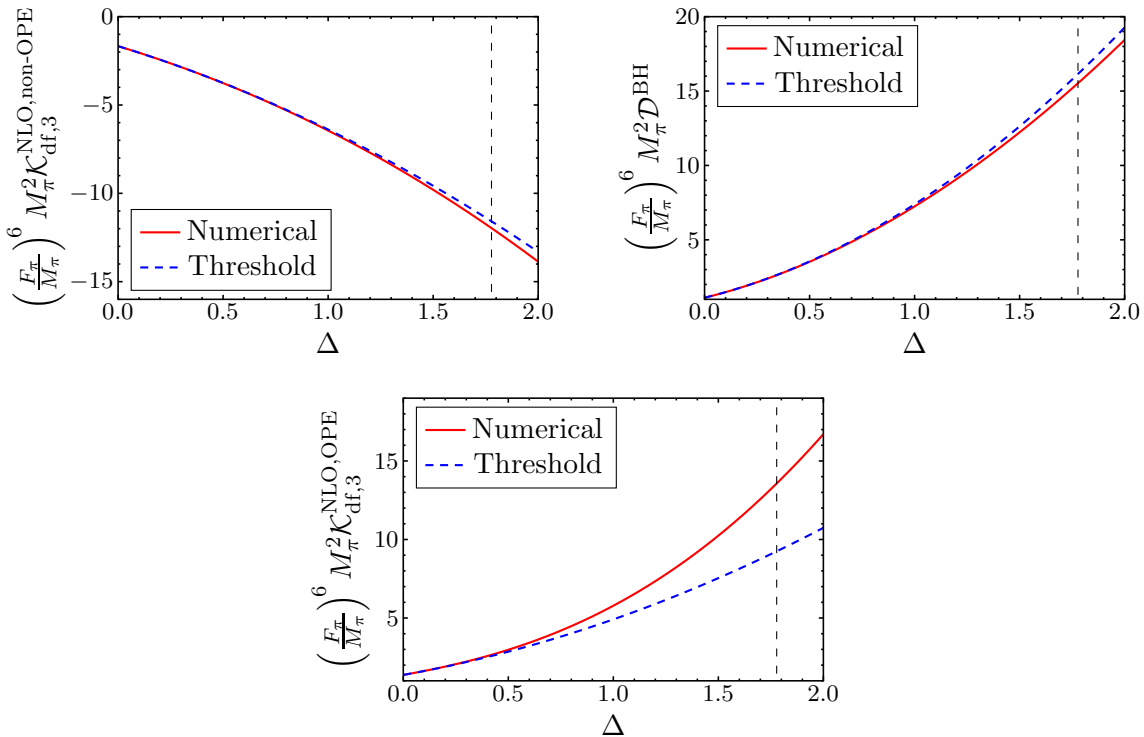

The determination of $\Kdf$ in \rcite{Blanton:2021llb} assumes that 
the threshold expansion truncated at quadratic order provides an accurate representation of the full amplitude up to the inelastic threshold at $E^*=5\m$.
Our results allow a test of this assumption, as we can compare the threshold expansion against the full result evaluated numerically for a particular kinematical configuration.
Throughout this section, we make use of momentum family 1 (see \cref{tab:families} in \cref{app:families}).
We have checked that the results for other kinematic families are comparable.

In \cref{fig:K_NLO} we perform this test for the total $\Kdf^\NLO$, both for the physical pion mass and for $\m=340$\;MeV, the latter being the heaviest pion mass used in \rcite{Blanton:2021llb}.
In both cases, we observe that the threshold expansion works reasonably well up to the inelastic threshold, where the discrepancy is $\sim$\,10\% for physical pions, and $\sim$\,20\% in the heavier case.
As expected, the convergence is better for smaller pion masses.

We can perform the same comparison separately for each of the three components appearing in \cref{eq:sketch} and \cref{fig:sketch}.
This is presented in \cref{fig:K_NLOparts} for $\m=340$\;MeV.
We recall from \cref{sec:off-shell} that each of the three components is independent of the renormalization scale in ChPT, although they do depend on the off-shell convention.
Of the three, only the BH subtraction depends on the cutoff function $H$.
We observe that the difference is $\lesssim$\,5\% at the inelastic threshold in the case of the non-OPE and the BH subtraction contribution, while for the OPE part it is $\sim$\,30\%.

\begin{figure}[th!]
\newcommand{\subfigurehspace}{.7\columnwidth}
    \centering
    \begin{minipage}{\subfigurehspace}
    \centering\hspace{5ex}
        \begin{tikzpicture}[trim axis left]
            \begin{axis}[
         		general plot,
                xmin=-0., xmax=2., xtick distance=0.5, restrict x to domain=-1:3,   
                ymin=-2, ymax=19, ytick distance=5, restrict y to domain=-6:19,
                xlabel={$\Delta$},         
                ylabel={$\left(\displaystyle\frac{\Fpi}{\Mpi}\right)^{\!6}\m^2\Kdf^\text{NLO,OPE}$}, ylabel shift=-0.8ex,
                legend pos = north west,
                every tick label/.append style={font=\normalsize},
                legend style={font=\normalsize},
                            y tick label style={
                    /pgf/number format/.cd,
                    fixed,
                 fixed zerofill,
                    precision=0,
                    /tikz/.cd
                    },
                x tick label style={
                    /pgf/number format/.cd,
                    fixed,
                 fixed zerofill,
                    precision=1,
                    /tikz/.cd
                    },
                ]
               	\addplot[verticalline] coordinates {(1.77778,19) (1.77778,-4)};  
               	\addplot[zeroline] coordinates {(0,0) (2,0)};  
    
                \addplot[allwaveline, mark=none, wide legend] 
                    table[x index=0, y index=1, col sep=tab] 
                    {FitData/OPEdata.txt};
                \addplot[swaveline, mark=none, wide legend] 
                    table[x index=0, y index=2, col sep=tab,] 
                    {FitData/OPEdata.txt};
                \addplot[dwaveline, mark=none, wide legend] 
                    table[x index=0, y index=3, col sep=tab,] 
                    {FitData/OPEdata.txt};
                \addplot[gwaveline, mark=none, wide legend] 
                    table[x index=0, y index=4, col sep=tab] 
                    {FitData/OPEdata.txt};

               	\legend{,,All $\ell$, $\ell=0$, $\ell=2$, $\ell=4$,}   
    
    		\end{axis}
        \end{tikzpicture}
    \end{minipage}
    \caption[]{
        Comparison of contributions to $\Kdf^{\NLO, \OPE}$ from different interacting pair partial waves, numerically evaluated for momentum family 1 (see \cref{tab:families}).
        Results are shown for $\Mpi=340\;$MeV, corresponding to the heaviest pion mass in \rcite{Blanton:2021llb}.
        The black solid line is the full result including all partial waves, and the dashed vertical line indicates the inelastic threshold.
        Contributions for $\ell\gtrsim 4 $ are negligible.}
\label{fig:K_NLO_OPE_l}
\end{figure}
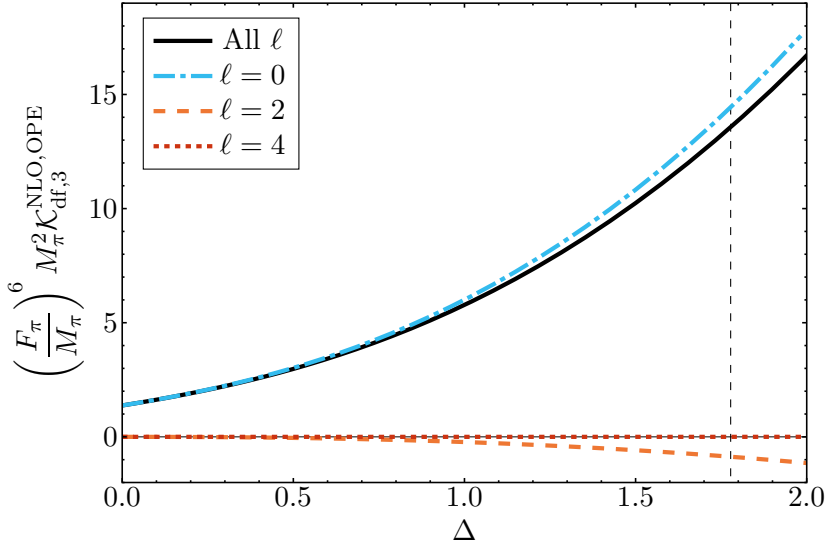

For the OPE contribution we can also study the convergence as higher partial waves of the interacting pair are included.
The threshold expansion contains only $s$- and $d$-waves, while the full result contains all even pair partial waves.%
\footnote{
    The decomposition into partial waves is obtained by evaluating $\Re \cM_{\df,3}^{\uu\NLO,\OPE}$ using \cref{eq:MdfuuNLOOPEdef}, and decomposing $\cM_{2,\ell,\off}^{\NLO}$ into partial waves by numerical integration.}
In \cref{fig:K_NLO_OPE_l} we show the contributions of the first three nonzero partial waves to the numerical result for the OPE contribution for $\m=340\,$MeV.
We find a rapid convergence with $\ell$.
In particular, the contribution is negligible below the inelastic threshold for $\ell\geq4$.
A similar result holds for smaller pion masses.

\section{Details of the calculation of $\Kdf^\NLO$}
\label{sec:calculation}

In this section, we go in detail through the calculation outlined in \cref{sec:outlineNLO} to obtain the results of \cref{sec:summary}.
Given the length of the expressions, the algebra is done with either Wolfram Mathematica~\cite{Mathematica} or FORM~\cite{Vermaseren:2000nd}, and is cross-checked independently by several of the authors.
Likewise, all numerical calculations are performed using Mathematica or \cpp, using CHIRON~\cite{Bijnens:2014gsa}, \looptools~\cite{Hahn:1998yk} and GSL~\cite{galassi2002gnu}, with independent cross-checks and preferably using several different methods.

\subsection{General form of the threshold expansion}
\label{sec:genthreshKdf}

Throughout the calculation of $\Kdf^\NLO$, it will often be useful to work with $\Kdfuu$, the unsymmetrized version of $\Kdf$.
These quantities are related by $\Kdf=\cS\big\{\Kdfuu\big\}$, where $\cS$ indicates symmetrization of both initial and final states, as in \cref{eq:Mdfdef}.
It is thus necessary to identify the form of the terms that can contribute to $\Kdfuu$ up to quadratic order.%
\footnote{
    Note that in this subsection, all statements concerning $\Kdf/\Kdfuu$ apply equally well to $\Mdf/\cM_{\df,3}^\uu$ or $\cD/\cD^\uu$, for those cases in which
    divergences are absent such that the threshold expansion is well-defined.}

The symmetries that constrain $\Kdfuu$ are the same as those for a $2+1$ theory (i.e., with two identical particles and one that is different).
Thus, we can use the threshold expansion for the latter theories derived in \rcite{Blanton:2021mih}:%
\footnote{
    Two operators are missed by the analysis of \rcite{Blanton:2021mih}, $\mathcal Q_{3-}$ and $\mathcal Q_{tu}$.
    Note that we have used a somewhat different basis than in \rcite{Blanton:2021mih}.}
\begin{multline}
    \Kdfuu 
        = c_0 + c_1 \Delta + c_2 \Delta_3^S + c_3 \tij{33} 
    \\
        + c_4 \Delta^2 + c_5 \Delta \Delta_3^S + c_6 \Delta \tij{33} 
        + c_7 \Delta_3 \Delta'_3 + c_8 (\Delta_3^S)^2 + c_9 \Delta_3^S \tij{33}+ c_{10} \tij{33}^{\,2}
    \\
        + c_{11} \cQ_{--} + c_{12} \cQ_{+-} + c_{13} \cQ_{3-} + c_{14} \cQ_{tu} + \cO(\Delta^3)\,.
    \label{eq:Kdfuu}
\end{multline}
Some quantities used above have already been defined in \cref{eq:Deltas_}.
Moreover, $\Delta_3^S \equiv \Delta_3+\Delta'_3$ and
\begin{equation}
    \begin{gathered}
        \cQ_{--} \equiv 4\left[\frac{p_-\cdot k_-}{9\m^2}\right]^2\,,
        \qquad
        \cQ_{+-} \equiv 2\left(\frac{p_+ \cdot k_-}{9\m^2}\right)^2 + 2\left(\frac{p_-\cdot k_+}{9\m^2}\right)^2\,,
        \\
        \cQ_{3-} \equiv 4\,\frac{p_-\cdot k_+}{9\m^2} \,\frac{p_-\cdot k_3}{9\m^2} 
            +4\,\frac{p_+\cdot k_-}{9\m^2} \,\frac{p_3\cdot k_-}{9\m^2} \,,
        \qquad
        \cQ_{tu} \equiv \tij{13}\tij{23}+\tij{31}\tij{32}\,,
    \label{eq:Qs}
    \end{gathered}
\end{equation}
with $p_\pm = p_1\pm p_2$ and $k_\pm=k_1\pm k_2$.
Here, the initial and final spectators are taken to have momenta $k_3$ and $p_3$, respectively.
We note that, in \cref{eq:Kdfuu}, only terms on the final line contain values of $\ell$ other than zero.

Symmetrizing $\Kdfuu$ then leads to the following contributions to $\Kdf$:
\begin{subequations}
    \label{eq:Kdf-sym}
    \begin{align}
        \Kiso &= 9 c_0\,,
        \label{eq:Kiso}
        \\
        \Kisoone &= 9 c_1 + 6 c_2 - 2 c_3\,,
        \label{eq:Kisoone}
        \\
        \Kisotwo &=  9 c_4 + 6 c_5 - 2 c_6 + c_7 + 5 c_8 - c_9 + c_{10} 
        + 4c_{11} + 2 c_{12} - 2 c_{13} + \tfrac12 c_{14}\,,
        \label{eq:Kisotwo}
        \\
        \KA &=  3 c_8 + c_9 
        - 3 c_{11} + c_{12} + 4 c_{13} + \tfrac12 c_{14}\,,
        \label{eq:KA}
        \\
        \KB &= c_{10} 
        + 9 c_{11} + 3 c_{12} - 6 c_{13} - c_{14}
        \label{eq:KB}
        \,.
    \end{align}
\end{subequations}
As expected, several terms from $\Kdfuu$ are often combined in each term in $\Kdf$ because of the larger symmetry of the latter.
Note that some of the terms that are purely $s$-wave in \cref{eq:Kdfuu} can lead to nontrivial angular dependence after symmetrization, and thus contribute to $\KA$ and $\KB$, because there can be nonzero angular momentum between the spectator and the pair.

We also make use of another basis of operators, in which all quantities are expressed in terms of $\tij{ij}$ [defined in \cref{eq:Deltas_}], for expanding the symmetrized $K$-matrix directly.
We can write
\begin{equation}
    \Kdf = c'_0+c'_1 \cQ_0'+ c'_2 \cQ_1' + c'_3 \cQ_2' + c'_4 \cQ_3' + \cO(\Delta^3)\,,
    \label{eq:KdfQexp}
\end{equation}
where 
\begin{equation}
    \begin{alignedat}{4}
        \cQ_0'&\equiv\cS[\tij{11}]
            &&= -2\Delta\,,\\
        \cQ_1'&\equiv\cS[\tij{11}\tij{11}]
            &&= \Delta^2+\DeltaB\,,\\
        \cQ_2'&\equiv\cS[\tij{11}\tij{12}+\tij{11}\tij{21}]
            &&= \tfrac{1}{2}(2\Delta^2+\DeltaA-2\DeltaB)\,,\\
        \cQ_3'&\equiv\cS[\tij{11}\tij{22}+\tij{21}\tij{12}]
            &&=\tfrac{1}{2}(2\Delta^2-\DeltaA+\DeltaB)\,.
    \end{alignedat}
\end{equation}
Here, we have indicated the relation to the basis used in \cref{eq:Kdfthrexp}.
Employing these relations, we find that the coefficients in \cref{eq:Kdfthrexp} are given by
\begin{equation}
    \begin{gathered}
        \Kiso       = c'_0\,,          \qquad 
        \Kisoone    = -2c'_1\,,        \qquad
        \Kisotwo    = c'_2+c'_3+2c'_4\,, \\
        \KA       = c'_3-c'_4\,,      \qquad
        \KB       = c'_2-2c'_3+c'_4\,.
    \end{gathered}
    \label{eq:KdfQcoeff}
\end{equation}

The requirement for the two methods of expansion to agree serves as a cross-check.
Specifically, \cref{eq:Kdf-sym,eq:KdfQcoeff} [or, equivalently, \cref{eq:KdfQexp,eq:Kdfthrexp}] must match.

\subsection{Threshold expansion of the non-OPE part of $\cM_3$}
\label{sec:threxpand-M3}

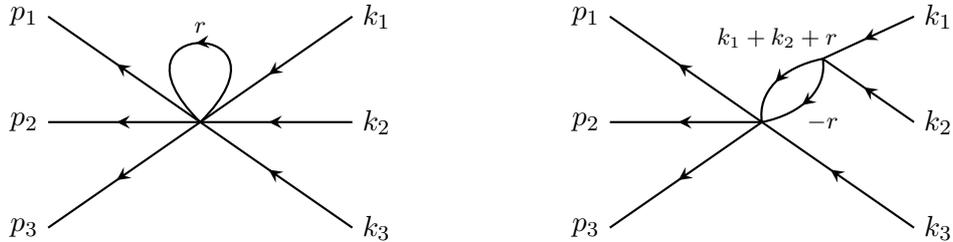
\begin{figure}[ht]
    \centering
    \begin{subfigure}{0.48\textwidth}
    \centering
        \begin{tikzpicture}[xscale=\diagramxscale,yscale=\diagramyscale]
            \makeexternallegs
            \coordinate (v) at (0,0);
            \draw[dprop] (k1) -- (v) -- (p1);
            \draw[dprop] (k2) -- (v) -- (p2);
            \draw[dprop] (k3) -- (v) -- (p3);
            \draw[dprop=.52] (v) .. controls (.7,1) and (-.7,1) .. (v) node[midway, above] {\footnotesize $r$};
        \end{tikzpicture}
    \end{subfigure}
    \begin{subfigure}{0.48\textwidth}
    \centering
        \begin{tikzpicture}[xscale=\diagramxscale,yscale=\diagramyscale]
            \makeexternallegs
            \coordinate (v1) at (+.4,+.6);
            \coordinate (v2) at (  0,  0);
            \draw[dprop] (k1) -- (v1) to[bend right=40] (v2) -- (p1);
            \draw[dprop] (k2) -- (v1) to[bend left=40] (v2)  -- (p2);
            \draw[dprop] (k3) -- (v2) -- (p3);
            \draw (v2) + (0.1,0.8) node {\footnotesize$k_1+k_2 + r$};
            \draw (v2) + (0.4,0) node {\footnotesize$-r$};
        \end{tikzpicture}
    \end{subfigure}
    \caption{Examples of non-OPE NLO diagrams for which no subtraction is needed.}
    \label{fig:nosubNLO}
\end{figure}

In this section, we focus on threshold-expanding the non-OPE piece of the amplitude in \cref{eq:sketch}.
This is the piece related to one-particle-irreducible diagrams, like those in \cref{fig:nosubNLO,fig:BHNLO}, for which the contribution to $\Kdf$ can be computed independently of any subtraction.
For the $I=3$ system, this contribution is given by 
\begin{align}
    \Mdf^{\NLO,\nonOPE} 
        &= A^{(4)}(k_1, -p_1, k_2, -p_2, k_3, -p_3)
        + A^{(4)}(k_1, -p_2, k_2, -p_1, k_3, -p_3)
        \notag\\
        &+ A^{(4)}(k_1, -p_1, k_2, -p_3, k_3, -p_2)
        + A^{(4)}(k_1, -p_2, k_2, -p_3, k_3, -p_1)
        \label{eq:fullamplitude}\notag\\
        &+ A^{(4)}(k_1, -p_3, k_2, -p_1, k_3, -p_2)
        + A^{(4)}(k_1, -p_3, k_2, -p_2, k_3, -p_1)\,,
\end{align}
where $A^{(4)}$ is defined in eq.~(35) of \rcite{Bijnens:2021hpq}, and we have converted to our conventions as described above \cref{eq:6pt}.

This part of the amplitude can be divided as
\begin{equation}
    A^{(4)} =A_C + A_J + A_\pi + A_L + A_l \,, \label{eq:A4parts}
\end{equation}
where the division is similar to that in  eq.~(35) of \rcite{Bijnens:2021hpq}, although here we group some terms together.
In particular, $A_C$ is the part that contains the $\overline C$ functions, i.e., $\overline C$, $\overline C_{11}$ and $\overline C_{21}$ ($\overline C_3$ does not contribute to $I=3$), $A_J$ contains all terms with $\bar J$ functions, and $A_\pi$, $A_L$, and $A_l$ are terms containing only factors of $\kappa$, $L$ or LECs, respectively (see \cref{sec:expressions,app:loops} for definitions).
Some of these amplitudes have imaginary parts, but we only need to calculate the real parts since the imaginary parts do not contribute to $\Kdf$ [see \cref{eq:master}].

The computation of $c'_i$ [following \cref{eq:KdfQexp}] is straightforward for $A_l$, $A_L$, and $A_\pi$.
The results for the latter two are given in \cref{eq:nonOPE-piL}.
The remaining parts, which contain loop integrals, require more care.
As is shown in the following subsections, they can be expanded in terms of squared momenta that are either small or close to threshold.
These can then be straightforwardly related to $\tij{ij}$ and $\Delta_i^{(\prime)}$ through \cref{eq:Deltas_}.
Alternatively, one can use (at least) three different particular kinematic configurations to numerically determine the corresponding coefficients in the $\Delta$ variables.
This cross-check is described in \cref{app:families}.

\subsubsection{Threshold expansion of $A_J$}
\label{sec:nonOPE-J}

Terms containing $\bar J(q^2)$ functions must be expanded about either $q^2=0$ or $q^2=4\Mpi^2$ (the two-particle threshold).
In the notation of \rcite{Bijnens:2021hpq}, $A_J=A_J^{(1)}+A_J^{(2)}$, where all $\bar{J}$ functions in $A_J^{(1)}$ are expanded about $q^2=4\Mpi^2$, while in $A_J^{(2)}$ both cases appear.
It is, however, easy to separate $A_J^{(2)}$ in two parts, each of which contains one single case.
Using again the notation of \rcite{Bijnens:2021hpq} [in particular eq.~(B9) thereof],
\begin{multline}
    A_J^{(2)} = (R_{132456}+R_{241356}+R_{152634}+R_{152634}+R_{261543}+R_{536412}+R_{645321}\\
          + R_{142356}+R_{231456}+R_{251634}+R_{162543}+R_{635421}+R_{546312})A_{J,0}^{(2)}\,,
\end{multline}
where $R_{ijklmn}$ indicates an operator, acting on $A_{J,0}^{(2)}$, that permutes $p_1\to p_i$, $p_2\to p_j$, etc., with $\{p_1,\ldots,p_6\}=\{k_1,k_2,k_3,-p_1,-p_2,-p_3\}$ as described above \cref{eq:6pt}.
It is easy to check that, after projecting to $I=3$, the first line above leads to terms that contain $\bar{J}$ functions that have to be expanded about the two-particle threshold, while the second has those that need to be expanded around $q^2=0$.

Using the definition~\eqref{eq:J}, the expansions in the two cases are
\begin{align}
    \frac1\kappa\Re\bar J(4\m^2+\bar s) &=
    2 - \frac12 \frac{\bar s}{\m^2}+ \frac1{12} \frac{\bar s^2}{\m^4} 
    - \frac1{60} \frac{\bar s^3}{\m^6}+ \cO(\bar s^4)\,,
    \\
    \frac1\kappa\bar J(t) &=
    \frac{1}6 \frac{t}{\m^2} + \frac{1}{60} \frac{t^2}{\m^4} + \frac{1}{420} \frac{t^3}{\m^6} + \cO(t^4)\,.
    \label{eq:Jbarexp}
\end{align}
Note that only $\Re\bar J$, not $\Im\bar J$, is smooth at threshold.
After conversion to threshold-expansion parameters, the results are given in \cref{eq:nonOPE-J}.

\subsubsection{Threshold expansion of $A_C$}
\label{sec:nonOPE-C}

All $\overline C$, which are functions of three pairs of momenta as defined in \cref{app:loops}, can be expanded either for all three pairs being small or one small and the other two near threshold.
Given that the expansions are analytic, for $\Re\Mdf^{\NLO,\nonOPE}$ the Feynman integrals can be performed na\"ively with a principal-value prescription.
For $C$, defined in \cref{eq:C}, the starting point is the Feynman-parameter representation
\begin{align}
    C\equiv
    C(p_1,p_2,\dots,p_6) = -\kappa\int_0^1 \d x\,\d y\,\d z\,\frac{\delta(1-x-y-z)}{\Mpi^2-xy q_1^2 - yz q_2^2 - zx q_3^2}\,,
\end{align}
with $q_1=p_1+p_2$, $q_2=p_3+p_4$ and $q_3=p_5+p_6$.
For the three $q_i^2$ small, we straightforwardly expand the denominator and then perform the Feynman integrals, arriving at
\begin{align}
    \frac{C}{\kappa}
    =-\frac{1}{2\Mpi^2}-\frac{1}{24\Mpi^4}\bigl(q_1^2+q_2^2+q_3^2\bigr)-\frac{1}{180\Mpi^6}\bigl(q_1^4+q_2^4+q_3^4+q_1^2 q_2^2+q_2^2 q_3^2+q_3^2 q_1^2\bigr)+\cdots
\end{align}
For the expansion about threshold, take for example $q_1^2$, $\bar s_2$, and $\bar s_3$ small, where $\bar s_2=q_2^2-4\Mpi^2$ and $\bar s_3=q_3^2-4\Mpi^2$.
We can write the Feynman parametrization of $C$ as
\begin{align}
    C 
    &= -\kappa\int_0^1 \d x\,\d y\,\d z\,\frac{\delta(1-x-y-z)}{\Mpi^2-4 \Mpi^2 (yz+zx)-xy q_1^2 - yz \bar s_2 - zx \bar s_3}
    \notag\\
    &= -\kappa\int_0^1 \d x\,\d y\,\d z\,\frac{\delta(1-x-y-z)}{\Mpi^2(1-2z)^2-xy q_1^2 - yz \bar s_2 - zx \bar s_3}\,.
    \label{eq:Cint2}
\end{align}
Since we know the integral is analytic above threshold, we expand na\" ively in $q_1^2$, $\bar s_2$, and $\bar s_3$ and perform the Feynman integrals, first doing the $x,y$ integrals and then the $z$ integral using the principal-value prescription, discarding the imaginary part.
In particular, after expanding and performing the $x,y$ integrals, one obtains integrals of the type
\begin{align}
    \principal\!\int_0^1\d z\,\frac{1}{(1-2z)^n}
    = \frac{1}{2}\,\principal\!\int_{-1}^{1}\d v\,\frac{1}{v^n}
    = \frac{1}{4}\int_\pi^0 e^{-in\theta}\,\d e^{i\theta}
    + \frac{1}{4}\int_\pi^0 e^{in\theta}\,\d e^{-i\theta}\,,
\end{align}
where we changed variables to $v\equiv 1-2z$ and implemented the principal-value ($\principal$) prescription as the average of the contours above and below the singularity at $v=0$ in the complex plane, letting $v=e^{\pm i\theta}$.%
\footnote{
    We refer readers unfamiliar with principal values for multiple poles to \rcite{Davies:1996gee}.}
The integrals vanish for odd values of $n$ and are equal to $-1/(n-1)$ for even  values.
This way, the result for \cref{eq:Cint2} becomes
\begin{multline}
     \frac{C}{\kappa}
    =\frac{1}{2\Mpi^2}+\frac{1}{\Mpi^4}\biggl(\frac{5}{72}q_1^2-\frac{1}{24}\bar s_2-\frac{1}{24}\bar s_3\biggr)\\
    +\frac{1}{\Mpi^6}\biggl[\frac{2}{225}q_1^4-\frac{1}{90}q_1^2(\bar s_2+\bar s_3)+\frac{1}{180}\bigl(\bar s_2^2+\bar s_3^2+\bar s_2\bar s_3\bigr)\biggr]+\cdots
\end{multline}
The other triangle integrals are expanded using the same methods.
We have checked the expansions numerically against \looptools~\cite{Hahn:1998yk}.
The contributions to $\Kdf$ from $A_C$ are given in \cref{eq:nonOPE-C}.

\subsubsection{The full non-OPE threshold expansion}
\label{sec:nonOPE-res}

Performing the expansion described above to second order, we find that the total contributions from $\Re\Mdf^{\NLO,\nonOPE}$ are
\begin{subequations}
    \begin{align}
        \frac{\F^6}{\m^6}\,\Kiso 
            & \supset 
            14 \kappa + 33 L + 36 (8 \lrI +  \lrIII -2 \lrIV)\,,
        \\
        \frac{\F^6}{\m^6}\,\Kisoone 
            & \supset 
            -\frac{35}2 \kappa + 12 L + 36 (20 \lrI + \lrII - 4 \lrIV)\,,
        \\
        \frac{\F^6}{\m^6}\,\Kisotwo 
            & \supset
            -\frac{9747}{50} \kappa - 216 L + 324 (2 \lrI + \lrII)\,,
        \\
        \frac{\F^6}{\m^6}\,\KA 
            & \supset
            \frac{576}5 \kappa - 54 L - 81 (2 \lrI - 3 \lrII)\,,
        \\
        \frac{\F^6}{\m^6}\,\KB 
            & \supset
            -\frac{13797}{50} \kappa - 162 L + 243 (2 \lrI + \lrII)\,.
    \end{align}
    \label{eq:fullnonOPE}%
\end{subequations}
For comparison, the LO results from the non-OPE diagram in \cref{fig:FeynmanLO:contact} are
\begin{equation}
    \frac{\F^4}{\m^4}\,\Kiso \supset -18 \,,
    \qquad
    \frac{\F^4}{\m^4}\,\Kisoone \supset  -26 \,.
    \label{eq:nonOPE-LO}
\end{equation}
This is specific to our off-shell convention (see \cref{sec:off-shell}).

The results for the separate parts of $A^{(4)}$ in \cref{eq:A4parts} are as follows.
The part stemming from $A_l$ can be directly read from \cref{eq:fullnonOPE} as the terms containing $\ell_i^\rr$.
The remaining parts not containing $\bar J$ or $\overline C$ loop functions, i.e., $A_\pi$ and $A_L$, give
\begin{multline}
\frac{\Fpi^6}{\m^4}\Kdf^{\NLO,\nonOPE} \supset
                    17\kappa
                    + 33L
    +\Delta\big(
                    10\kappa
                    + 48L
                    \big)
    +\Delta^2\biggl(-\frac{387}{2}\kappa
                    -\frac{351}{2}L
                    \biggr)
                    \\
    +\DeltaA\biggl(-\frac{495}{8}\kappa
                    -\frac{675}{8}L
                    \biggr)
    +\DeltaB\biggl(-\frac{1323}{8}\kappa
                    -\frac{1215}{8}L
                    \biggr)\,.
  \label{eq:nonOPE-piL}
\end{multline}
The part $A_J$ containing solely $\bar J$ functions, using the results from \cref{sec:nonOPE-J}, reads
\begin{multline}
 \frac{\Fpi^6}{\m^4}\Kdf^{\NLO,\nonOPE} \supset  -72 \kappa
     + \Delta   (-223 \kappa)
     + \Delta^2 \biggl(-\frac{243}4 \kappa\biggr) \\
     + \DeltaA  \biggl(\frac{1233}{4} \kappa\biggr)
     + \DeltaB  \biggl(-\frac{117}2 \kappa\biggr)\,,
    \label{eq:nonOPE-J}
\end{multline}
and the $\overline C$ functions of $A_C$, calculated employing the methods presented in \cref{sec:nonOPE-C}, all together give
\begin{multline}
\frac{\Fpi^6}{\m^4}\Kdf^{\NLO,\nonOPE} \supset 69 \kappa
      + \Delta  \biggl(- 36 L + \frac{391}2 \kappa\biggr)
      + \Delta^2 \biggl(- \frac{81}2 L + \frac{5931}{100} \kappa\biggr)\\
      + \DeltaA \biggl(\frac{243}8 L - \frac{5247}{40} \kappa\biggr)
      + \DeltaB \biggl(-\frac{81}8 L - \frac{10413}{200} \kappa\biggr)\,.
    \label{eq:nonOPE-C}
\end{multline}
The sum of \cref{eq:nonOPE-piL,eq:nonOPE-J,eq:nonOPE-C}, plus the $\ell^\rr_i$ terms, gives \cref{eq:fullnonOPE}.

\subsection{Bull's head subtraction contribution}
\label{sec:bullhead}

\begin{figure}[ht]
    \centering
    \begin{subfigure}{0.48\textwidth}
    \centering
        \begin{tikzpicture}[xscale=\diagramxscale,yscale=\diagramyscale]
            \makeexternallegs
            \coordinate (v1) at (+.5,+.5);
            \coordinate (v2) at (-.5,+.5);
            \coordinate (v3) at (  0,-.7);
            \draw[dprop] (k1) -- (v1) -- (v2) node[midway, above] {\footnotesize$r$} -- (p1);
            \draw[dprop] (k2) -- (v1) -- (v3) -- (v2) -- (p2);
            \draw[dprop] (k3) -- (v3) -- (p3);
            \draw (+0.6,-.3) node {\footnotesize$k_1{+}k_2{-}r$};
            \draw (-0.6,-.3) node {\footnotesize$p_1{+}p_2{-}r$};
        \end{tikzpicture}
        \caption{The ``bull's head'' diagram.}
        \label{fig:BHNLO:regular}
    \end{subfigure}
    \begin{subfigure}{0.48\textwidth}
    \centering
        \begin{tikzpicture}[xscale=\diagramxscale,yscale=\diagramyscale]
            \makeexternallegs
            \coordinate (v1) at (+.5,+.5);
            \coordinate (v2) at (-.5,+.5);
            \coordinate (v3) at (  0,-.7);
            \draw[dprop=.8] (v1) .. controls ($ (k1)!.5!(v1) $) .. (p1);
            \draw[line width=3pt, white] (k1) .. controls ($ (p1)!.5!(v2) $) .. (v2);
            \draw[dprop=.2] (k1) .. controls ($ (p1)!.5!(v2) $) .. (v2);
            \draw[dprop] (v2) -- (v3) -- (v1);
            \draw[dprop] (k2) -- (v1) -- (v2) node[midway, above] {\footnotesize$r$}-- (p2)  ;
            \draw[dprop] (k3) -- (v3) -- (p3);
            \draw (+0.6,-.3) node {\footnotesize$p_1{-}k_2{+}r$};
            \draw (-0.6,-.3) node {\footnotesize$k_1{-}p_2{+}r$};
        \end{tikzpicture}
        \caption{The ``crossed bull's head'' diagram.}
        \label{fig:BHNLO:crossed}
    \end{subfigure}
    \caption{
        Two configurations of the triangle-loop diagram, of which only (a) requires subtraction [in the form of $\DuuBH$, \cref{eq:DuuBH}].
        There are a total of 15 diagrams with the triangle topology, of which 9 correspond to the configuration (a) [so their sum corresponds to the symmetrization of~(a)]  and 6 to the configuration~(b).}
   \label{fig:BHNLO}
\end{figure}
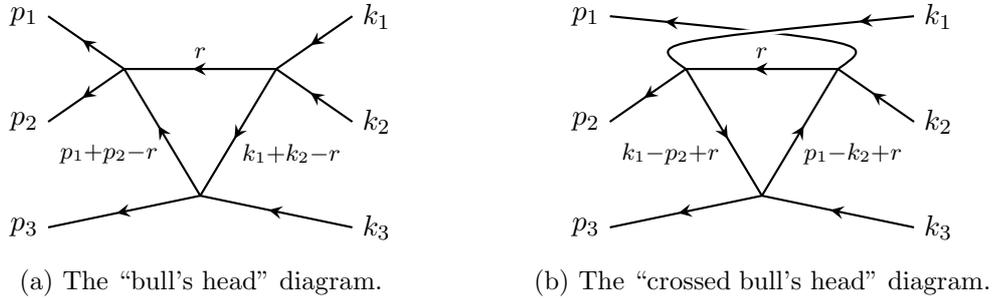

The BH piece of the subtraction, $\cD^\BH$ in \cref{eq:sketch}, concerns the symmetrization of the final term in \cref{eq:Duu-expand}, which subtracts the ``bull's head'' diagram in \cref{fig:BHNLO:regular}.
This is the only cutoff-dependent contribution to $\Kdf^\NLO$, and due to the integral over $H(x)$, it requires numerical evaluation.

Since there are only $s$-wave contributions, the subtraction term reduces to
\begin{align}
    \DuuBH(\bm p_3, \bm k_3) &= -\frac{1}{\F^6}(2p_1\cdot p_2)\, I(\bm p_3,\bm k_3) (2k_1\cdot k_2)\,,\label{eq:DuuBH_}\\
   I(\bm p_3,\bm k_3) &\equiv \int_r \frac{H(x_r) \big[(P-r)^2 - 2\m^2]H(x_r)}{\big[(\pp-r)^2-\m^2+i\epsilon\big]\big[(\kk-r)^2-\m^2+i\epsilon\big]}\,,
   \label{eq:DuuBH}
\end{align}
where $\pp\equiv p_1+p_2$, $\kk\equiv k_1+k_2$, and $x_r \equiv (P-r)^2/(4\m^2)$.
For brevity, we define $G(r;\pp,\kk)$ (not to be confused with $G^\infty$ above) such that
\begin{equation}
    \DuuBH=-\frac{1}{\F^6}(2p_1\cdot p_2)(2k_1\cdot k_2)\int_r H^2(x_r)G(r;\pp,\kk)\,.
    \label{eq:DuuBH-G}
\end{equation}
We stress that $r$ is on-shell and that the integral is Lorentz-invariant.

\subsubsection{Threshold expansion}
\label{sec:BHexpand}

The integrand of \cref{eq:DuuBH} can be expressed to any order in the threshold expansion using $\Delta,\Delta_3,\Delta'_3$, and $\tij{33}$ only.
Since the amplitude is free of singularities in the real part, it follows that the real part of the subtraction is also singularity-free.
However, due to possible singularities in the integrand, some manipulations need to be done before the coefficients can be evaluated.

Starting with \cref{eq:DuuBH}, we stay in the CMF of the whole system, where $\bm P =\bm 0$ and $(P-r)^2=E^2-2E\omega_r+\m^2$, and set $\d^3 r = r^2\d r\,\d\!\cos\theta\,\d\phi$.
We rewrite the first denominator as
\begin{equation}
    (\pp-r)^2-\m^2 = \pp^2-2\pp\cdot r = \pp^2-2E_{\pp}\omega_r +2\bpp\cdot \bm r\equiv 4 \m^2-4 \m\omega_r+\Delta_{\pp}\,,
\end{equation}
which defines $\Delta_{\pp}$; $\Delta_{\kk}$ is defined similarly from the second denominator.
Note that $\pp=(E_{\pp},\bpp)$ is the momentum of the two-pion system.
We then expand na\"ively in $\Delta_{\pp}$ and $\Delta_{\kk}$, which are both $\cO(\sqrt{\Delta})$, so 
\begin{equation}
    I(\bm p_3,\bm k_3) = \int_r H^2(x_r)\big[E^2-2E\omega_r-\m^2\big]\sum_{a,b=0}^\infty \frac{\Delta_{\pp}^a\Delta_{\kk}^b}{(4\m^2-4\m\omega_r)^{a+b+2}}\,,
\end{equation}
and cut off the sums at a suitable order.
After that, we can perform all angular integrals using
\begin{equation}
    \int \d\!\cos\theta\,\d\phi\: (\bpp\cdot \bm r)(\bkk\cdot\bm r)
    = 4\pi\, \frac{r^2}{3}\, \bpp\cdot\bkk
\end{equation}
and its generalizations.%
\footnote{
    The general case can be compactly written as
    \begin{equation*}
        \int \d\!\cos\theta\,\d\phi \prod_{a=1}^{2n} (\bm p_a\cdot\bm r) = 4\pi\, \frac{r^{2n}}{(2n+1)!!}\, \delta^{i_1i_2\cdots i_{2n}} p_1^{i_1}p_2^{i_2}\cdots p_n^{i_{2n}}
    \end{equation*}
    (the case with an odd number of $\bm p_a$ vanishes by symmetry).
    Repeated indices are summed, and $\delta^{i_1\cdots i_n}$ generalizes the Kronecker $\delta$ to the totally symmetric tensor obtained by summing all distinct ways of distributing the indices $i_1,\ldots,i_{2n}$ among $n$ Kronecker $\delta$'s, i.e., $\delta^{ijkl}=\delta^{ij}\delta^{kl}+\delta^{ik}\delta^{jl}+\delta^{il}\delta^{jk}$, etc.}
Finally, we expand the explicit and implicit [via $E=3\m\sqrt{1+\Delta}$ and $x_r=(E^2-2E\omega_r+\m^2)/4\m^2$] dependence on $\Delta$ to the order needed.
We will show in \cref{sec:BHhadamard} that na\"ively expanding $H(x_r)$ this way is valid, despite the non-analyticity of $H$.

The remaining integral over $r$ can be rewritten as an integral over $\omega_r$.
Introducing the variable $z$, defined via $\omega_r=\m(1+2z^2)$ so that $r^2\d r/[2\omega_r(2\pi)^3] = \m^2z^2\sqrt{1+z^2}\,\d z/(2\pi^3)$, leads to integrals of the type
\begin{equation}
     H_{m,n} \equiv \frac{1}{\pi^2}\int_0^{1/\sqrt{3}}\d z~\frac{\sqrt{1+z^2}}{z^{m}} \frac{\d^n}{\d x^n}\big[H^2(x)\big]\,.
    \label{eq:defHn}
\end{equation}
Since $x = 1-3z^2$, the integration limits are $x=1$ at $z=0$ and $x=0$ at $z=1/\sqrt3$.
Due to $H$ and all its derivatives vanishing at $x=0$, the $E$-dependence in the upper limit of the integral is not captured by the expansion, meaning this is an expansion of asymptotic nature.
However, as can be seen in the upper-right panel of \cref{fig:K_NLOparts}, the expansion to quadratic order approximates the full numerical result very well for the range of energies of interest.

We will return to the evaluation of $H_{m,n}$ in \cref{sec:BHhadamard}.
For now, we use it to state the result, which to second order in the threshold expansion is
\begin{align}
    \frac{\F^6}{\m^4}\Re{}&\DuuBH = \bigl(\tfrac{3}{2}H_{0,0}-\tfrac{1}{4}H_{2,0}\bigr)
    + \bigl(\tfrac{21}{4}H_{0,1}+\tfrac{1}{8}H_{4,0}\bigr)\Delta
    + \bigl(\tfrac{81}{16}H_{0,0}-\tfrac{9}{32}H_{2,0}-\tfrac{3}{32}H_{4,0}\bigr)\Delta_3^S\notag\\&
    + \bigl(\tfrac{9}{16}H_{0,0}+\tfrac{15}{32}H_{2,0}-\tfrac{3}{32}H_{4,0}\bigr)\tij{33}
    + \bigl(\tfrac{63}{8}H_{0,2}-\tfrac{33}{64}H_{0,1}-\tfrac{13}{256}H_{6,0}-\tfrac{19}{64}H_{4,0}\bigr)\Delta^2\notag\\&
    + \bigl(\tfrac{63}{4}H_{0,1}+\tfrac{3}{64}H_{6,0}+\tfrac{3}{32}H_{4,0}\bigr)\Delta\Delta_3^S
    + \bigl(\tfrac{9}{128}H_{6,0}-\tfrac{63}{128}H_{4,0}+\tfrac{27}{64}H_{2,0}\bigr)\Delta\tij{33}\notag\\&
    + \bigl(-\tfrac{9}{32}H_{4,0}+\tfrac{81}{32}H_{2,0}-\tfrac{81}{16}H_{2,0}\bigr)\bigl(\Delta_3^S\bigr)^2
    + \bigl(\tfrac{891}{32}H_{0,0}-\tfrac{243}{64}H_{2,0}-\tfrac{9}{64}H_{4,0}\bigr)\Delta_3 \Delta_3^\prime\notag\\&
    + \bigl(-\tfrac{9}{128}H_{6,0}+\tfrac{9}{64}H_{4,0}+\tfrac{189}{128}H_{2,0}+\tfrac{81}{64}H_{0,0}\bigr)\Delta_3^S\tij{33}\notag\\&
    + \bigl(-\tfrac{27}{640}H_{6,0}+\tfrac{27}{160}H_{4,0}+\tfrac{297}{640}H_{2,0}+\tfrac{81}{320}H_{0,0}\bigr)\tij{33}^{\,2}\,.
    \label{eq:BHexp}
\end{align}
This is the same expansion as in \cref{eq:Kdfuu}, but only a subset of the terms is needed.
The integration-by-parts relation
\begin{multline}
    H_{m,n+1}+H_{m-2,n+1} =\frac{1}{6} \bigl[(2-m)H_{m,n}-(m+1) H_{m+2,n}\bigr]\\
        -\frac16\bigg[\big(f_{m-1}'(z)+f_{m+1}'(z)\big)\frac{\d^n}{\d x^n}H^2(x)\bigg]_0^{1/\sqrt3}\,,
    \label{eq:Hrelation}
\end{multline}
where $f_n$ is defined in \cref{eq:fdefinition}, has been used extensively in simplifying \cref{eq:BHexp}; the ``surface terms'' [the second line of \cref{eq:Hrelation}] vanish identically, but are revisited in \cref{sec:BHanalytic}.
After symmetrization over all 9 possibilities, analogous to \cref{eq:Kdf-sym}, we get 
\begin{align}
    \frac{\F^6}{\m^4}\Re\cD^\BH &= \Big[\tfrac{27}{2} H_{0,0} - \tfrac 94 H_{2,0}\Big]
        +\Delta\Big[\tfrac{117}{4} H_{0,0} - \tfrac{21}{8} H_{2,0} + \tfrac 34 H_{4,0} + \tfrac{189}{4} H_{0,1}\Big]\notag\\
        &+\Delta^2\Big[ \tfrac{243}{160} H_{0,0} + \tfrac{2241}{320} H_{2,0} - \tfrac{423}{160} H_{4,0} - \tfrac{369}{1280} H_{6,0} + \tfrac{5751}{64} H_{0,1} + \tfrac{567}{8} H_{0,2}\Big]\notag\\
        &+\DeltaA\Bigl[ -\tfrac{891}{64} H_{0,0} + \tfrac{1161}{128} H_{2,0} - \tfrac{45}{64} H_{4,0} - \tfrac{9}{128} H_{6,0}\Bigr]\notag\\
        &+\DeltaB\Big[ \tfrac{81}{320} H_{0,0} + \tfrac{297}{640} H_{2,0} + \tfrac{27}{160} H_{4,0} - \tfrac{27}{640} H_{6,0}\Big]\,,
    \label{eq:result-BH}
\end{align}
where $\cD^\BH=\cS\{\DuuBH\}$ is the bull's head term in \cref{eq:sketch}.

\subsubsection{Hadamard finite-part integration}
\label{sec:BHhadamard}

\Cref{eq:defHn} for $n=0$ and $m>0$ has a troublesome singularity in the endpoint $z=0$, so it is not possible to na\"ively apply the Cauchy principal value to evaluate $H_{m,n}$.
However, it is possible to use the \emph{Hadamard finite-part} prescription.
Let us present its (rather simple) application before turning to the question of its validity.
We define the functions $f_m(z)$ via
\begin{equation}\label{eq:fdefinition}
    \frac{\d}{\d z}f_m(z) = \frac{1}{\pi^2} \frac{\sqrt{1+z^2}}{z^{m}}\,.
\end{equation}
Using partial integration on \cref{eq:defHn}, we arrive at
\begin{align}
    \label{eq:regularHn}
    H_{m,0} = f_m(z) H^2(x)\big|_{z=0}^{1/\sqrt3}-\int_{0}^{1/\sqrt3} \d z\: f_m(z) (-6z)\, \frac{\d}{\d x}H^2(x)\,,
\end{align}
an expression that is independent of the integration constant in $f_m(z)$.
The $z=0$ limit of the first term is singular, whereas the second term (the integral) is non-singular since derivatives of $H(x)$ vanish exponentially as $z\to0$ or $x\to1$.
The $z=1/\sqrt3$ or $x=0$ limit vanishes by construction since $H$ and its derivatives all vanish at $x=0$.
The Hadamard finite part of $H_{m,0}$ is obtained by dropping the singular $z=0$ limit, and if the prescription is valid, $\DuuBH$ is obtained by replacing all divergent $H_{m,n}$ with their finite parts.

It is at first not obvious that the Hadamard finite-part prescription is valid in our case, since the standard proofs involve complex integrals that break down due to the non-analyticity of $H(x)$.
However, \rcite{hadamard} presents a proof using only smoothness criteria, which our integrands do satisfy.
It also requires $m>1$, but $H_{0,0}$ is non-singular, and we do not need $H_{1,0}$ (which only has an integrable singularity).
Lastly, the prescription of course requires $\DuuBH$ to be finite, but we know from \cref{sec:threxpand-M3} that such is the case, at least for the real part: $\Re\cM_3$ lacks the divergences that $\DuuBH$ would subtract.

As a last remark, we note that the Hadamard finite-part integration validates the na\"ive threshold expansion used to obtain \cref{eq:defHn}, which involved Taylor expanding $H(x)$.
This Taylor series converges for $0<x<1$, but the convergence is extremely poor in the vicinity of the essential singularities at the endpoints.
However, after applying \cref{eq:regularHn}, all integrands contain a derivative of $H(x)$, which is exponentially suppressed near those endpoints.
Therefore, nothing remains that is sensitive to the Taylor expansion in the region where it converges poorly.
Thus, the non-analyticity of $H(x)$ causes no problems for our method.

\subsubsection{Analytic approximation}
\label{sec:BHanalytic}

The result \cref{eq:regularHn} is sufficient to obtain a numerical result for $\cD^\BH$:
One simply evaluates the Hadamard finite part of $H_{m,n}$ numerically.
However, it is possible, at least for a wide class of functions $H(x)$, to find an analytic approximation to \cref{eq:defHn} that approximates $\cD^\BH$ well, leaving a cutoff-dependent residue that must be evaluated numerically.
Doing this, we are able to express $\Kdf$ almost entirely as an analytic, cutoff-in\-de\-pen\-dent expression, with small numerical cutoff-dependent corrections.

Letting $H^2(x) \equiv 1 + \widetilde H^2(x)$, we get
\begin{equation}
    H_{m,n} = \widetilde H_{m,n} + \delta_{n,0}\,\frac{1}{\pi^2}\int_0^{1/\sqrt{3}}\d z~\frac{\sqrt{1+z^2}}{z^{m}}\,,
    \label{eq:Hmn-analytic}
\end{equation}
where $\widetilde H_{m,n}$ is obtained by substituting $H(x)\to\widetilde H(x)$ in \cref{eq:defHn}, and the remaining integral can be evaluated analytically, taking the Hadamard finite part when $m\neq0$.
Thus, making the choice $f_0(0)=0$, we have
\begin{equation}
    H_{m,n} = \widetilde H_{m,n} + f_m(1/\sqrt3)\, \delta_{n,0} \,.
    \label{eq:Hmnf}
\end{equation}
The analytic approximation is then obtained by setting $\widetilde H_{m,n} = 0$, which allows everything to be expressed in terms of $f_m\equiv f_m(1/\sqrt3)$.
We may thus rewrite \cref{eq:result-BH} as follows:
\begin{align}
    \frac{\F^6}{\m^4}\Re\cD^\BH &= \Big[\tfrac{27}2f_0-\tfrac94f_2 + \Diso\Big]
        + \Delta\big[36f_0-6f_2-\tfrac{15}{16}f_4 + \Disoone\big]
        \notag\\
        &+ \Delta^2\big[\tfrac{2313}{160}f_0+\tfrac{171}{320}f_2-\tfrac{1677}{640}f_4-\tfrac{519}{1280}f_6 + \Disotwo\big]
        \notag\\
        &+ \DeltaA\bigl[-\tfrac{891}{64}f_0+\tfrac{1161}{128}f_2-\tfrac{45}{64}f_4-\tfrac9{128}f_6 + \DA\bigr]
        \notag\\
        &+ \DeltaB\big[\tfrac{81}{320}f_0+\tfrac{297}{640}f_2+\tfrac{27}{160}f_4-\tfrac{27}{640}f_6 + \DB\big]\,.
    \label{eq:analytic-BH}
\end{align}
Here, $\cD_X$ are $H$-dependent numerical corrections stemming from $\widetilde H_{m,n}$.
They are defined by the requirement that \cref{eq:analytic-BH} equals \cref{eq:result-BH}, and their values are given in \cref{eq:Dthrexp}.
In \cref{app:Hdependence}, we investigate the dependence of the $\cD_X$ upon the choice of cutoff function.
Note that $\widetilde H_{m,n}\to0$ is not a good approximation for individual $H_{m,n}$, but as the smallness of $\cD_X$ shows, it clearly works at the level of $\cD^\BH$ for our choice of $H$.

We stress that \cref{eq:analytic-BH} is \emph{not} obtained by substituting $H_{m,n}\to f_m\,\delta_{n,0}$ in \cref{eq:result-BH}, because the surface terms in \cref{eq:Hrelation} do not vanish for $\widetilde H_{m,n}$ and $f_m\,\delta_{n,0}$ separately.
In other words, $H_{m,n}\to f_m\,\delta_{n,0}$ must be applied before \cref{eq:Hrelation}.

However, looking at \cref{eq:Hrelation} for the $f_m\,\delta_{n,0}$ terms alone, we find that
\begin{equation}
    (m+1)f_{m+2}(z)+(m-2)f_m(z)=-(1+z^2)f_{m+1}'(z)\,.
    \label{eq:frelation}
\end{equation}
This allows \cref{eq:Hrelation} to be restated in a more symmetrical form as 
\begin{multline}
    H_{m,n+1}+H_{m-2,n+1} =\frac{1}{6} \bigl[(2-m)H_{m,n}-(m+1) H_{m+2,n}\bigr]\\
        -\frac16\bigg[\big[
        (2-m)f_m(z)-(m+1)f_{m+2}(z)
        \big]
    \frac{\d^n}{\d x^n}H^2(x)\bigg]_0^{1/\sqrt3}\,,
   \label{eq:better-Hrelation}
\end{multline}
but more importantly, since $f'_{m+1}(1/\sqrt{3})$ is just a rational number times $\kappa$, we can use \cref{eq:frelation} to reduce \cref{eq:analytic-BH} entirely in terms of $\kappa$ and $f_0\equiv f_0(1/\sqrt{3})=\frac43\kappa(4+3\log3)$:
\begin{multline}
    \frac{\F^6}{\m^4}\Re\cD^\BH = 
    \Big[96\kappa + 9f_0 + \Diso\Big]
        + \Delta\big[296\kappa + 24 f_0 + \Disoone\big]
        + \Delta^2\big[\tfrac{5661}{50}\kappa + \tfrac{621}{40}f_0 + \Disotwo\big]\\
        + \DeltaA\bigl[-\tfrac{1764}5\kappa + \tfrac{135}{32}f_0 + \DA\bigr]
        + \DeltaB\bigl[-\tfrac{612}{25}\kappa + \tfrac{189}{160}f_0 + \DB\bigr]\,.
   \label{eq:better-analytic-BH}
\end{multline}

\subsubsection{Direct numerical evaluation}
\label{sec:BH-num}

For a cross-check of the above results and for determining $\DuuBH$ further away from threshold, it is necessary to evaluate \cref{eq:DuuBH} numerically without expansion.
It is possible to evaluate it directly using numerical integration with a suitably chosen small $\epsilon>0$, but the singularities in the integrand can lead to poor convergence or errors.
Nevertheless, we have successfully performed this direct integration with 20 digits of accuracy, more than sufficient to reproduce \cref{eq:Dthrexp}.
In this subsection, however, we present a less numerically demanding approach, which complements and validates the direct evaluation.
Another approach, making no assumptions about the smoothness of $\cM_3$, is presented in \cref{app:MminusD}.

We note that the singularities in the integrand of \cref{eq:DuuBH} occur only where $H(x_r)=1$,%
\footnote{
    This follows from the construction of $\Kdf$, but can also be proven directly as follows.
    At the pole where $(\pp-r)^2=\m^2$,
    \begin{equation*}
        4\m^2 x_r = (P-r)^2 = (p_3+\pp-r)^2 = p_3^2 + (\pp-r)^2 + 2p_3\cdot (\pp-r) = 2\m^2 + 2\omega_{p_3}\omega_{\pp-r}\,,
    \end{equation*}
    where in the last equality we dropped the spatial part by going to the rest frame of either momentum.
    We have $\omega_{p_3}\geq m$ and $|\omega_{\pp-r}|\geq m$, so as long as $(\pp-r)$ has positive energy, this proves that $x_r\geq 1$.
    At threshold, $\omega_{\pp-r} = 2m - m$ is positive, and since it is an analytic function of the kinematics, there is a path from threshold to any other configuration, and such a path clearly does not involve a switch to the negative-energy branch.
    The same holds for $(\kk-r)^2=\m^2$.}
so we separate
\begin{equation}
    \int_r H^2(x_r) G(r;\pp,\kk)
    = \int_r \widetilde H^2(x_r)\, G(r; \pp, \kk)\,\theta(R-|\bm r|) + \int_r\, G(r;\pp,\kk)\,\theta(R-|\bm r|)\,,
    \label{eq:D-split}
\end{equation}
for some suitable cutoff $R$ such that $H(x_r) = 0$ when $|\bm r| > R$.
Here, $\theta$ is the Heaviside step function and $\widetilde H^2(x_r)\equiv H^2(x_r)-1$ is the same as in \cref{sec:BHexpand}.
The first term on the right-hand side is free from singularities and safe to evaluate numerically, while the second term, now free from $H(x_r)$, admits further simplification.
This is easier in the CMF of $\pp+\kk$, similar to the Breit frame in scattering, which we mark by $\breit$.
Setting up the $\bm r^\breit$-integration in spherical coordinates with suitably aligned axes, i.e.,
\begin{equation}
    \begin{gathered}
        \bm r^\breit = (r\sin\theta\cos\phi,r\sin\theta\sin\phi,r\cos\theta)\,,\\
        -\bpp^\breit = \bkk^\breit = (0,0,q)\,,\qquad 
        \bm P^\breit = (0,Q\sin\gamma,Q\cos\gamma)\,,
    \end{gathered}
\end{equation}
where we reuse the symbol $r$ as $|\bm r^\breit|$, we find
\begin{equation}
    G^\breit(r;\pp,\kk) = \frac{a_1 + a_2\cos\theta + a_3\sin\theta\sin\phi}{(b_1 - c\cos\theta)(b_2 + c\cos\theta)}\,,
    \label{eq:Giframe}
\end{equation}
where
\begin{equation}
    \begin{alignedat}{6}
        a_1 &= E^2-\m^2-2E^\breit\omega_r\,,\qquad&
        a_2 &= 2rQ\cos\gamma\,,\qquad&
        a_3 &= 2rQ\sin\gamma\,,\\
        b_1 &= \pp^2 -2{p_{+0}^\breit}\,\omega_r + i\epsilon\,,\qquad &
        b_2 &= \kk^2 -2{k_{+0}^\breit}\,\omega_r + i\epsilon\,,\qquad &
        c &= 2 r q\,.
    \end{alignedat}
    \label{eq:abc}
\end{equation}
\Cref{eq:Giframe} is a convenient parametrization for evaluating the first term in \cref{eq:D-split}, which must be performed in the same frame as the second in order for the integration limits to match.
More importantly it allows the angular integrals in the second term to be performed analytically, leaving
\begin{equation}
    \int_r G^\breit(r;\pp,\kk)\,\theta(R-r) = \int_0^R \frac{2\pi r^2 \d r}{2\omega_r^\breit (2\pi)^3}\: g(a_1,a_2; b_1,b_2; c)\,,
    \label{eq:G-1D}
\end{equation}
where
\begin{equation}\label{eq:g}
    g(a_1,a_2; b_1,b_2; c) = 
    \begin{cases}
        \displaystyle
        \frac{2 a_1}{b_1b_2}\,,   
            &   \text{if $c = 0$\,,}\\[1ex]
        \displaystyle\sum_{i=1,2}
        \frac{A_i}{c(b_1+b_2)}\bigl[\log(b_i+c)-\log(b_i-c)\bigr]
        \qquad
            &   \text{otherwise}\,,
    \end{cases}
\end{equation}
with
\begin{equation}
    A_1 = a_1 + \frac{a_2b_1}{c}\,,\qquad A_2 = a_1 - \frac{a_2b_2}{c}\,.
\end{equation}
There are singularities at $b_1+b_2=0$ and $b_i=\pm c$, both of which are regulated by the $+i\epsilon$ from the propagators.
Even without regulation, the singularities are integrable, except at threshold where actual divergences develop at $b_1+b_2\to 0$ and $r\to0$.
These cancel to keep $\DuuBH$ finite, but nevertheless present a numerical problem.

We have used three successful approaches to numerically evaluating \cref{eq:G-1D}, providing cross-checks against each other, against the brute-force evaluation of \cref{eq:DuuBH}, and against the semi-analytic threshold expansion in the previous sections.

One method is to keep $\epsilon$ small but finite, giving an integrand with narrow spikes rather than singularities, which is numerically manageable with the right precautions.
For sufficiently small $\epsilon$, the $\epsilon$-dependence of the result is very weak, and the $\epsilon\to 0$ limit is well approximated.
It suffers stability issues near threshold, and we find that $\sqrt{\epsilon}\lesssim\Delta$ is required.

Another method is to deform the integration path into the complex plane, using, e.g., $r= z - i\alpha z(1-z)$ for $z\in[0,R]$, where $\alpha>0$ is arbitrary.
With reasonably large $\alpha$ (but smaller than $4\m$ to avoid the $\omega_r=0$ pole at $r=i\m$), the path avoids the singularities by a wide margin, allowing $\epsilon=0$ and giving a smooth integrand that is easy to integrate.
The deformation crosses some branch cuts if near thresholds, so it is only usable for the real part, and it cannot reach the threshold limit due to the divergence at $r\to 0$.

The third method is to express the $\epsilon\to0$ limit in terms of Cauchy principal values, i.e., we replace the arguments of the logarithms in \cref{eq:g} with their absolute values.
This approach is full of subtleties, requiring careful consideration of the details in \rcite{Davies:1996gee}.
Basically, in addition to the principal-value part, we need to add the na\"ive double-pole contribution multiplied by a factor of 2.

\subsubsection{The full bull's head subtraction}

Evaluating $f_0$ in \cref{eq:better-analytic-BH}, the contributions of $\cD^\BH$ to $\Kdf$ are
\begin{subequations}
    \begin{align}
        \frac{\F^6}{\m^6}\,\Kiso 
            & \supset 
            -36\kappa(4 + \log3)
            - \Diso\,,
        \\
        \frac{\F^6}{\m^6}\,\Kisoone 
            & \supset 
            -8\kappa(53+12\log 3)
            -\Disoone\,,
        \\
        \frac{\F^6}{\m^6}\,\Kisotwo 
            & \supset 
            -\frac{27\kappa}{50}(363+115\log 3)
            -\Disotwo\,,
        \\
        \frac{\F^6}{\m^6}\,\KA 
            & \supset 
            \frac{9\kappa}{40}(1468-75\log 3)
            -\DA\,,
        \\
        \frac{\F^6}{\m^6}\,\KB 
            & \supset 
            \frac{9\kappa}{200}(404-105\log 3)
            -\DB\,.
    \end{align}
    \label{eq:fullBH}%
\end{subequations}
The numerical values of these contributions are given in \cref{tab:breakdown}.
From there, we find that $\cD_X$ are indeed quite small compared to the full bull's head contributions: With the exception of $\cD_2$, they differ by more than an order of magnitude.

It is perhaps unexpected that the analytic approximation, which effectively amounts to the invalid cutoff choice $H(x)=\theta(x-1)$, should be so accurate.
However, as seen in \cref{sec:BH-num}, the integral is dominated by the pole at small $r$, and $H(x_r)$ is very close to 1 in the vicinity of this pole.
Conversely, the region where $H(x_r)$ differs meaningfully from 1 contributes very little.
\Cref{app:Hdependence} looks more closely at this.

\subsection{OPE diagrams}
\label{sec:OPE}

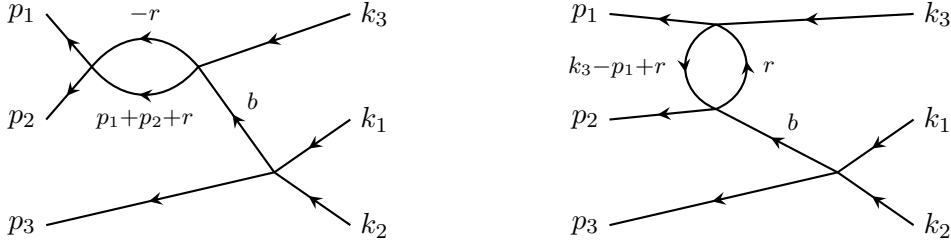
\begin{figure}[tbh]
    \centering
    \begin{subfigure}{0.48\textwidth}
    \centering
        \begin{tikzpicture}[xscale=\diagramxscale,yscale=\diagramyscale]
            \makeexternallegsshifted
            \coordinate (v1) at (+.5,-.5);
            \coordinate (v2) at (  0,+.5);
            \coordinate (v3) at (-.7,+.5);
            \draw[dprop] (k1) -- (v2) to[bend right=60, looseness=1.5] (v3) -- (p1);
            \draw[dprop] (k2) -- (v1);
            \draw[dprop] (v1) -- (v2) node [midway, above right] {\footnotesize$b$};
            \draw[dprop] (v2) to[bend left=60, looseness=1.5] (v3) -- (p2);
            \draw[dprop] (k3) -- (v1) -- (p3);
            \draw (-.35,1) node {\footnotesize $-r$};
            \draw (-.35,0) node {\footnotesize $p_1{+}p_2{+}r$};
        \end{tikzpicture}
    \end{subfigure}
    \begin{subfigure}{0.48\textwidth}
    \centering
        \begin{tikzpicture}[xscale=\diagramxscale,yscale=\diagramyscale]
            \makeexternallegsshifted
            \coordinate (v1) at (+.5,-.5);
            \coordinate (v2) at (-.3,+.1);
            \coordinate (v3) at (-.3,+.9);
            \draw[dprop] (k1) -- (v3) -- (p1);
            \draw[dprop] (k2) -- (v1);
            \draw[dprop] (v1) -- (v2) node [midway, above right] {\footnotesize$b$};
            \draw[dprop] (v2) to[bend right=60] (v3) to[bend right=60] (v2) -- (p2);
            \draw[dprop] (k3) -- (v1) -- (p3);
            \draw (+.05,.5) node {\footnotesize $r$};
            \draw (-.95,.5) node {\footnotesize $k_3{-}p_1{+}r$};
        \end{tikzpicture}
    \end{subfigure}
    \caption{
        Examples of OPE NLO diagrams.
        There are also diagrams where the loop appears on the lower right.}
    \label{fig:OPENLO}
\end{figure}

The contributions of the OPE diagrams, such as those in \cref{fig:OPENLO}, can be computed analytically by directly performing the pertinent subtractions.
Unlike in the LO case described in \cref{sec:explicitLO}, an additional challenge here is how to deal with the higher two-pion partial waves in the two-pion scattering subamplitude.
While all partial waves appear in this subamplitude, it is possible to show that only two types of contributions are relevant at quadratic order: purely $s$-wave ones (\mbox{$\ell=\ell'=0$}) and those with $d$-wave on one side (\mbox{$\ell=2$, $\ell'=0$} or vice versa).

The starting point is the unsymmetrized version of the ``master equation''~\eqref{eq:master}, which we now restrict to OPE contributions:
\begin{equation}
    \Kdf^{\uu\NLO,\OPE}(\bm p_3, \bm k_3)_{\ell' m',\ell m} 
        = \Re\Mdf^{\uu\NLO,\OPE}(\bm p_3, \bm k_3)_{\ell' m', \ell m} \,.
    \label{eq:master-uu}
\end{equation}
In turn, we need the definition of the divergence-free three-particle amplitude
\begin{multline}
    \Mdf^{\uu\NLO,\OPE}(\bm p_3, \bm k_3)_{\ell' m',\ell m}  
    \\
        = \cM_3^{\uu\NLO,\OPE}(\bm p_3, \bm k_3)_{\ell' m',\ell m}
        - \cD^{\uu\NLO,\OPE}(\bm p_3, \bm k_3)_{\ell' m',\ell m}\,,
    \label{eq:MdfuuNLOOPEdef}
\end{multline}
where $\cM_3^{\uu\NLO,\OPE}$ is given by one-particle-reducible diagrams like those shown in \cref{fig:OPENLO}.
Before projection onto pair angular momenta, we have 
\begin{align}
    \cM_3^{\uu\NLO,\OPE}
    =  -&\cM_{2,\off}^\NLO(\bar s'_2,t'_2,u'_2) \frac1{b^2-\m^2+i\epsilon} \cM_{2,\off}^\LO(\bar s_2,t_2,u_2)
    \\
    -\;&\cM_{2,\off}^\LO(\bar s'_2,t'_2,u'_2) \frac1{b^2-\m^2+i\epsilon} \cM_{2,\off}^\NLO(\bar s_2,t_2,u_2)\,,
    \label{eq:MuuNLOOPE}
\end{align}
where $b$ is the momentum of the exchanged particle and
\begin{equation}
    \begin{gathered}
        \bar s'_2\equiv(p_1+p_2)^2-4\m^2\,,
        \qquad t'_2 = (p_1-k_3)^2\,, 
        \qquad u'_2 = (p_2-k_3)^2\,,
        \\
        \bar s_2 = (k_1+k_2)^2-4\m^2\,,
        \qquad t_2=(k_1-p_3)^2\,, 
        \qquad u_2=(k_2-p_3)^2\,.
    \end{gathered}
\end{equation}
The off-shell amplitudes have a single leg off shell (that of the exchanged particle), thus
\begin{equation}
    \bar s_2+t_2+u_2 = b^2 - \m^2 \equiv \bar b^2 = \bar s'_2+t'_2 + u'_2\,.
    \label{eq:sturel}
\end{equation}
The quantities $\bar s_2$, $\bar s'_2$, and $\bar b^2$ are defined so that they vanish at threshold (as do $t_2$, $u_2$, $t'_2$ and $u'_2$), and are thus convenient in a threshold expansion.

For the subtraction term, $\cD^{\uu\NLO,\OPE}$, we must use the $\{\ell, m\}$ basis, where (keeping the indices implicit) we have
\begin{multline}
    \cD^{\uu\NLO,\OPE}(\bm p_3, \bm k_3)\\
        = -\cM_{2}^\NLO(\bm p_3) G^\infty(\bm p_3, \bm k_1) \cM_{2}^\LO(\bm k_3)
        - \cM_{2}^\LO(\bm p_3) G^\infty(\bm p_3, \bm k_1) \cM_{2}^\NLO(\bm k_3)\,,
    \label{eq:DuuNLOOPE}
\end{multline}
recalling \cref{eq:M2def,eq:Ginfdef} for relevant definitions.
The absence of the subscript ``off'' on the factors of $\cM_2$ in \cref{eq:DuuNLOOPE} indicates that these amplitudes are all evaluated on shell.

Next, we note that $\cM_2^\LO$ contains only $s$-waves and is purely real,
\begin{equation}
    \F^2 \cM_{2,\off}^\LO(s_2,t_2,u_2) = - 2 \m^2 - \bar s_2 + \bar b^2\,.
    \label{eq:M2offLO}
\end{equation}
Since the poles at $\bar b^2=0$ in $\cM_3^{\uu}$ and $\cD^{\uu}$ cancel, we are able to set $\epsilon\to 0$.
Thus, to obtain the real part of $\Mdf^\uu$, we can use the expressions above with $\cM_2^\NLO$ replaced by its real part.
Furthermore, since we are matching to the threshold expansion of $\Kdf^\uu$, keeping up to quadratic terms, we can expand $\Re\cM_2^\NLO$ about threshold and drop terms of higher-than-cubic order in the quantities that vanish at threshold.
Cubic order is required because of the $\bar b^2$ in the denominator coupled with the fact that $\cM_2^\LO$ does not vanish at threshold.
It then turns out, as shown explicitly below, that the only term that contains other than $s$-waves is that proportional to $t_2 u_2$.
Thus, we write
\begin{equation}
    \F^4 \Re \cM_{2,\off}^\NLO(\bar s_2, t_2, u_2) = 
    \F^4 \Re \cM_{2s,\off}^\NLO(\bar s_2, t_2, u_2) + e_{tu} t_2 u_2\,  + e'_{tu} \frac{(\bar b^2 - \bar s_2)}{M_\pi^2} t_2 u_2  \,,
\end{equation}
where the first term on the right-hand side contains the purely $s$-wave contributions.
We treat the purely $s$-wave and the ``$t_2 u_2$'' terms separately.
The coefficients $e_{tu}$ and $e'_{tu}$ are real by construction.

\subsubsection{Expression for \texorpdfstring{$\Re\cM_{2,\off}^\NLO$}{ReM{2,off}NLO}}

As explained in \cref{sec:off-shell}, we use the choice of off-shell two-particle amplitude given in \rcite{Bijnens:2021hpq}.
For the $I=2$ channel, the NLO amplitude is
\begin{align}
    \F^4 \cM_{2,\off}^\NLO(s,t,u) &= A^{(4)}(t_2,u_2,s_2) + A^{(4)}(u_2,s_2,t_2)\,,
\end{align}
where 
\begin{align}
    A^{(4)}(s_2,t_2,u_2) 
        &= d_1 (t_2-u_2)^2 + d_2 \m^2 s_2 + d_3 s_2^2 + d_4 \m^4 \notag\\
        &\qquad + f_1(s_2) \bar J(s_2) + f_2(s_2,t_2) \bar J(t_2) + f_2(s_2,u_2) \bar J(u_2)\,,
    \label{eq:A4NLO}
    \\
    f_1(s) &= d_5 s_2^2 + d_6 \m^2 s_2 + d_7 \m^4\,,
    \\
    f_2(s,t) &= d_8 t_2^2 + d_9 t_2 \m^2 + d_{10} s_2 \m^2 + d_{11} s_2t_2 + d_{12} \m^4\,,
\end{align}
with the function $\bar J$ defined in \cref{eq:J}.
We revert to the standard Mandelstam variable $s_2$ rather than $\bar s_2=s_2-4\m^2$ in order to simplify the comparison with \rcite{Bijnens:2021hpq}.
The constants in the above expressions are~\cite{Bijnens:2021hpq}
\begin{equation}
    \begin{gathered}
        \begin{aligned}
        d_1 &=-\tfrac5{36}\kappa - \tfrac16 L + \tfrac12 \lrII\,, 
        \\
        d_2 &= (N-\tfrac{29}9)\kappa + (N-\tfrac{11}3)L  - 8 \lrI + 2 \lrIV\,,
        \\
        d_3 &= (\tfrac{11}{12} - \tfrac12 N)\kappa + (1-\tfrac12 N) L + 2 \lrI + \tfrac12 \lrII\,,
        \\
        d_4 &= (\tfrac{20}9 - \tfrac12 N)\kappa + (\tfrac83 - \tfrac12 N) L + 
        8 \lrI + 2 \lrIII - 2 \lrIV\,,
        \end{aligned}\\
    d_5 = \tfrac12 N-1\,,\qquad d_6=(3-N)\,, \qquad d_7=\tfrac12 N-2\,,
    \\
    d_8 = \tfrac13\,,\qquad d_9=-\tfrac53\,,\qquad d_{10}=-\tfrac23\,, \qquad d_{11}=\tfrac16\,, \qquad d_{12}=\tfrac73\,,
    \end{gathered}
\end{equation}
with $N=3$, and remaining definitions are given below \cref{eq:fullnonOPE}.
Imaginary parts arise only from the $\bar J$ function.
Its real part is analytic above threshold, and the expansions that we need are given in \cref{eq:Jbarexp}.
Combining these results, we obtain
\begin{align}
    \begin{split}
        \F^4 \Re\cM_{2,\off}^\NLO(\bar s_2, t_2, u_2) 
        &= e_0 \m^4 + e_1 \m^2 \bar s_2 + e_2 \bar s_2^2 + e_3 \m^2 \bar b^2 \\&+ e_4 \bar s_2 \bar b^2 + e_5 (\bar b^2)^2 + e_{tu}  t_2u_2
        \\
        &+ e_6 \frac{\bar s_2^2 \bar b^2}{\m^2} + e_7 \frac{\bar s_2 (\bar b^2)^2}{\m^2} 
        + e_8 \frac{(\bar b^2)^3}{\m^6} + e'_{tu} \frac{(\bar b^2- \bar s_2)}{\m^2} t_2u_2\,,
        \label{eq:ReM2offNLO}
    \end{split}
\end{align}
where the constants $e_i$ and $e_{tu}$ are known in terms of the $d_i$.
Cubic terms $\bar s_2^3$ are not needed for the quadratic threshold expansion, since they give rise only to higher-order terms.
The on-shell amplitude is obtained by setting $\bar b\to0$.

\subsubsection{Decomposition of $t_2u_2$}
\label{sec:tudecomp}

As noted above, only the $t_2u_2$ term in \cref{eq:ReM2offNLO} contains nonzero angular momenta in the pair CMF.
To show this explicitly, we consider the final-state pair (with momenta $p_1$ and $p_2$) for which
\begin{align}
    t'_2 u'_2 &= \tfrac14 (\bar s'_2-\bar b^2)^2 -4 (\bm a^{\,*}_{p}\cdot \bm k_p^*)^2 \,,
    \label{eq:tu}
\end{align}
where $\bm a^{\,*}_p$ and $\bm k_p^*$ are the three-momenta $\bm p_1$ and $\bm k_3$ boosted to the CMF of the final-state pair.
Their magnitudes are given by
\begin{equation}
    |\bm a^{\,*}_p|^2 = \qpsq = \tfrac14 \bar s'_2\,,
    \qquad
    k_p^{*2} = \frac{(s'_2 -  \bar b^2)^2}{4 s'_2 }- \m^2 \,.
\end{equation}
To pull out the $d$-wave part, we use
\begin{align}
    (\bm a^{\,*}_{p}\cdot \bm k^*)^2 &= \qpsq  k^{*2} \biggl[ 
        \frac{8\pi}{15} \sum_{m} Y^*_{2 m}(\bmh a^{\,*}_{p}) Y_{2 m}(\bmh k_p^*) 
        + \frac13 
        \biggr]\,.
        \label{eq:tudsdecomp}
\end{align}
Thus, we can separate $s$- and $d$-wave parts as $t'_2u'_2 = [t'_2 u'_2]_s + [t'_2 u'_2]_d$, where
\begin{equation}
    [t'_2 u'_2]_s = \tfrac14 (\bar s'_2-\bar b^2)^2 -\tfrac43 q_{2,p}^{*2} k_p^{*2}\,,
    \qquad
    [t'_2u'_2]_d = q_{2,p}^{\,*2}  k_p^{*2} \frac{8\pi}{15}  \sum_{m} Y^*_{2 m}(\hat a^{\,*}_{p}) Y_{2 m}(\hat k_p^*) \,.
    \label{eq:tu-sd}
\end{equation}
An analogous expression holds for $t_2 u_2$.

In the following, we will need the expansion of $k_p^{\,*2}$ about threshold, which is given by
\begin{equation}
    k_p^{\,*2} = \tfrac14 \bar s'_2 - \tfrac12 \bar b^2 
    +  \tfrac1{16} (\bar b^2)^2 
    + \dots\,,
\end{equation}
where the ellipsis contains terms of higher order that do not contribute at the order we work.
Note that the partial-wave decomposition of the term ${(\bar b^2- \bar s_2)} t_2u_2$ is analogous since $(\bar b^2- \bar s_2)$ is purely $s$-wave.

\subsubsection{$s$-wave contributions}
\label{sec:swaveOPE}

Including the $[t_2u_2]_s$ and ${(\bar b^2- \bar s_2)} [t_2u_2]_s$ terms, the $s$-wave part of the real part of the NLO amplitude becomes 
\begin{multline}
\F^4 \Re\cM_{2s,\off}^\NLO(\bar s_2, \bar b^2) =
e_0 \m^4 + e_1 \m^2 \bar s_2 + e'_2 \bar s_2^2 + e_3 \m^2 \bar b^2 
+ e'_4 \bar s_2 \bar b^2 + e'_5 (\bar b^2)^2 
\\
+ e'_6 \frac{\bar s_2^2 \bar b^2}{\m^2} + e'_7 \frac{\bar s_2 (\bar b^2)^2}{\m^2}
+ e'_8 \frac{(\bar b^2)^3}{\m^2}
+\dots\,,
\end{multline}
where 
\begin{equation}
    \begin{alignedat}{3}
        e'_2 &= e_2 + \tfrac16 e_{tu} \,,
        \qquad&
        e'_4 &= e_4 - \tfrac13 e_{tu} \,,
        \qquad&
        e'_5 &= e_5 + \tfrac14 e_{tu} \,,
        \\
        e'_6 &= e_6 + \tfrac12 e'_{tu} \,,
        \qquad&
        e'_7 &= e_7 - \tfrac1{48} e_{tu} - \tfrac7{12} e'_{tu}\,,
        \qquad&
        e'_8 &= e_8 + \tfrac14 e'_{tu}\,.
    \end{alignedat}
\end{equation}
We thus have 
\begin{align}
    \cR_{s}^\NLO(\bar s_2)&\equiv
    \F^4 \Re\cM_{2s,\on}^\NLO(\bar s_2) 
    = e_0 \m^4 + e_1 \m^2 \bar s_2 + e'_2 \bar s_2^2 + \dots\,,
    \\
    \delta_{s}^\NLO(\bar s_2, \bar b^2) &\equiv
    \F^4 \Re\cM_{2s,\off}^\NLO(\bar s_2) - \F^4 \Re\cM_{2s,\on}^\NLO(\bar s_2) 
    \notag\\
    &= \bar b^2 \bigg(e_3 \m^2 + e'_4 \bar s_2 + e'_5 \bar b^2 +
    e'_6 \frac{\bar s_2^2}{\m^2} + e'_7 \frac{\bar s_2 \bar b^2}{\m^2} 
    + e'_8 \frac{(\bar b^2)^2}{\m^2} +  \dots \bigg) \,.
\end{align}
The LO result from \cref{eq:M2offLO} gives
\begin{align}
    \cR_{s}^\LO(\bar s_2) 
        &\equiv \F^2 \cM_{2s,\on}^\LO(\bar s_2) = - 2 \m^2 - \bar s_2\,,
    \\
    \delta_{s}^\LO(\bar b^2)
        &\equiv \F^2  \cM_{2s,\off}^\LO(\bar s_2,\bar b^2) - \F^2 \cM_{2s,\on}^\LO(\bar s_2) 
        = \bar b^2\,.
\end{align}

We can now perform the required subtraction.
Since we are considering purely $s$-wave terms, the projection onto pair angular momenta is trivial, and we can work with Mandelstam variables.
Thus, the contribution of $s$-wave two-particle amplitudes is
\begin{alignat}{4}
    -&\bar b^2\F^6 \Kdf^{\uu\NLO,\OPE,s}
    \notag\\
    &\qquad=
         \cR_s^\NLO(\bar s'_2) \delta_s^\LO(\bar b^2)
        &&+ \delta_s^\NLO(\bar s'_2,\bar b^2) \cR_{s}^\LO(\bar s_2)
        &&+ \delta_s^\NLO(\bar s'_2,\bar b^2) \delta_{s}^\LO(\bar b^2)
    \notag\\
    &\qquad+ \cR_s^\LO(\bar s'_2) \delta_s^\NLO(\bar s_2,\bar b^2)
        &&+ \delta_s^\LO(\bar b^2) \cR_{s}^\NLO(\bar s_2)
        &&+ \delta_s^\LO(\bar b^2) \delta_{s}^\NLO(\bar s_2,\bar b^2)
    \,.
\end{alignat}
Substituting the results above, we obtain
\begin{multline}
    \F^6 \Kdf^{\uu\NLO,\OPE,s} 
        = g_0^s \m^4 + g_1^s \m^2(\bar s'_2+\bar s_2) 
        + g_2^s \bar s'_2 \bar s_2 + g_3^s (\bar s'_2+\bar s_2)^2
    \\
        + g_4^s \m^2 \bar b^2 + g_5^s \bar b^2 (\bar s'_2+\bar s_2)
        + g_6^s (\bar b^2)^2 +\dots\,,
    \label{eq:kdfuuNLOOPEs2}
\end{multline}
where the coefficients $g_i^s$ are known in terms of the $e_i$, and the ellipsis indicates higher-order terms that are not needed.

The final step is to convert the variables to those used in the threshold expansion for $\Kdfuu$, \cref{eq:Kdfuu}, using the results
\begin{equation}
    \bar s'_2+\bar s_2 = 9 \m^2 \Delta_3^S\,,
    \qquad
    \bar s'_2 \bar s_2 = 81 \m^4 \Delta_3' \Delta_3\,,
    \qquad
    \bar b^2 = 9 \m^2 (\Delta_3^S- \Delta - \tij{33})
    \,.
\end{equation}
In this way, we obtain the contributions to the coefficients $c_1$--$c_{10}$ in \cref{eq:Kdfuu} from the $s$-wave parts of $\cM_2$.

\subsubsection{$d$-wave contributions}
\label{sec:dwaveOPE}

First, we consider the $e_{tu}$ term, which gives a contribution of $[tu]_d$ to the NLO matrix element.
Here, we must use the $\{k,\ell,m\}$ basis for $\Kdf$ since the subtraction is given in this basis.
Specifically, the $d$-wave contribution to the subtraction of \cref{eq:DuuNLOOPE} is, after recombining with spherical harmonics,
\begin{multline}
    \Re\cD^{\uu\NLO, \OPE, d}(\{p_i\},\{k_i\})
    \\
    =- \sum_{m'}\sqrt{4\pi}Y_{2 m'}(\hat a_p^*)\Re\bigl\{ \cM_{2,\ell'=2}^\NLO(\bm p_3) \bigr\}
    \biggl(\frac{ k_p^*}{q_{2,p}^*}\biggr)^{\!2}\,
    \frac{\sqrt{4\pi} Y_{2 m'}(\hat k_p^*)}{\bar b^2}
    \,\cM_{2s}^\LO(\bm k_3)
    \;+ \leftrightarrow \,.
    \label{eq:ReDuuNLOOPEd}
\end{multline}
Here, $\leftrightarrow$ indicates the term in which the roles of the LO and NLO vertices are interchanged and the notation is as in \cref{sec:tudecomp}.
The factor of $(k_p^*/q_{2,p}^*)^2$ arises from $G^\infty$, \cref{eq:Ginfdef}.
This contribution to $\cD^{\uu}$ is to be subtracted from
\begin{multline}
    \Re\cM_3^{\uu\NLO, \OPE, d}(\{p_i\},\{k_i\})
    \\
    =- \sum_{m'} \sqrt{4\pi}Y_{2 m'}(\hat a_p^*)\Re\big\{ \cM_{2,\ell'=2,\off}^\NLO(\bm p_3) \big\}
    \,\frac{\sqrt{4\pi} Y_{2 m'}(\hat k_p^*)}{\bar b^2}
    \,\cM_{2s,\off}^\LO(\bm k_3)
    \;+ \leftrightarrow \,.
    \label{eq:ReDuuNLOOPEL}
\end{multline}
The key observation now is that, using the decomposition of the $t u$ term given by
\cref{eq:tu,eq:tudsdecomp}, 
\begin{equation}
    \cM_{2,\ell'=2,\off}^\NLO(\bm p_3) = - \frac8{15} q_{2,p}^{*2} k_p^{*2}
    \qquad \text{and}\qquad
    \cM_{2,\ell'=2}^\NLO(\bm p_3) = - \frac8{15} q_{2,p}^{*4}\,,
\end{equation}
implying that
\begin{equation}
    \cM_{2,\ell'=2}^\NLO(\bm p_3) \biggl(\frac{ k_p^*}{q_{2,p}^*}\biggr)^{\!2} 
    = \cM_{2,\ell'=2,\off}^\NLO(\bm p_3)\,.
\end{equation}
In other words, the barrier factor from $G^\infty$ converts the on-shell amplitude appearing in the subtraction term into exactly the off-shell amplitude.
Thus, the subtraction only picks out the difference between on- and off-shell values of $\F^2\cM_2^\LO$, given by $\delta_s^\LO(\bar b^2)=\bar b^2$, and simply cancels the pole.
One therefore obtains the contribution
\begin{equation}
    \F^6 \Kdf^{\uu\NLO,\OPE,d} \supset - e_{tu} \bigl( [t'_2u'_2]_d + [t_2 u_2]_d \bigr)\,.
    \label{eq:kdftuuNLOd1}
\end{equation}
The two terms on the right-hand side contain $\{\ell', \ell\}=\{2,0\}$ and $\{0,2\}$, respectively.
To convert to our standard basis, we use
\begin{equation}
    [tu]_d = tu - [tu]_s\,,
\end{equation}
with $[tu]_s$ given by \cref{eq:tu-sd}, and observe that 
$t'_2 u'_2+t_2 u_2 = 81 \m^4 Q_{tu}$ [see \cref{eq:Qs}],
so that
\begin{equation}
    \F^6 \Kdf^{\uu\NLO,\OPE,d} \supset e_{tu} 
    \bigl(- 81 \m^4 \cQ_{tu} + [t'_2u'_2]_s + [t_2 u_2]_s \bigr)\,.
    \label{eq:kdftuuNLOd2}
\end{equation}
The $[tu]_s$ terms can be expanded in powers of $\bar s'_2$, $\bar s_2$, and $\bar b^2$, and lead to additional contributions of the form of \cref{eq:kdfuuNLOOPEs2}, with $g_i^s \to g_i^d$.
These can then be converted to the standard variables of the threshold expansion as explained above and thus contribute to $c_1-c_{10}$ in \cref{eq:Kdfuu}.

Next, we consider the $e'_{tu}$ term, which has the form $(\bar b^2-\bar s_2) [tu]_d$.
The analysis for the $\bar s_2 [tu]_d$ part is the same as for $[tu]_d$ alone, leading to
\begin{equation}
\F^6 \Kdf^{\uu\NLO,\OPE,d} \supset - e'_{tu} \frac{\bar s_2}{\m^2}
\bigl(- 81 \m^4\cQ_{tu} + [t'_2u'_2]_s + [t_2 u_2]_s \bigr)\,.
\label{eq:kdftuuNLOd3}
\end{equation}
Recalling that the subtraction has already been done, we note that all terms are of too high order to contribute.

Finally, we consider the $\bar b^2 [tu]_d$ part of the NLO amplitude.
Since this vanishes on shell, there is no subtraction term and we easily find
\begin{equation}
\F^6 \Kdf^{\uu\NLO,\OPE,d} 
\supset 2 e'_{tu} \bigl( [t'_2u'_2]_d + [t_2 u_2]_d \bigr) + \dots \,,
\label{eq:kdftuuNLOd}
\end{equation}
where the overall factor of $2$ comes from the value of the LO amplitude at threshold.
The remainder of the analysis is as for the $[tu]_d$ term above, except that $e_{tu} \to 2 e'_{tu}$.

\subsubsection{The full OPE contribution}

Combining $s$- and $d$-wave contributions computed in \cref{sec:swaveOPE,sec:dwaveOPE}, we end up with expressions for $c_1$ through $c_{10}$ and $c_{14}$ in terms of the coefficients $d_i$; there are no contributions to $c_{11}$, $c_{12}$, and $c_{13}$.
We can now symmetrize using \cref{eq:Kiso,eq:Kisoone,eq:Kisotwo,eq:KA,eq:KB}.
We find that the contributions of the OPE diagrams at NLO are
\begin{subequations}
    \begin{align}
        \frac{\F^6}{\m^6}\,\Kiso 
            & \supset 
            25 \kappa + 78 L - 72(8 \lrI + 6 \lrII + \lrIII - 2\lrIV) \,,
        \\
        \frac{\F^6}{\m^6}\,\Kisoone 
            & \supset 
            \tfrac{6831}{20} \kappa + 372 L -18(74 \lrI + 67 \lrII - 14 \lrIV) \,,
        \\
        \frac{\F^6}{\m^6}\,\Kisotwo 
            & \supset
            \tfrac{230481}{280} \kappa + 576 L -108(10 \lrI + 11 \lrII) \,,
        \\
        \frac{\F^6}{\m^6}\,\KA 
            & \supset
            - \tfrac{53199}{560} \kappa + 45 L + \tfrac{27}{2}(14 \lrI - 17 \lrII) \,,
        \\
        \frac{\F^6}{\m^6}\,\KB 
            & \supset
            \tfrac{54171}{140} \kappa + 216 L - 324(2\lrI + \lrII) \,.
    \end{align}
    \label{eq:fullOPE}%
\end{subequations}
For comparison, the LO results from the OPE diagram in \cref{fig:FeynmanLO:OPE} are
\begin{equation}
    \frac{\F^4}{\m^4}\,\Kiso \supset 36 \,,
    \qquad
    \frac{\F^4}{\m^4}\,\Kisoone \supset  63 \,.
    \label{eq:OPE-LO}
\end{equation}
Like \cref{eq:nonOPE-LO}, this is specific to our off-shell convention (see \cref{sec:off-shell}).

\section{Conclusions and outlook}
\label{sec:conclusions}

This work presents the NLO ChPT result for the isospin-3 three-particle $K$-matrix, $\Kdf$, which parametrizes three-particle interactions in the RFT three-particle finite-volume formalism~\cite{Hansen:2014eka,Hansen:2015zga}.
In particular, we have focused on the leading five terms in the threshold expansion.
To determine $\Kdf$, we have used the three-pion amplitude calculated in \rcite{Bijnens:2021hpq} combined with the relation between this amplitude and $\Kdf$ derived in \rcite{Hansen:2015zga}.
The main results of this work are summarized in \cref{sec:expressions} and, in particular, in \cref{eq:results}.

Various simplifications play an important role in obtaining these results.
The first is the result of \cref{eq:master} that, at NLO in ChPT, $\Kdf$ is simply given by $\Re\cM^\NLO_\text{df,3}$, rather than requiring the solution of integral equations.
The second simplification is that, although various contributions to $\cM^\NLO_\text{df,3}$ can be singular at threshold, these singularities are absent in the real part of the total result.
This allows us to obtain analytic results for the threshold expansion for almost all parts, the exception being the cutoff-dependent parts of the integrals appearing in the BH subtraction.
The latter turn out to be numerically small.

One of the motivations for this work was to address the substantial discrepancy between lattice results for $\Kdf$ and the LO ChPT prediction~\cite{Blanton:2019vdk,Fischer:2020jzp,Blanton:2021llb}.
 Focusing on the results for the first two terms in the expansion of $\Kdf$, namely $\Kiso$ and $\Kisoone$, we find that the NLO corrections are able to resolve the large disagreement between lattice QCD and LO ChPT, thus increasing confidence on the extractions of $\Kdf$ from lattice calculations.
We observe, however, that NLO effects are somewhat large in these two quantities.
Regarding the term in the threshold expansion of $\Kdf$ that couples to $d$-waves, $\KB$, we find a sign disagreement between the lattice QCD result and NLO ChPT result.
While we do not have a definitive answer, we stress that the NLO ChPT contribution is the leading effect for $\KB$.
Potentially, NNLO effects could be large and account for the discrepancy.

Since we are using an expansion of $\Kdf$ about threshold, it is important to verify its convergence.
This is necessary since at NLO in ChPT, $\Kdf$ has contributions to all orders in the threshold expansion due to the presence of loop integrals.
To do so, in \cref{sec:validity}, we address the validity of the truncation of $\Kdf$ at quadratic order.
We find that for a pion mass of $M_\pi=340$\;MeV (the heaviest used in \rcite{Blanton:2021llb}), the corrections beyond quadratic order account only for a 20\% of the total at the $5\pi$ inelastic threshold.
The corrections are even smaller for lighter masses.
Similarly, two-pion partial waves with $\ell>2$ in OPE diagrams, which do not enter the threshold expansion at quadratic order, only add a negligible contribution to the full $\Kdf$.
We conclude that truncating the threshold expansion at quadratic order provides a good approximation to $\Kdf$.

Another timely question that we address is the cutoff dependence of $\Kdf$.
All lattice QCD calculations using the RFT formalism have adopted the same choice of cutoff function.
However, this choice is not unique.
In \cref{app:Hdependence}, we discuss how the NLO ChPT result varies for different cutoffs.
Overall, we find that for a wide set of cutoff functions, the dependence is small, provided that the function does not drop to zero very rapidly below the two-pion threshold.

The results of this work can be extended to other systems, higher orders, or other EFTs.
For instance, in preparation for future lattice QCD calculations, $\Kdf$ for other three-pion isospin channels could be derived since \rcite{Bijnens:2021hpq} provides results for the six-pion amplitude for general isospin.
A potential issue is that, due to the presence of resonances in two-particle subchannels and in the three-particle channel itself, the convergence of ChPT might be poor.
Examples include $\sigma$ and $\rho$ in two-particle and $\omega$ and $h_1$ in three-particle channels.
$\Kdf$ could also be computed for other systems of mixed mesons at maximal isospin, such as $\pi^+\pi^+ K^+$ and $K^+K^+\pi^+$.
In this case, the full amplitude is not available yet, and the results from \rcite{Bijnens:2021hpq} would need to be extended to SU(3) ChPT.
This is a very compelling follow-up in light of the recent lattice QCD results for such systems~\cite{Draper:2023boj} and the observed tension with the LO ChPT prediction.

As shown by this work, the combination of EFTs and lattice QCD continues to be a potent tool for studying the hadron spectrum.
This synergy has already yielded valuable insights into the three-hadron problem and will certainly keep contributing in the future.

\subsection*{Acknowledgments}

The work of JBB was supported by the Spanish MU grant FPU19/04326. Additionally, JBB received support from the European project H2020-MSCA-ITN-2019//860881-HIDDeN and the staff exchange grant 101086085-ASYMMETRY, and from the Spanish AEI project PID2020-113644GB-I00/AEI/10.13039/501100011033.
The work of FRL was supported in part by the U.S. Department of Energy (USDOE), Office of Science, Office of Nuclear Physics, under grant Contract Numbers DE-SC0011090 and DE-SC0021006. FRL also acknowledges financial support by the Mauricio and Carlota Botton Fellowship.
The work of JB, TH and MS was supported by the Swedish Research Council grants contract numbers 2016-05996 and 2019-03779.
TH also acknowledges support from Charles University Research Center (UNCE/SCI/013), Czech Republic.
The work of SRS was supported in part by the USDOE grant No.~DE-SC0011637.

JBB and FRL would like to thank the Physics Department at the University of Washington for its hospitality during a visit in which this work was initiated.

\appendix

\section{Dependence on the cutoff}
\label{app:Hdependence}

The cutoff function $H(x)$ is arbitrary so long as it is smooth and satisfies $H(x)=0$ if $x\leq 0$ and $H(x)=1$ if $x\geq 1$.
Throughout this paper, we have been using the standard version
\begin{equation}
    H(x) = \exp\Bigl[-\tfrac1x\, \exp\bigl(-\tfrac{1}{1-x}\bigr)\Bigr]\,,   \quad\text{when}\;\; 0 < x < 1\,,
    \label{eq:H-standard}
\end{equation}
but \rcite{Briceno:2017tce} and others consider a generalization thereof, corresponding to the replacement
\begin{equation}
    x \to 1+\frac4{3-\alpha}(x-1)\,,\qquad
    -1\leq \alpha < 3\,,
    \label{eq:alpha}
\end{equation}
with $\alpha=-1$ corresponding to~\cref{eq:H-standard}.
Larger $\alpha$ give sharper cutoffs, with the limit $\alpha\to3$ being a step function, $H(x)\to\theta(x-1)$.
There is also the symmetric version
\begin{equation}
    H(x) = \Big[1 + \exp\bigl(\tfrac{1}{x} - \tfrac{1}{1-x}\bigr)\Big]^{-1}\,,\quad\text{when}\;\; 0 < x < 1\,,
    \label{eq:H-symmetric}
\end{equation}
which commonly appears in other areas, and which is more numerically well-behaved due to the lack of nested exponentials.

\begin{figure}[thp]
    \centering
    \begin{subfigure}{0.8\textwidth}
    \begin{tikzpicture}
         \begin{axis}[
            general plot,
            width=1\textwidth,
            height=0.65625\textwidth,
            xmin=-1.4, xmax=2,
            ymin=-.3, ymax=.9, restrict y to domain=-1:1,
            xlabel={$\alpha$},
            ylabel={$\cD_X$}, ylabel shift=-1.5ex,
            legend pos=north east,
             every tick label/.append style={font=\normalsize},
            legend style={font=\normalsize},
            ylabel style = {rotate=-90}, ylabel shift=-2ex,
            x tick label style={
                /pgf/number format/.cd,
                fixed,
             fixed zerofill,
                precision=1,
                /tikz/.cd
                },
            ]
            
            \tikzset{
                symmetricH/.style={domain=-1.4:-1},
            }
            
            \addplot[zeroline] {0};
            \addplot +[black,mark=none] coordinates {(-1,-.3) (-1, .9)};
        
            \addplot[D0line, symmetricH] {1.115279179633-1.162344};
            \addplot[D1line, symmetricH] {3.460517898442-3.352888};
            \addplot[D2line, symmetricH] {2.256705230899-1.673344};
            \addplot[DAline, symmetricH] {-1.855605159609+1.974249};
            \addplot[DBline, symmetricH] {-0.04222573649282+0.08225416};
            
            \addplot[D0line, mark=none, wide legend] 
                table[x index=0, y index=1, col sep=tab] 
                {H-dependence/alpha_remainder.dat};
            \addplot[D1line, mark=none, wide legend] 
                table[x index=0, y index=2, col sep=tab] 
                {H-dependence/alpha_remainder.dat};
            \addplot[D2line, mark=none, wide legend] 
                table[x index=0, y index=3, col sep=tab] 
                {H-dependence/alpha_remainder.dat};
            \addplot[DAline, mark=none, wide legend] 
                table[x index=0, y index=4, col sep=tab] 
                {H-dependence/alpha_remainder.dat};
            \addplot[DBline, mark=none, wide legend] 
                table[x index=0, y index=5, col sep=tab] 
                {H-dependence/alpha_remainder.dat};

            \legend{,,,,,,,$\Diso$,$\Disoone$,$\Disotwo$,$\DA$,$\DB$}
        \end{axis}
    \end{tikzpicture}
    \end{subfigure}
    \caption{
        The numerical remainders $\cD_X$ as functions of the parameter $\alpha$, employing \cref{eq:H-standard} together with \cref{eq:alpha}.
        All $\cD_X$ diverge as $\alpha\to3$, while the standard $H(x)$ and the numerical values of \cref{eq:Dthrexp} are recovered at $\alpha=-1$.
        For comparison, the values obtained with the symmetryic $H$, \cref{eq:H-symmetric},  are shown to the left of $\alpha=-1$.
        The line $\cD_X=0$ corresponds to the analytic approximation in \cref{sec:BHexpand} being exact.}
    \label{fig:alpha}
\end{figure}
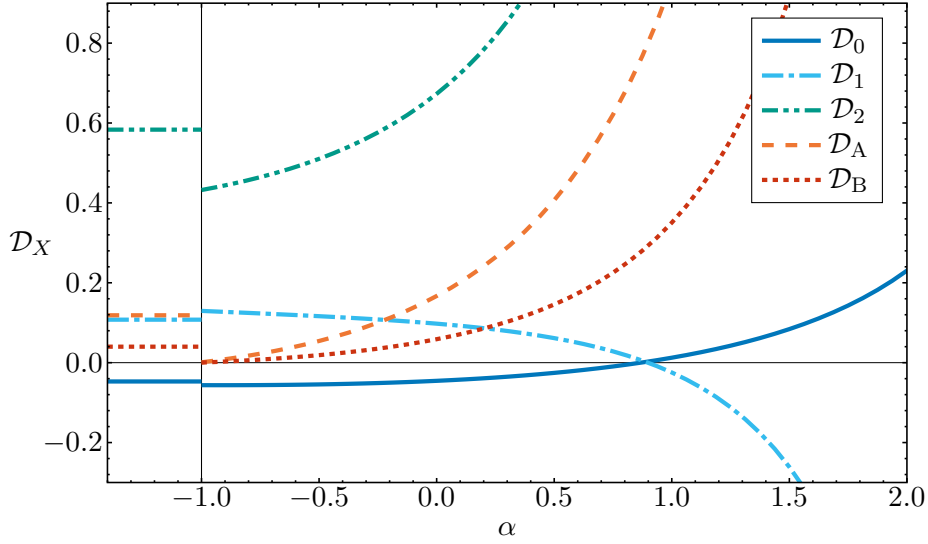

At NLO in ChPT, $\Kdf$ depends on $H(x)$ through the coefficients $\cD_X$, defined in \cref{eq:analytic-BH}.
These are effectively the remainders of $\cD^\BH$ after removing the analytic approximation obtained with $H(x)=\theta(x)$.
With the stardard cutoff choice, \cref{eq:H-standard}, we obtain the values in \cref{eq:Dthrexp}, which we restate here:
\begin{equation}
    \begin{gathered}
        \Diso       \approx -0.0563476589\,,\qquad
        \Disoone    \approx 0.129589681\,,\qquad
        \Disotwo    \approx 0.432202370\,,\\
        \DA         \approx 9.07273890\cdot10^{-4}\,,\qquad
        \DB         \approx 1.62394747\cdot10^{-4}\,.
    \end{gathered}
\end{equation}
For comparison,  with \cref{eq:H-symmetric} one instead obtains
\begin{equation}
    \begin{gathered}
        \Diso       \approx -0.0470650424\,,\qquad
        \Disoone    \approx 0.107630347\,,\qquad
        \Disotwo    \approx 0.583361673\,,\\
        \DA         \approx -0.118643915\,,\qquad
        \DB         \approx -0.0400284275\,.
    \end{gathered}
\end{equation}
In \cref{fig:alpha} we show the result for these quantities as a function of $\alpha$ in \cref{eq:alpha}.

A few features can be noted.
Using the standard $H$, \cref{eq:H-standard}, the analytic approximation is very good, in the sense that $|\cD_X|\ll|\cK_X|$.
It is especially good for $X=\mathrm A,\mathrm B$.
With larger $\alpha$, the approximation of the quadratic order in the threshold expansion ($X=2,\mathrm A,\mathrm B$) quickly grows worse, but in the leading orders ($X=0,1$) it grows better:
Near $\alpha=0.875$, $\Kiso$ and $\Kisoone$ are almost exactly approximated.
All $\cD_X$ diverge as $\alpha\to 3$, corresponding to extremely sharp cutoffs.

\section{Loop integrals}
\label{app:loops}

Let us bring up some basic definitions.
Our notation, which follows \rcite{Bijnens:2021hpq}, differs from the standard Passarino--Veltman integrals by extra factors of $\kappa=1/(16\pi^2)$.
For the bubble integrals we use
\begin{align}
\begin{split}
B(q^2)
    &=\frac{1}{i}\int\frac{\d^d\ell}{(2\pi)^d}
        \frac{1}{\left(\ell^2-\Mpi^2\right)\left[(\ell-q)^2-\Mpi^2\right]}\\
    &=\kappa\,\frac1{\tilde\epsilon}-\kappa-L+\bar J(q^2)\,.
\end{split}
\label{eq:AB}
\end{align}
Following ChPT conventions, dimensional regularization in $4-2\epsilon$ dimensions uses
\begin{equation}
    \frac{1}{\tilde\epsilon} \equiv \frac{1}{\epsilon} - \gamma_\mathrm{E} + \log4\pi - \log\mu^2 + 1\,,
\end{equation}
where $\gamma_\mathrm{E}$ is the Euler--Mascheroni constant.
We employ the standard definition for $\bar J(q^2)$:
\begin{equation}
    \bar J(q^2)
    \equiv\kappa\bigg(2+\beta\log\frac{\beta-1}{\beta+1}\bigg)\,,
    \label{eq:J}
\end{equation}
with $\beta\equiv\beta(q^2)=\sqrt{1-\frac{4\Mpi^2}{q^2}}$.
Regarding the scalar triangle integrals, we have
\begin{align}
\label{eq:C}
  C(p_1,p_2,\dots,p_6)
  &=\frac{1}{i}\int\frac{\d^d\ell}{(2\pi)^d}\frac{1}{(\ell^2-\Mpi^2)[(\ell-q_1)^2-\Mpi^2][(\ell+q_2)^2-\Mpi^2]}\,,
\end{align}
where $q_1=p_1+p_2$, $q_2=p_3+p_4$; for completeness, we also define $q_3=p_5+p_6$.
The expressions for $\cM_3^\NLO$ also include the tensor integrals $C_{11}$ and $C_{21}$ (and $C_3$, which does not contribute at $I=3$); their definitions and properties are found in appendix~A of \rcite{Bijnens:2021hpq}.
In all cases, $\overline C_X$ denotes the UV-finite part of $C_X$, although among these integrals, only $C_{21}$ is UV-divergent.

For the numerical evaluation of the triangle integrals we use \looptools~\cite{Hahn:1998yk}.
We also use the following analytic expression, which can be utilized once the tensor integrals are reduced to the scalar ones through the Passarino--Veltman reduction.
The one-loop triangle function $C$ (related to the standard $C_0$) can be written in terms of 12 dilogarithms as
\begin{multline}
    \frac1\kappa\,C(p_1,p_2,\dots,p_6)
    =C_0(q_1^2,q_2^2,q_3^2,\Mpi^2,\Mpi^2,\Mpi^2)\\
    =\sum_{\alpha_1,\alpha_2\in\{-1,1\}}
    \frac{\alpha_1}{\xi_1\xi_2}\,\operatorname{Li}_2
    \biggl\{
    \Big(1-\alpha_1\xi_2\Big)
    \bigg[
    \frac1{1-\alpha_2\beta_1\xi_2}
    +i\alpha_2\operatorname{sgn}\{q_1^2\}\epsilon
    \bigg]
    \biggr\}+\text{cycl.}\,,
\end{multline}
with $\lambda\equiv\lambda(q_1^2,q_2^2,q_3^2)>0$ the K\"all\'en triangle function, $\beta_1=\sqrt{1-\frac{4\Mpi^2}{q_1^2}}$, \mbox{$\xi_1\equiv q_1^2-q_2^2-q_3^2$}, \mbox{$\xi_2\equiv\sqrt{1-{4q_2^2q_3^2}/{\xi_1^2}}$}, and ``cycl.'' standing for cyclic permutations in $\{q_1^2,q_2^2,q_3^2\}$.

\section{Cancellation of imaginary parts}
\label{app:noimag}

To confirm that the formalism is working as it should, we would like to check that $\Kdf^\NLO$, defined by \cref{eq:inteqNLO,eq:MdfdefNLO}, is indeed real, so that it is correct to use the master equation~\eqref{eq:master}.
This should be possible at a diagram-by-diagram level, with each small set of contributions to $\cM_3^\NLO$ matched with corresponding subtraction terms in $\Mdf$ and corresponding $\rho$ terms in \cref{eq:inteqNLO}.
In the following subsections, we show that this holds as long as the contributing amplitudes, both on and off shell, satisfy unitarity.
This is the case if a consistent off-shell convention is used throughout (see \cref{sec:off-shell}).

All cancellations proven here have been verified using direct calculation at a selection of kinematic configurations detailed in \cref{app:families}.

\subsection{Bull's head diagram}
\label{app:noimagBH}

The BH diagram contributing to $\cM_3$ (\cref{fig:bullhead} and its crossings) leads to imaginary contributions due to the presence of two two-particle cuts in the $s$-channel.
Here, we address the issue of how these imaginary contributions are canceled in the transition to $\Kdf$.
This requires both the subtraction term $\cD^{\uu\NLO}$ and the $\rho$ terms in \cref{eq:inteqNLO}, and can be broken down diagrammatically in such a way that the cancellation is straightforward.

We consider the BH diagram in \cref{fig:BHNLO:regular} as a contribution to the unsymmetrized amplitude, $\cM_{3}^{\uu\NLO}(\bm p,\bm k)$, since the cancellation can be seen before symmetrization.
We start with the NLO subtraction.
The relevant part of this term is the one corresponding to the BH diagram,
\begin{align}
    \cD^{\uu\NLO}(\bm p,\bm k) &\supset \cD_{ss}^\BH(\bm p,\bm k)
    \equiv \cM^\LO_{2s}(\bm p) \int_r \left\{G^\infty_{ss}(\bm p,\bm r) \cM^\LO_{2s}(\bm r) G_{ss}^\infty(\bm r,\bm k)\right\}
    \cM^\LO_{2s} (\bm k)\,.
\end{align}
In the main text, all quantities were implicit matrices in $\ell, m$ space.
Here, since $\cM_2^\LO$ is purely $s$-wave, we can drop the $\ell, m$ indices in all quantities and leave it implicit that $\cD_{ss}^\BH$ is nonzero only for $\ell'=\ell=0$ (as indicated by the subscripts).
This means that $G^\infty_{ss}$ is different here than in \cref{eq:Gss}:
\begin{equation}
    G^\infty_{ss}(\bm p,\bm r) = \frac{H(x_p)H(x_r)}{b_{pr}^2 - \m^2 + i \epsilon}\,,\qquad
    b_{pr} = (P - p - r)^2\,.
\end{equation}
Note that in our calculation, $H(x_p)=1$ because $p$ is a momentum for which both particles in the interacting pair can be on shell.
Since $\cM_2^\LO$ is real, the imaginary part of $\cD_{ss}^\BH$ arises only from the $i\epsilon$ in $G^\infty_{ss}$, and can be pulled out using the standard Cauchy principal value,
\begin{equation}
\frac1{z+i\epsilon} = \principal \frac1z - i \pi \delta(z)\,.
\end{equation}

To proceed, we assume that $\bm p$ and $\bm k$ are chosen such that the poles in $G^\infty_{ss}(\bm p,\bm r)$ and $G^\infty_{ss}(\bm r,\bm k)$ occur for non-overlapping values of $\bm r$.
Then we get two distinct contributions to the imaginary part, one from the left-hand cut and the other from the right-hand cut.
The case where the poles overlap can be handled with a more careful application of the principal value, following \rcite{Davies:1996gee}, and we have done so as a cross-check.
However, it can be circumvented with the following argument: An infinitesimal deformation of $\bm p$ and $\bm k$ is enough to remove the overlap, so by the smoothness of $\Kdf$, if the imaginary parts cancel in the deformed case, they must also cancel without the deformation.

Thus, assuming that the poles do not overlap, we can change to the pair CMF variables $\bm r_p^*$ and $\bm r_k^*$ (in the notation of \cref{sec:rolekdf}) {\em separately} for the two delta-function contributions, and it is then straightforward to evaluate them.
We focus on the left-hand (LH) cut contribution, as all the following holds separately for the right-hand (RH) cut.

Although the explicit form of the imaginary part is not needed in the following, it is still instructive to compute it.
We find that the LH cut contribution to the imaginary part is
\begin{equation}
    i \Im\cD_{ss}^\mathrm{BH,LH}(\bm p,\bm k) =
    \cM_{2s}^\LO(\bm p) \frac{(- i \pi)}{(2\pi)^3} \frac{q_{2,p}^*}{4 E_{2,p}^*}
    \int \d\Omega(\bmh r_p^*)\,\big\{ \cM_{2s}^\LO(\bm r_{\on}) G^\infty_{ss} (\bm r_{\on},\bm k) \big\}\,
    \cM_{2s}^\LO(\bm k)\,,
\end{equation}
where $\bm r_{\on}$ is the on-shell projection of $\bm r$ obtained using the prescription of \rcite{Hansen:2014eka}.
The integral in this equation runs over the directions of $\bm r_p^*$, and both $\cM_2^\LO(\bm r_{\on})$ and $G^\infty_{ss}(\bm r_{\on},\bm k)$ depend on this direction.
The no-overlap assumption implies that the latter quantity is real, i.e., we do not have simultaneous contributions from both delta functions.
We now observe that
\begin{equation}
    \frac{q_{2,p}^*}{E_{2,p}^*} = 16 i \pi  \rho(\bm p)\,,
\end{equation}
so the contribution from the LH cut has the form of an $\cM_2\rho\Kdf$ term, namely
\begin{equation}
    i \Im \cD_{ss}^\mathrm{BH,LH}(\bm p,\bm k) =
    2 \cM_{2s}^\LO(\bm p) \rho(\bm p)
    \int \frac{\d\Omega(\bmh r_p^*)}{4\pi} \left\{ \cM_{2s}^\LO(\bm r_{\on}) G^\infty_{ss} (\bm r_{\on}, \bm k) \right\}
    \cM_{2s}^\LO(\bm k)\,.
    \label{eq:DBHLH}
\end{equation}

Next, we consider the following unsymmetrized term appearing in \cref{eq:inteqNLO}, which has the same $\cM_2\rho\Kdf$ form,
\begin{equation}
    \frac13 \cM_{2s}^\LO(\bm p)\,\rho(\bm p)\,\Mdf^\LO(\bm p,\bm k)\,.
    \label{eq:M2rhoKdf}
\end{equation}
Again, this is purely $s$-wave (since $\cM_2^\LO$ and $\Mdf^\LO$ are), so we just need to consider the $\ell=\ell'=0$ part.
Only a subset of the terms in $\Mdf^\LO$ contribute to the cancellation of the imaginary part of the BH diagram.
These are all contained in the OPE contribution
\begin{equation}
    \Mdf^{\LO,\OPE}(\bm p,\bm k) = - \cS\bigg\{
    \cM_{2,\off}^\LO(\bm p) \frac1{b_{pk}^2-\m^2+i\epsilon} \cM_{2,\off}^\LO(\bm k)
    -
    \cM_{2s}^\LO(\bm p) G^\infty_{ss}(\bm p,\bm k) \cM_{2s}^\LO(\bm k) \bigg\}\,.
    \label{eq:MdfLOOPE}
\end{equation} 
Here, as in the main text, $b_{pk} = P-p-k$ and the subscript ``off" indicates that the $b_{pk}$ leg is off shell.
Note it is important that a consistent off-shell convention is used throughout the calculation.

We now note that the initial-state symmetrization in \cref{eq:MdfLOOPE} (i.e., that over $k_1$, $k_2$, $k_3$) will be repeated when the $\cM_2\rho\Kdf$ term is symmetrized.
Thus, one can drop the initial-state symmetrization in \cref{eq:MdfLOOPE} and remove the factor of $1/3$ in \cref{eq:M2rhoKdf}.
The final-state symmetrization then yields three terms,
\begin{align}
    \Mdf^{\LO,\OPE}(\bm p,\bm k) &\supset \Mdf^{\uu\LO}(\bm p,\bm k) + \Mdf^{\uu\LO}(\bm a_p,\bm k) + \Mdf^{\uu\LO}(\bm b_p,\bm k)\,,
    \label{eq:MdfLOOPE-sym}
\end{align}
where $-\Mdf^{\uu\LO}(\bm p,\bm k)$ is the expression in braces in \cref{eq:MdfLOOPE} and $\bm a_p,\bm b_p$ are the momenta of the final-state interacting-pair particles.
Note that while the symmetrized $\Mdf^{\LO,\OPE}$ is a function of the total center-of-mass energy alone, the individual terms are not.
The first term, $\Mdf^{\uu\LO}(\bm p,\bm k)$, will contribute to the cancellation of the imaginary part of the NLO OPE diagram (to be discussed in the following subsection), while the other two contributions, $\Mdf^{\uu\LO}(\bm a_p,\bm k)$ and $\Mdf^{\uu\LO}(\bm b_p,\bm k)$, are needed for the cancellation of the imaginary part of the BH diagram.

In fact, since $\cM_2^\LO(\bm p)$ is purely $s$-wave, there is an implicit angular integral over the first argument in both these BH contributions, arising from the projection onto $\ell'=0$.
It then follows from $\bm a^*_p = -\bm b^*_p$ that $\Mdf^{\uu\LO}(\bm a_p,\bm k)=\Mdf^{\uu\LO}(\bm b_p,\bm k)$.
We thus keep only one of these two terms and multiply by a factor of 2.
This leads to the final contribution to the unsymmetrized left-hand cut part of $[\cM_2 \rho \Kdf]_\BH$:
\begin{align}
    - &2 \cM_{2s}^\LO(\bm p) \rho(\bm p) \int \frac{\d\Omega(\bmh a_p^*)}{4\pi}
    \cM_{2,\off}^\LO(\bm a_{p,\on}) \frac1{b_{kr}^2-\m^2+i\epsilon} \cM_{2,\off}^\LO(\bm k)
    \notag\\
    +\;&2 \cM_{2s}^\LO(\bm p) \rho(\bm p) \int \frac{\d\Omega(\bmh a_p^*)}{4\pi}
    \cM_{2s}^\LO(\bm a_{p,\on}) G^\infty_{ss}(\bm a_{p,\on},\bm k) \cM_{2s}^\LO(\bm k)\,.
    \label{eq:finalform}
\end{align}
This is purely imaginary, as the pole in the $b_{kr}$ propagator is not crossed, given our assumptions about $p$ and $k$.
Since $\cD$ enters with a minus sign, we see that the second term in \cref{eq:finalform} exactly cancels the imaginary part of $\cD^\mathrm{BH,LH}(\bm p,\bm k)$ given in \cref{eq:DBHLH}.
This leaves the first term in \cref{eq:finalform}, which itself exactly cancels the LH cut contribution to the imaginary part of the full BH diagram in the amplitude $\cM_3$ (using the cutting rules).

Exactly analogous arguments hold for the RH cut part, in which one must use the $\Kdf \rho \cM_2$ term from the relation between $\cM_3$ and $\Kdf$.
Thus, altogether, we have seen how the imaginary parts must cancel in the full BH contributions to $\Kdf$.

\subsection{OPE diagrams}

We now consider the OPE diagrams in which the initial interaction is of NLO in ChPT and the final one of LO.
The arguments are identical for the ``flipped'' time ordering.

Since we are here interested in the imaginary part, several simplifications occur.
By construction, the OPE pole is canceled in $\Mdf$, so the only source of a imaginary part is $\cM_2^\NLO$, and this is present only in the $s$-wave, so only the $ss$ part of $G^\infty$ contributes.
The contribution to the unsymmetrized $\Mdf$ is thus given by
\begin{align}
    i\Im\Mdf^{\uu\NLO}(\bm p,\bm k)
    \supset
    -i\Im&\big[\cM_{2,\off}^\NLO(\bm p)\big] \frac1{b_{pk}^2 - \m^2 + i \epsilon} \cM_{2,\off}^\LO(\bm k)
    \notag\\
    +\;i\Im&\big[\cM_{2s}^\NLO(\bm p)\big] G^\infty_{ss}(\bm p,\bm k) \cM_{2s}^\LO(\bm k)
    \,,
    \label{eq:Imone}
\end{align}
where, as usual, $\bm p$ and $\bm k$ are final and initial spectator momenta, respectively, and ``off" indicates that the $b_{pk}$ leg is off shell.
As above, here we are using the notation without implicit $\ell m$ indices.
To this must be added the contribution from the $\cM_2 \rho\Kdf$ term in \cref{eq:inteqNLO}, in which the OPE part of $\Kdf^\LO=\Mdf^\LO$, i.e., the first term of \cref{eq:MdfLOOPE-sym}, is included.
This contribution, which is purely imaginary, is
\begin{align}
    - &\cM_{2s}^\LO(\bm p) \rho(\bm p) \cM_{2,\off}^\LO(\bm p) \frac1{b_{pk}^2-\m^2+i\epsilon} \cM_{2,\off}^\LO(\bm k)
    \notag\\
    +\;&\cM_{2s}^\LO(\bm p) \rho(\bm p) \cM_{2s}^\LO(\bm p) G^\infty_{ss}(p,k) \cM_{2s}^\LO(\bm k)\,.
    \label{eq:Imtwo}
\end{align}
Now, we use unitarity and cutting rules to obtain
\begin{equation}
    i \Im\cM_{2,\off}^\NLO(\bm p) = - \cM_{2s}^\LO(\bm p) \rho(\bm p) \cM_{2,\off}^\LO(\bm p)\,,
\end{equation}
which applies also for the on-shell amplitude.
Using this, we find that the sum of \cref{eq:Imone,eq:Imtwo} vanishes.

\subsection{Remaining diagrams}

The remaining diagrams with an imaginary part involve a LO six-point vertex and an $s$-channel loop closed by a LO four-point vertex, either in the initial or final state.
An example is shown in \cref{fig:nosubNLO}.
These diagrams are divergence-free by themselves.
Thus, the imaginary parts must be canceled by $\cM_2 \rho \Kdf$-like terms appearing in the six-point vertex contribution to $\Kdf$.
That this is the case follows from unitarity, which introduces a factor of $-\rho$, and the double symmetrization, which cancels the $1/3$.
No off-shell amplitudes appear in the imaginary parts, so the cancellation is independent of the off-shell convention.

\section{Threshold expansion using single-parameter kinematic configurations}
\label{app:families}

In this appendix, we explain a method that we use to cross-check several of the calculations presented in the main text, and also for plotting the numerical behavior of the full NLO contribution in \cref{sec:validity}.
In particular, it provides an alternative analytic approach for obtaining the contributions of $A_J$, $A_\pi$, $A_L$, and $A_l$ to $\Mdf^{\NLO,\nonOPE}$, which are discussed in \cref{sec:threxpand-M3}.
It has also been use to perform a numerical check of all other contributions to $\Kdf$.

As described in the main text, $\Kdf$ is a function of eight kinematic degrees of freedom, so it is not straightforward to explore its general momentum dependence.
Near threshold, however, its behavior is characterized by a few parameters, five if we work to quadratic order in $\Delta$ [see \cref{eq:Kdfthrexp}].
In order to determine these coefficients from a given contribution to $\Kdf$, one approach is to use families of momenta, each of which is a one-dimensional projection of the full momentum dependence.
If one uses enough such families and controls the momentum dependence of $\Kdf$ to high-enough order for each family, then the coefficients $\cK_X$ of the threshold expansion can be determined.

Each family depends on a single parameter $p$ that has dimension of momentum and, by design, vanishes at threshold.
We use the five families listed in \cref{tab:families}.
While this is more than the minimum number of families needed for our applications, using this number provides redundancy and cross-checks.
Family 1 is the one used in \rcite{Bijnens:2021hpq}, while family 2 is a variant thereof, with the momenta arranged as equilateral triangles.
Families 3--5 use isosceles triangles instead.
\begin{table}[t]
    \centering
    {\renewcommand{\arraystretch}{1.3}
    \begin{tabular}{c*{4}{@{$\quad\big($}r@{, }r@{, }r@{, }r@{$\big)$}}}
        \toprule
        \multicolumn{1}{c@{$\quad\phantom{\big(}$}}{$a$}
                &    \multicolumn{4}{c}{$p_1^{(a)}(p)$}
                &    \multicolumn{4}{c}{$p_2^{(a)}(p)$}
                &    \multicolumn{4}{c}{$k_1^{(a)}(p)$}
                &    \multicolumn{4}{c}{$k_2^{(a)}(p)$}\\      
        \midrule
        1       &   $\omega_p$  &   $p$ &   0   &   0   
                &   $\omega_p$  &   $-\tfrac12p$    &   $\tfrac{\sqrt3}{2}p$ &   0   
                &   $\omega_p$  &   0   &   0   &   $-p$   
                &   $\omega_p$  &   $\tfrac{\sqrt3}{2}p$ & 0 &   $\tfrac12p$
                \\
        2       &   $\omega_p$  &   $p$ &   0   &   0   
                &   $\omega_p$  &   $-\tfrac12p$    &   $\tfrac{\sqrt3}{2}p$ &   0   
                &   $\omega_p$  &   $-p$ &   0   &   0   
                &   $\omega_p$  &   $\tfrac12p$    &   $\tfrac{\sqrt3}{2}p$ &   0
                \\
        3       &   $\omega_{2p}$  &   $2p$ &   0   &   0   
                &   $\Omega_{7}$  &   $-p$    &   $\tfrac{\sqrt3}{2}p$ &   0   
                &   $\omega_{2p}$  &   0 &   0   &   $-2p$  
                &   $\Omega_{7}$  &  $\tfrac{\sqrt3}{2}p$ &  0 &  $p$ 
                \\
        4       &   $\omega_{2p}$  &   $2p$ &   0   &   0   
                &   $\Omega_{3}$  &   $-p$    &   $\sqrt2 p$ &   0   
                &   $\omega_{2p}$  &   0 &   0   &   $-2p$  
                &   $\Omega_{3}$  &  $\sqrt2 p$ &  0 & $p$ 
                \\
        5       &  $\Omega_{5}$    &   $p$ &   $-2p$&  0
                &  $\omega_{2p}$   &   0   &   $2p$&   0
                &  $\Omega_{5}$    &   0   &   $p$ &   $2p$
                &  $\omega_{2p}$   &   0   &   0   &   $-2p$
                \\
        \bottomrule
    \end{tabular}}
    \caption[]{
        The five families used in the calculations, labeled by $a$.
        We use $\omega_p\equiv \sqrt{p^2+\m^2}$, and, for brevity, $\Omega_3\equiv \omega_{\sqrt3 p}$, $\Omega_5\equiv \omega_{\sqrt5 p}$ and $\Omega_7\equiv \omega_{\sqrt7 p/2}$.
        All momenta are on-shell with invariant mass $\m$, and in all cases, $\bm P=\bm p_1+\bm p_2+\bm p_3 = \bm k_1+\bm k_2+\bm k_3 = \bm 0$.
        For compactness, $p_3$ and $k_3$ have been omitted but are easily inferred using $\bm P=\bm 0$ and the on-shell condition.}
    \label{tab:families}
\end{table}

To use the families to determine the coefficients in the threshold expansion, we note that
\begin{equation}
    \Kdf\big(\family_a\big) = c_0^a + c_1^a p^2 + c_2^a p^4 + \cO(p^6) \,,
\end{equation}
where $a$ labels the family and $\family_a\equiv\big\{p_i^{(a)}(p),k_i^{(a)}(p)\big\}$.
The coefficients $c_i^a$ can be determined numerically or analytically.
We also need the expansions
\begin{equation}
    \begin{gathered}
        \Delta\big(\family_a\big) = d_1^a\, p^2 + d_2^a\, p^4 + \cO(p^6)\,,
        \\
        \DeltaA\big(\family_a\big) =  d_\mathrm{A}^a \,p^4 + \cO(p^6)\,,
        \qquad
        \DeltaB\big(\family_a\big) = d_\mathrm{B}^a \,p^4 + \cO(p^6)\,.
    \end{gathered}
\label{eq:Deltaexp}
\end{equation}
A closely related approach replaces the expansion in $p^2$ with one in $E^{*2}$.
We have used both and checked that the results agree.

A single family is sufficient to determine
\begin{equation}
    \Kiso = c_0^a\,, \qquad \Kisoone = c_1^a/d_1^a\,,
    \label{eq:Kisoexp}
\end{equation}
with other families providing cross-checks.
To obtain the quadratic constants, we need three families of momenta, from which we can construct the matrix
\begin{equation}
    Q = \begin{pmatrix} 
        (d_1^{a_1})^2 & d_\mathrm{A}^{a_1} & d_\mathrm{B}^{a_1} \\
        (d_1^{a_2})^2 & d_\mathrm{A}^{a_2} & d_\mathrm{B}^{a_2} \\
        (d_1^{a_3})^2 & d_\mathrm{A}^{a_3} & d_\mathrm{B}^{a_3} \\
    \end{pmatrix}.
\end{equation}
We also collect the quadratic coefficients of $\Kdf$ into a vector and subtract the $p^4$ term arising from $\Kisoone$:
\begin{equation}
    V = \begin{pmatrix}
        c_2^{a_1}-c_1^{a_1} d_2^{a_1}/d_1^{a_1} \\
        c_2^{a_2}-c_1^{a_2} d_2^{a_2}/d_1^{a_2} \\ 
        c_2^{a_3}-c_1^{a_3} d_2^{a_3}/d_1^{a_3} \\
    \end{pmatrix}.
\end{equation}
Then,
\begin{equation}
    \begin{pmatrix} 
        \Kisotwo \\ 
        \KA \\ 
        \KB
    \end{pmatrix} 
    = Q^{-1} V\,.
\end{equation}
This assumes that $Q$ is invertible, which is true for some triplets of families.
In particular, for the expansion of the matrix element, it is convenient to use the triplets of families $\{1,2,3\}$ and $\{1,2,4\}$.
For numerical cross-checks, on the other hand, the triplet $\{1,4,5\}$ turned out to be the most convenient.

\section{An integration method for less well-behaved $\cM_3$}
\label{app:MminusD}

\Cref{sec:BH-num} covers the method used to calculate $\DuuBH$ without first performing a threshold expansion.
The applicability of this calculation relies on the finiteness of $\DuuBH$, which in turn follows from the finiteness of the corresponding part of the amplitude, $\MuuBH$.
If that were not the case, one would have to regularize the divergences on both sides in a consistent way before the subtraction can take place, and it is not obvious how to do that.
The same situation also invalidates the threshold expansion of \cref{sec:BHexpand,sec:BHhadamard,sec:BHanalytic} unless the divergent parts can be isolated first.

In this appendix, we present an alternative approach, which computes the difference $\MuuBH-\DuuBH$ without explicitly dealing with the individual terms.
In the present case of NLO  scattering at maximum isospin, this is nothing but an overly complicated cross-check procedure (and, in earlier stages, a contingency in case the finiteness turned out to be wrong), but it is conceivable that when the scope is generalized, one will eventually encounter a sufficiently pathological subtraction that this approach becomes worthwhile.
Despite this more general outlook, we will present it as it would be applied to the present calculation, so that the technical details can be shown in full.

The goal is to write $\MuuBH$ as similarly as possible to $\DuuBH$, and then manipulate its expression to obtain a piece identical to $\DuuBH$ plus compensatory terms, which must then equal $\MuuBH-\DuuBH$.
Since this quantity is always divergence-free, its evaluation should be unproblematic.
To specify $\MuuBH$, which is parametrization-dependent, it turns out that the most convenient parametrization for our purposes is the 5th among the ones presented in \rcite{Bijnens:2021hpq}, for which the $\OO(N+1)/\OO(N)$ Lagrangian is
\begin{equation}
    \cL
        =\frac{\F^2}2\,\partial_\mu\Phi^\mathsf{T}\partial^\mu\Phi+\F^2\chi^\mathsf{T}\Phi\,,
    \qquad 
    \Phi
    =\Phi_5
        =\frac1{1+\frac14\frac{\pmb{\phi}^\mathsf{T}\pmb{\phi}}{\F^2}}\biggl(1-\frac14\frac{\pmb{\phi}^\mathsf{T}\pmb{\phi}}{\F^2},\,\frac{\pmb{\phi}^\mathsf{T}}F\biggr)^{\!\mathsf{T}}\,,
\end{equation}
where $\pmb\phi$ is a real vector of fields transforming linearly under the unbroken part of the symmetry group.
In this representation, the 4-point vertex for flavors $f_i$ and incoming off-shell momenta $p_i$ reads%
\footnote{
    With the momenta on-shell, this of course reduces to the parametrization-independent LO 4-particle amplitude, but the off-shell form presented here is precisely the 4-point vertex in the $\Phi_5$ parametrization.
    Note that this is a different off-shell convention than anywhere else in this paper.
    They are compatible, however, since our main convention only separates OPE from non-OPE, whereas the one presented here separates the BH contribution from the rest of the non-OPE part (and is more convenient for that purpose).}
\begin{align}
    \F^2\cM_{(5)}^\LO(p_1,f_1;p_2,f_2;p_3,f_3;p_4,f_4)
    &=\delta_{f_1f_2}\delta_{f_3f_4}\big(\Mpi^2+p_1\cdot p_2+p_3\cdot p_4\big)\notag\\
    &+\delta_{f_1f_3}\delta_{f_2f_4}\big(\Mpi^2+p_1\cdot p_3+p_2\cdot p_4\big)\notag\\
    &+\delta_{f_1f_4}\delta_{f_2f_3}\big(\Mpi^2+p_1\cdot p_4+p_2\cdot p_3\big)\,.
\end{align}

\begin{figure}[tb]
    \centering
    \begin{tikzpicture}[xscale=\diagramxscale,yscale=\diagramyscale]
        \makeexternallegcoordinates
        \draw (k1) node [right] {$p_4$};
        \draw (k2) node [right] {$p_3$};
        \draw (k3) node [right] {$p_6$};
        \draw (p1) node [left] {$p_1$};
        \draw (p2) node [left] {$p_2$};
        \draw (p3) node [left] {$p_5$};
        \coordinate (v1) at (+.5,+.5);
        \coordinate (v2) at (-.5,+.5);
        \coordinate (v3) at (  0,-.7);
        \draw[dprop] (k1) -- (v1);
        \draw[dprop] (k2) -- (v1);
        \draw[dprop] (p1) -- (v2);
        \draw[dprop] (p2) -- (v2);
        \draw[dprop] (k3) -- (v3);
        \draw[dprop] (p3) -- (v3);
        \draw[dprop] (v3) -- (v1) -- (v2) -- (v3);
        \draw (   0,+.65) node {\footnotesize$r$};
        \draw (+0.6,-.3) node {\footnotesize$r{-}p_3{-}p_4$};
        \draw (-0.6,-.3) node {\footnotesize$r{+}p_1{+}p_2$};
    \end{tikzpicture}
    \caption{
        The bull's head diagram, showing the momentum routing used in \cref{eq:MBH}.
        All momenta are ingoing, following the conventions of \rcite{Bijnens:2021hpq}; the routing of \cref{fig:BHNLO} is straightforwardly obtained through crossing.}
    \label{fig:bullhead}
\end{figure}
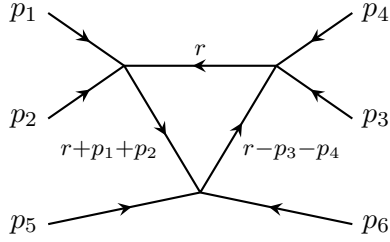

Now, we let $\MuuBH$ be precisely the contribution from the diagram \cref{fig:bullhead} in this parametrization.
Assembling the diagram and fixing all external flavors to
\begin{equation}
    |\pi^+\rangle = |\pi^-\rangle^* = \frac{|1\rangle+i|2\rangle}{\sqrt{2}}\,,
\end{equation}
we get considerable cancellation among the terms in the sum over internal flavors, leaving
\begin{multline}
    \MuuBH
        =\frac1{\F^6}\,(2p_1\cdot p_2)(2p_5\cdot p_6)\\
        \times\frac{1}{i}\int\frac{d^4 r}{(2\pi)^4}
        \frac{r^2-\Mpi^2-2r\cdot p_3}{\big[r^2-\Mpi^2\big]\big[(r+p_1+p_2)^2-\Mpi^2\big]\big[(r-p_3-p_4)^2-\Mpi^2\big]}\,.
    \label{eq:MBH}
\end{multline}
Replacing $\{p_1,\ldots,p_6\}$ by $\{k_1,k_2,k_3,-p_1,-p_2,-p_3\}$ as described above \cref{eq:6pt}, this bears a striking resemblance to \cref{eq:DuuBH}.
Indeed, comparing to \cref{eq:DuuBH-G} reveals
\begin{equation}
    \MuuBH
    = \frac{\ppkk}{\F^6}\frac1i\int\frac{d^4 r}{(2\pi)^4}
    \frac{G(r;\pp, \kk)}{r^2-\m^2+i\epsilon}\,,\qquad
    \DuuBH= -\frac{\ppkk}{\F^6}\int_r H^2(x_r)G(r;\pp,\kk)\,,
    \label{eq:MBH-G}
\end{equation}
where $\ppkk\equiv (2 p_1\cdot p_2)(2 k_1\cdot k_2)$ for brevity.

The key step is now to place the $r^0$ integral of $\MuuBH$ in the complex plane, and close the integration contour from below.
This picks up three poles, one for each of the three propagators going on-shell with positive energy (closing it above would pick up the negative-energy poles).
The residue at $r^2=\m^2$ contributes $-\Omega \F^{-6}  \int_r G(r;\pp,\kk)$; that is, it is precisely $\DuuBH$ except that there is no $H(x_r)$.
However, this absence is not straightforward to compensate for, since without a cutoff the on-shell integrals become UV-divergent.
Let us therefore look closer at $G$, whose numerator is
\begin{equation}
    (P-r)^2 - 2\m^2 = (r^2-\m^2) + (P^2 -\m^2) - 2P\cdot r\,.
    \label{eq:G-numerator}
\end{equation}
The first term on the right-hand side vanishes on-shell, while in $\MuuBH$ it cancels one propagator and gives a simple, UV-divergent $B$ integral [see \cref{eq:J}].
The second term is UV-finite.
The third is also UV-finite for the purposes of $\MuuBH$, but under an $\int_r$ integral it diverges logarithmically.
These divergences must ultimately cancel, but those cancellations are very difficult to handle numerically.
Therefore, we apply tensorial Passarino--Veltman reduction, replacing $r$ with $\pp$ and $\kk$:
\begin{equation}
    \begin{aligned}
    -2P\cdot r \quad\longrightarrow & \phantom{\:+\:} 
          \Big(\big[(\pp-r)^2-\m^2]-[r^2-\m^2]-\pp^2\Big)\xi(\pp,\kk)\\
        &+ \Big(\big[(\kk-r)^2-\m^2]-[r^2-\m^2]-\kk^2\Big)\xi(\kk,\pp)\,,
    \end{aligned}
    \label{eq:Pr-PV}
\end{equation}
where we define $\xi$ to also handle the case $\pp=\kk$:
\begin{equation}
   \xi(p,k) \equiv \begin{cases}
        \displaystyle
        \frac{p^2(P\cdot k) - (p\cdot k)(P\cdot p)}{p^2k^2 - (p\cdot k)^2}\,,  &   \text{if}\;\;p\neq k\,,\\
        \displaystyle
        \frac{P\cdot p}{2p^2}\,,  &   \text{if}\;\;p=k\,.
    \end{cases}
\end{equation}
Each term in square brackets in \cref{eq:Pr-PV} cancels a propagator in $\MuuBH$ and gives another $B$ integral.

With this in mind, we define the fully UV-safe integrand
\begin{equation}
    \widetilde G(r;p,k) \equiv \frac{P^2-\m^2 - \pp^2\xi(\pp,\kk)-\kk^2\xi(\kk,\pp)}{\big[(p-r)^2 - \m^2 + i\epsilon\big]\big[(k-r)^2 - \m^2 + i\epsilon\big]}\,
    \label{eq:Gtilde}
\end{equation}
(note that it is $p,k$ in the denominator but $\pp,\kk$ in the numerator; this will be important later).
Compensating for the modified numerator results in
\begin{multline}
    \MuuBH
    = \frac{\ppkk}{\F^6}\bigg\{
        \frac1i \int\frac{d^4 r}{(2\pi)^4}
        \frac{\widetilde G(r;\pp, \kk)}{r^2-\m^2+i\epsilon} + \big[1 - \xi(\pp,\kk) - \xi(\kk,\pp)\big]B\big((\pp-\kk)^2\big)\\
        + \xi(\pp,\kk)B(\kk^2) + \xi(\kk,\pp)B(\pp^2)\bigg\}\,,
    \label{eq:MBH-Gtilde}
\end{multline}
where it is now manifest what carries the UV divergence [namely $B\big((\pp-\kk)^2\big)$, which matches the result obtained by evaluating \cref{eq:MBH} the standard way], while
\begin{equation}
    \DuuBH
        = -\frac{\ppkk}{\F^6}\int_r H^2(x_r)\bigg\{ \widetilde G(r; \pp, \kk) 
        + \frac{\xi(\pp,\kk)}{(\kk-r)^2-\m^2+i\epsilon} + \frac{\xi(\kk,\pp)}{(\pp-r)^2-\m^2+i\epsilon}\bigg\}\,.
    \label{eq:D-Gtilde}
\end{equation}
Here, each term in the integral is individually convergent: We have moved the problematic cancellations to the extra $B$'s in \cref{eq:MBH-Gtilde}, where they are no problem at all.
The extra $\xi$ terms have at most simple poles.

Now, we apply the contour integration discussed above \cref{eq:G-numerator} to the first term in \cref{eq:MBH-Gtilde}.
The $r=\m$ residue now contributes $-\Omega \F^{-6} \int_r \widetilde G(r;\pp,\kk)$, which mostly cancels against the first term in $\DuuBH$, while the remaining residues (from $\widetilde G$) can either come from two simple poles or one double pole.
Covering both cases, their contribution is%
\footnote{
    There is no kinematic configuration that gives a triple pole; the $r=\m$ pole is always separate from the others.
    In particular, the $\pp=\kk$ version of $G_P$ is appropriate at threshold.}
\begin{equation}
    \label{eq:GP}
    G_P(r) \equiv
    \begin{cases}
        \widetilde G(r;-\kk,\pp-\kk) + \widetilde G(r;\kk-\pp,-\pp)\,, &   \text{if}\;\;\pp\neq\kk\,,\\
        \displaystyle-\widetilde G(r;-\pp,-\pp)\Big[1 + \frac{2r\cdot\pp + \pp^2 + 2\omega_r{\pp}{}_0}{2\omega_r^2}\Big]\,,  &    \text{if}\;\;\pp=\kk\,,
    \end{cases}
\end{equation}
so that 
\begin{equation}
    \frac{\ppkk}{\F^6}\int\frac{d^4 r}{(2\pi)^4} \frac{\widetilde G(r;\pp, \kk)}{r^2-\m^2+i\epsilon} - \DuuBH = -\frac{\ppkk}{\F^6}\int_r \big[G_H(r) + G_P(r)\big]\,,
    \label{eq:BH-sub-master}
\end{equation}
where we defined
\begin{align}
     G_H(r) 
        &\equiv \widetilde G(r;\pp,\kk) - H^2(x_r)G(r;\pp,\kk)\notag\\
        & = \big[1 - H^2(x_r)\big]\widetilde G(r;\pp,\kk) - \frac{H^2(x_r)\xi(\pp,\kk)}{(\kk-r)^2-\m^2+i\epsilon} - \frac{H^2(x_r)\xi(\kk,\pp)}{(\pp-r)^2-\m^2+i\epsilon}\,.
    \label{eq:GH}
\end{align}
\Cref{eq:BH-sub-master}, along with \cref{eq:MBH-Gtilde}, is the master formula for this subtraction.

The right-hand side of \cref{eq:BH-sub-master} can be dealt with by modifying the methods of \cref{sec:BH-num}.
First, we numerically evaluate
\begin{equation}
    \breitint_{r<R}\big[1 - H^2(x_r)\big]\left\{\widetilde G(r;\pp,\kk) + \frac{\xi(\pp,\kk)}{(\kk-r)^2-\m^2+i\epsilon} + \frac{\xi(\kk,\pp)}{(\pp-r)^2-\m^2+i\epsilon}\right\},
    \label{eq:GH-H}
\end{equation}
where $(\breit)$ indicates that the integral is evaluated in the Breit frame, i.e., $\breitint\widetilde G(r;p,k)$ is taken in the CMF of $p+k$.
By design, the integrand is entirely free from singularities.
Then,
\begin{multline}
    \int_r G_H(r) = \eqref{eq:GH-H}+ \breitint_{r>R} \widetilde G(r;\pp,\kk) \\
    - \breitint_{r<R}\left\{\frac{\xi(\pp,\kk)}{(\kk-r)^2-\m^2+i\epsilon} + \frac{\xi(\kk,\pp)}{(\pp-r)^2-\m^2+i\epsilon}\right\},
    \label{eq:GH-1D}
\end{multline}
where the angles can be integrated out of the remaining integrals as in \cref{eq:G-1D}.
Note, however, that $a_{1,2,3}$ are now different than in \cref{eq:abc}, while for $\xi$ we get
\begin{equation}
    \breitint_{r<R} \frac{\xi(p,k)}{(k-r)^2 - \m^2 + i\epsilon} = \int_0^R \frac{2\pi r^2\d r}{2\omega_r(2\pi)^3}\: \xi(p,k)\, g_\xi(k^2-2k_0^\breit\omega_r + i\epsilon; c)\,,
\end{equation}
where $c$ is given by \cref{eq:abc}, and
\begin{equation}
    g_\xi(b;c) \equiv
    \begin{cases}
        \displaystyle\frac2b\,,     &   \text{if}\;\;c=0\,,\\[2ex]
        \displaystyle\frac1c\bigl[\log(b-c) - \log(b+c)\bigr]\qquad    &   \text{otherwise}\,.
    \end{cases}
\end{equation}
Likewise, the angles can be integrated out of $\int_r G_P$.
Note, however, that each individual $\widetilde G$ in \cref{eq:GP} is integrated in a different frame, as described below \cref{eq:GH-H}.
With that clarified, and with \cref{eq:abc} used in each frame separately,
\begin{subequations}
    \begin{align}
        \breitint_r\widetilde G(r;p,k) &= \int \frac{2\pi r^2\d r}{2\omega_r(2\pi)^3}\,N\, g(1, 0;  b_1, b_2; c)\,,\\
        \breitint_r\widetilde G(r;-\pp,-\pp) N'
       &= \int \frac{2\pi r^2\d r}{2\omega_r(2\pi)^3}\,N\,g\bigg[1 + \frac{\pp^2+4{p_{+0}^\breit}\,\omega_r}{2\omega_r^2}, -\frac{2rq}{2\omega_r^2};  b_1, b_2; c\bigg]\,,
    \end{align}
    \label{eq:GP-1D}%
\end{subequations}
where [taken from \cref{eq:Gtilde,eq:GP}, respectively] 
\begin{equation}
    \begin{aligned}
        N   &\equiv P^2-\m^2 - \pp^2\xi(\pp,\kk)-\kk^2\xi(\kk,\pp)\,,\\
        N'  &\equiv 1 + \frac{2r\cdot\pp + \pp^2 + 2\omega_r{\pp}{}_0}{2\omega_r^2}\,.
    \end{aligned}
\end{equation}
Most of the integrands in \cref{eq:GH-1D,eq:GP-1D} have singularities, but they are, in a sense, less severe than those encountered in \cref{sec:BH-num}, and all give finite integrals at threshold (this is obvious for $g_\xi$ but quite subtle for $G_P$).
Our efforts with the numerator also ensure that the integrals to infinity can be safely done using a suitable numerical method.
Assembling all these pieces completes the subtraction.

\addcontentsline{toc}{section}{References}
\renewcommand\raggedright{}
%\bibliography{references}

\providecommand{\href}[2]{#2}\begingroup\raggedright\endgroup

\end{document}